\documentclass[floatfix,reprint,nofootinbib,amsmath,amssymb,aps,prd,
]{revtex4-1}
\usepackage{mathrsfs}
\usepackage{physics}
\usepackage{graphicx}
\usepackage{caption}
\usepackage{subcaption}
\usepackage{empheq}
\usepackage{dcolumn}
\usepackage{bm}
\usepackage{diagbox}
\usepackage{mathtools}
\usepackage[normalem]{ulem}
\usepackage{xcolor}
\definecolor{colorLink}{rgb}{0.9,0,0} 
\definecolor{colorCite}{rgb}{0,0.7,0} 
\definecolor{colorURL} {rgb}{0,0,0.8} 
\usepackage[colorlinks=true,linktocpage=true,linkcolor=colorLink,citecolor=colorCite,urlcolor=colorURL]{hyperref}
\usepackage{relsize}
\usepackage{tikz}
\usepackage{lipsum}

\let\vec\mathbf

\let\oldhat\hat
\renewcommand{\hat}[1]{\oldhat{\mathbf{#1}}}

\def\fun#1#2{\lower3.6pt\vbox{\baselineskip0pt\lineskip.9pt
        \ialign{$\mathsurround=0pt#1\hfill##\hfil$\crcr#2\crcr\sim\crcr}}}

\newcommand{\beq}{\begin{equation}}
\newcommand{\eeq}{\end{equation}}
\newcommand{\beqa}{\begin{eqnarray}}
\newcommand{\eeqa}{\end{eqnarray}}

\def\x{{\vec{x}}}
\def\u{{\vec{u}}}
\def\hx{\hat{x}}
\def\k{{\vec{k}}}
\def\p{{\vec{p}}}
\def\q{{\vec{q}}}

\def\g2{\gamma_2}
\def\dD{\delta_{\rm D}}

\def\mnras{MNRAS}

\def\aap{A \& A}

\def\jcap{JCAP}

\def\prd{Phys. Rev. D}
\def\physrep{Phys. Rep.}

\begin{document}

\title{Wide-angle and Relativistic Effects in Fourier-Space Clustering Statistics}

	\author{Milad Noorikuhani}
	\email{mn2317@nyu.edu}
	\author{Roman Scoccimarro}
	\email{rs123@nyu.edu}
	\affiliation{Center for Cosmology and Particle Physics, Department of Physics, New York University, New York, NY 10003, USA}

\begin{abstract} 
		Galaxy power spectrum and bispectrum signals are distorted by  peculiar velocities and other relativistic effects arising from a perturbed spacetime background. In addition, study of correlation functions of tracers in Fourier space is often done in the plane-parallel approximation under which it is assumed that line-of-sight (LOS) vectors are parallel. In this work we show that a simple perturbative procedure can be employed for a fast evaluation of beyond plane-parallel (wide-angle) corrections to the power spectrum  and bispectrum. We also show that evolution of linear matter density fluctuations in a relativistic context can be found from a simple method. For the power spectrum at linear level, we compare leading order wide-angle contributions to multipoles of the galaxy power spectrum with those from non-integrated and integrated relativistic corrections and estimate their possible contamination on local $f_\text{NL}$ measurements to be of order a few. We also compute wide-angle corrections in the presence of nonlinear terms at one-loop order. For the bispectrum, we show that wide-angle effects alone, even with fully symmetric choices of LOS, give rise to imaginary, odd-parity multipoles of the galaxy bispectrum (dipole, octupole, etc.) which are in many cases larger than previously known ones of relativistic origin. We calculate these contributions and provide an estimator for measuring the leading order bispectrum dipole from data, using a symmetric LOS definition. Finally, we calculate the leading order corrections to multipoles of real plane-parallel bispectrum multipoles and estimate the apparent local $f_\text{NL}$ induced to be of order unity. 
\end{abstract}
\maketitle
	
\section{Introduction}
	
	Dark matter inhomogeneities at large scales obey simple conservation laws, which can be described by  matter overdensities and peculiar velocities responding to gravity. Evolution of the matter density contrast and velocity fields at large scales has been extensively studied at both linear and weakly nonlinear regimes (see e.g. \cite{Pee80,2002PhR...367....1B, Bal20}). Biased tracers like galaxies follow the peculiar motions of dark matter leading to the addition of a redshift/blueshift term to their Hubble redshift along the line-of-sight (LOS). This results in distortion of their observed number density contrast and anisotropies in their clustering statistics  in galaxy redshift  surveys~\cite{10.1093/mnras/227.1.1, 1992ApJ...385L...5H}.  
	
	In fact, the redshift-space comoving distance of an object, $\vec{s}$, differs from that in real space, $\vec{x}$, by $\vec{s} = \vec{x} +\hat{x}\, ( \vec{v}_{o} \cdot \hat{x} )/ \mathcal{H}$ where $\vec{v}_{o}$ is the peculiar velocity of the object, $\mathcal{H}$ is the conformal Hubble parameter ($\mathcal{H} = a H$) and $\hat{x}$ denotes the LOS direction. Tracers follow the large scale bulk motion of dark matter with almost the same peculiar velocity due to the equivalence principle~\cite{2018PhR...733....1D},  ($\vec{v}_{0} \approx \vec{v}_{DM}$). Then one can write $\vec{v}_{DM} = - \mathcal{H} f \vec{u}$ and study $\nabla \cdot \vec{u}$ which is a scalar quantity that, assuming negligible vorticity, contains all the information about the velocity field and equals density perturbation $\delta$ in linearized standard perturbation theory (SPT~\cite{2002PhR...367....1B}).  In such a context, the growth rate $f \equiv d \,\text{ln} D_{+} / d\, \text{ln}a$, where $D_{+}$ represents the linear growth factor of perturbations.  

	The distortion of observed distances breaks the translational symmetry due to being dependent on the LOS direction. At linear level in SPT, assuming uniform selection function, it can be shown that the leading redshift space distortion (RSD) contribution to observed galaxy number density contrast is:
	\begin{equation}
		\label{delta-lin}
		\delta_{g, s}(\vec{x})  = \delta_{g}(\vec{x}) + f \ \hat{x}\cdot \nabla_{x} (\vec{u}_{g}\cdot \hat{x}) + ... \,,
	\end{equation}
	where subscript $``s"$ labels redshift space quantities and dots denote nonlinear/higher-order terms and other subleading corrections. The well known Kaiser multipoles \cite{10.1093/mnras/227.1.1, 1992ApJ...385L...5H} can be obtained from this equation in the so-called \emph{plane-parallel} limit where all $\hat{x}$ vectors are assumed to be parallel, i.e $\hat{x} \equiv \hat{z} \approx \text{const}$. In such a limit the linear galaxy power spectrum in redshift space becomes: $P^{\text{lin}}_{g, s}(k, \mu) = (1 + \beta \mu^{2})^{2} P^{\text{lin}}_{g}(k)$ where $P^{\text{lin}}_{g}(k)$ is the linear (isotropic) power in absence of RSD: $ \langle \delta_{g} (\vec{k}) \,\delta_{g} (\vec{k'})  \rangle = P^{\text{lin}}_{g}(k)\, \delta_{D} (\vec{k} + \vec{k'})$,  $\mu \equiv \hat{k}\cdot\hat{z}$ and $\beta \equiv f/b_{1}$ with $b_{1}$ being the linear bias coefficient. This expression can be expanded in terms of Legendre polynomials, $\mathcal{L}_{\ell}(\mu)$, giving rise to Kaiser multipoles.
	
	The analysis of RSD provides opportunities for testing cosmological and gravitational models \cite{2011MNRAS.418.1725B, 2013A&A...557A..54D, 2013MNRAS.436...89R, 2012MNRAS.426.2719R, 2014MNRAS.443.1065B, 2016MNRAS.461.3781C, 2014MNRAS.439.3504S, 2014MNRAS.440.2692S, 2020arXiv200503751B, 2020arXiv200712607Z, 2020arXiv200708994G} as they have different predictions for the growth rate of perturbations. In addition, RSD impact our cosmological inferences based on $N$-point correlation functions of galaxy overdensities from redshift surveys. These applications often use the \emph{plane-parallel}/\emph{infinite-observer} approximations (hereafter simply referred to as plane-parallel approximation) in which the line of sight (LOS) is a constant direction and next-to-leading-order inverse distance corrections to the structures observed are neglected. In this paper, we use the terms \emph{wide-angle} and \emph{inverse-distance} for those contributions beyond plane-parallel and infinite-distance limit respectively.  
	
	In addition to RSD, there are sub-leading corrections to the observed galaxy number density contrast including those with  \emph{relativistic} origins \cite{J.Yoo_2009, 2009PhRvD..80h3514Y,2010PhRvD..82h3508Y, 2011PhRvD..84d3516C, 2011PhRvD..84f3505B,2012PhRvD..85b3504J, 2014PhRvD..90b3513Y, 2014JCAP...09..037B, 2019JCAP...04..050D, 2020JCAP...09..058D} which stem from the change in redshift, light trajectories, volume elements, etc.~because of the perturbed background space-time. Taking these into account in statistical analyses can lead to constrains on relativistic parameters. Usually, inverse-distance contributions are calculated along with other relativistic corrections but the plane-parallel approximation is still considered. However, wide-angle effects can be as large as other relativistic corrections and in some cases even larger (as we will explicitly show below).  

	Wide-angle and inverse-distance contributions to two-point statistics at linear level have been extensively studied so far \cite{1998ApJ...498L...1S, 1999ApJ...516..507B, 2000ApJ...535....1M, 2004ApJ...614...51S, 2008MNRAS.389..292P, 2010MNRAS.409.1525R, 2015MNRAS.447.1789Y,2016JCAP...01..048R,  2018MNRAS.476.4403C, Castorina_2020, 2021MNRAS.503L...6S}. Also the impact of a combination of full sky (angular) or wide-angle effects with relativistic terms on galaxy/halo two-point functions have also been studied in configuration space \cite{Bertacca_2012, raccanelli2016doppler, Tansella_2018, Beutler_2020}. Furthermore, there has been recent works on the investigation of relativistic  effects (including observer terms)  \cite{Grimm.et.al.2020} and also relativistic, wide-angle and observer effects \cite{Castorina-Dio-2021, Elkhashab.et.al2021} on multipoles of galaxy power spectrum at linear level. Evaluations of  such corrections to galaxy two-point functions in Fourier space are often done numerically. It is also useful to do similar evaluations with analytic or simple relations which make it easier to compare different contributions separately and also have a quick idea about dependence of each contribution on various parameters.  In this work, we provide a procedure for obtaining such relations. 
	
	It is also well known that local primordial non-Gaussianities (PNG) can lead to scale dependent corrections to power spectrum \cite{Dalal.et.al.2007, Slosar.et.al.2008} that scale similarly with relativistic and wide-angle effects (see e.g. \cite{Bruni.et.al.2012, 2012PhRvD..85b3504J, Camera.et.al.2015, Kehagias.et.al.2015, Lorenz.et.al.2018, Contreras.et.al.2019, Bernal.et.al.2020, Wang.et.al.2020, Maartens.et.al.2021, Viljoen.et.al.2021} for  an investigation of the bias on local PNG parameter caused by ignoring relativistic corrections). Therefore,  it is useful to have a simple and fast modeling of wide-angle and relativistic effects on redshift space correlation functions as it can make cosmological inferences more accurate. This is especially motivated in the era of precision cosmology with upcoming redshift surveys covering larger patches of sky with increased precision \cite{2016arXiv161100036D, 2020arXiv200708991E, 2018LRR....21....2A, 2015arXiv150303757S, Abbott_2020, 2014arXiv1412.4872D}.
	
	It is worth noting that although angular power spectra (based on redshifts and angles of tracers) use the observables in the most direct way, the estimators in the Fourier basis are simplest from the theoretical point of view and are enough to illustrate the size of the effects we are interested in. In addition, the vast majority of measurements have so far been done in this basis, although as the surveys probe larger scales it is more natural to change to a more suitable basis.
	
	One useful approach for this purpose is to first perform perturbative expansions based on appropriate parameters that quantify wide-angle effects in configuration space and then take the Fourier transform of the result \cite{2016JCAP...01..048R,  2018MNRAS.476.4403C}. In this work, we also use a perturbative procedure but in a way that makes calculations simpler and, more importantly, easily extendable to nonlinear regimes and higher-order statistics. For the galaxy power spectrum at linear level, we utilize this procedure to find simple formulas for leading order wide-angle corrections and compare them with ``non-integrated" and integrated relativistic terms at large scales. In this work. we also provide a simple procedure for finding the evolution of dark matter density contrast and other fluctuations in a relativistic context. This procedure can be extended to other situations, e.g. including large-distance modifications of gravity. 
	
	At the nonlinear level for two-point statistics, wide-angle corrections have been investigated in the Zel'dovich approximation (first order in Lagrangian perturbation theory) \cite{2018MNRAS.479..741C, Taruya_2019}. Also nonlinear RSD effects have been studied in spherical harmonic space \cite{gebhardt2020nonlinear}. However, to the best of our knowledge, wide-angle corrections to nonlinear power spectra in SPT have not been calculated so far . The perturbative procedure to deal with wide-angle corrections used in this work can significantly simplify such calculations to any order. As an example, we keep galaxy number density contrast and velocity fields up to third order in SPT and find leading order wide-angle and inverse-distance corrections to one-loop galaxy power spectrum around BAO scales and larger, assuming $\Lambda$CDM cosmology. We also compare, approximately, the size of \emph{Fingers-of-God} (FoG) contributions with that of wide-angle effects at those scales. From these results, we draw a conclusion about the importance of wide-angle and relativistic corrections at large scales in the presence of non-linearities.  
	
	Regarding higher-order statistics, three-point correlation functions (3PCFs) in redshift space have been largely studied in the plane-parallel approximation in both Fourier (bispectra) and configuration spaces \cite{1999ApJ...517..531S, 2016JCAP...06..014T, 2017MNRAS.469.2059S, 2017PhRvD..96d3526H, 2018JCAP...12..035D, 2019MNRAS.483.2078Y, 2020MNRAS.493..594B, 2020arXiv200507066M}. Extensions to this include full sky analysis of bispectrum/3PCF in angular space and/with inclusion of relativistic corrections \cite{2018PhRvD..97b3531B, 2019JCAP...04..053D, 2019MNRAS.486L.101C, 2019arXiv190605198J, 2020JCAP...03..065M, 2020arXiv200802266D, 2020arXiv200906197J, 2020arXiv201103503G}. However, to the best of our knowledge, a full calculation of wide-angle contributions to bispectra and higher order correlations in Cartesian Fourier space have not been carried out yet. The perturbative procedure followed in this work can be easily extended to facilitate such computations. We make use of that to calculate leading order wide-angle effects on galaxy bispectrum multipoles and compare them with inverse-distance/relativistic contributions. See also a recent work, \cite{Pardede.et.al.2023}, where authors find wide-angle effects in the galaxy bispectrum using a formulation based on an expansion in spherical tensors.

	It has been realized that relativistic effects in the plane-parallel approximation can result in non-zero odd-parity multipoles such as a dipole for bispectrum \cite{2019MNRAS.486L.101C, 2019arXiv190605198J, 2020JCAP...03..065M, 2020arXiv200906197J} which are also detectable in upcoming surveys. In this paper we show that wide-angle effects alone, with symmetric LOS definitions, also lead to such multipoles and that these are as important as (and in many cases more important than) relativistic contributions and should not be neglected for constraining relativistic parameters from data. Our perturbative procedure helps us separate different contributions in an efficient way. We also provide an estimator for the measurement of leading order dipole signal from data which is based on a symmetric LOS definition. 
	
	The paper is organized as follows: Section~\ref{power-spectrum} is devoted to perturbative calculations of wide-angle and relativistic corrections to multipoles of the galaxy power spectrum in redshift space. It begins with general definitions and continues in the following subsections that are assigned to finding the relativistic evolution of fluctuations, computing relativistic and wide-angle corrections and investigating their effects on local primordial non-Gaussianity measurements. Finally, wide-angle and inverse-distance corrections in the presence of non-linearities at large scales are investigated.  In Section~\ref{bispectrum}, the perturbative procedure is extended to the  calculation of wide-angle effects on the lowest higher-order statistic, namely the galaxy bispectrum. After some general definitions, in~\ref{b-imaginary} we show how  imaginary odd-parity multipoles emerge from wide-angle effects and we compare them to relativistic effects. In addition, bispectrum dipole moment signal-to-noise ratios are estimated for upcoming surveys. In~\ref{b-radial}, leading order wide-angle and inverse-distance corrections to even-parity (real) multipoles of bispectrum are obtained and compared with each other. Finally, in Section~\ref{conclusions} we conclude with some discussions. 
	
	Throughout the paper, we adopt the following conventions for integrals in configuration and Fourier spaces: \phantom{fg f ggh gh ghgh g }
	
	\begin{eqnarray}
		\label{conv1}
		\underset{\vec{x}}{\int} \enspace \equiv \enspace \int \frac{d^{3}x}{(2 \pi)^{3}} \, , \ \ \ \ \ \ \ \ \ \ 
		\underset{\vec{k}}{\int} \enspace \equiv \enspace \int d^{3}k
	\end{eqnarray}

\section{The Power Spectrum}
\label{power-spectrum}
	
\subsection{General Strategy}

Calculation of anisotropic $N$-point correlation functions in Fourier space beyond the plane-parallel approximation can become involved due to broken translational symmetry. However, a helpful approach is to define ``local" quantities such as the local power spectrum, $P_{g}(\vec{k}, \vec{x})$, and bispectrum, $B_{g}(\vec{k}_{1}, \vec{k}_{2},\vec{k}_{3}, \vec{x} )$, which are defined at each position $\vec{x}$ (a suitable choice of LOS) by taking the Fourier transform with respect to the relative position vectors of galaxies \cite{2015PhRvD..92h3532S} (see also \cite{2016JCAP...01..048R}). Then, these local quantities can be inserted into estimators which integrate over position vectors and yield the final multipoles in Fourier space \cite{2006PASJ...58...93Y, 2015PhRvD..92h3532S}. 
	
Definition of the LOS in these local quantities is not unique. For the power spectrum it can be defined as the bisector or midpoint of two galaxy displacement vectors or can also be chosen to be aligned with one of the galaxy position vectors (end-point LOS) \cite{2011MNRAS.415.2892B, 2014MNRAS.443.1065B, 2015MNRAS.452.3704S}. The latter makes it possible to evaluate estimators via Fast Fourier transform techniques \cite{2015PhRvD..92h3532S,BiaGilRug1510}. In this work we use the midpoint for the power spectrum and centroid for the bispectrum which are symmetric under exchange. 
	
To illustrate the main ideas, consider the general formula for calculating the galaxy power spectrum multipoles from the density perturbations:

	\begin{widetext}
		\begin{equation}
		\label{window}
		P_{g}^{\ell}(k)= \frac{(2 \pi)^3}{V_{s}}(2 \ell+1)\int \frac{d \Omega_{k}}{4 \pi}\underset{\vec{x}_{1}}\int \underset{\vec{x}_{2}}\int e^{{-i\vec{k}\cdot\vec{x}_{12}}} W(\vec{x}_{1}) W(\vec{x}_{2}) \mathcal{L}_{\ell}(\hat{k}\cdot\hat{x}_{c})\left\langle \begin{matrix} \delta_{g} \left( \vec{x}_{1}\right)\delta_{g} \left( \vec{x}_{2}\right) \end{matrix} \right\rangle.
	\end{equation}
	\end{widetext}
   where $\vec{x}_{1}$ and $\vec{x}_{2}$ integrals can be taken over  an infinite volume with window functions $W(\vec{x})$ appropriately defined to force the integrals to be confined to the survey region with volume $V_{s}$. Here $\vec{x}_{c} = (\vec{x}_{1} + \vec{x}_{2})/2 $  is the midpoint of the pair which is chosen to define the LOS. This vector is measured with respect to the observer, i.e $x_{c} = 0$ at the observor's location. We have: $\vec{x}_{1} = \vec{x}_{c} + \vec{x}_{12}/2$  and $\vec{x}_{2} = \vec{x}_{c} - \vec{x}_{12}/2$. The above integral can be re-written as:  
   	\begin{widetext}
   	\begin{equation}
   		\label{window2}
   		P_{g}^{\ell}(k)= \frac{(2 \pi)^3}{V_{s}}(2 \ell+1)\int \frac{d \Omega_{k}}{4 \pi}\underset{\vec{x}_{c}}\int \mathcal{L}_{\ell}(\hat{k}\cdot\hat{x}_{c}) \underset{\vec{x}_{12}}\int e^{{-i\vec{k}\cdot\vec{x}_{12}}} W(\vec{x}_{1}) W(\vec{x}_{2}) \left\langle \begin{matrix} \delta_{g} \left( \vec{x}_{1}\right)\delta_{g} \left( \vec{x}_{2}\right) \end{matrix} \right\rangle
   	\end{equation}
   \end{widetext}
   In general, it is a tedious task to evaluate this integral including all relevant contributions. In this work we aim to provide a simple and fast way of such a calculation to especially determine the importance of different corrections and their relative sizes. This can help one to decide which contribution should be kept and which can be neglected. 
   
   The window function in Eq. \eqref{window2} enforces the limitations in volume (and its shape). Therefore, we can always take $\vec{x}_{12}$ integral over infinity and take the $\vec{x}_{c}$ integral over the survey volume. Since we mainly do the calculations for scales that are well within the survey volume, we can just set the window function to unity for those modes. In section \ref{window-geometry}, we will discuss how our results can be generalized to include effects of the window function and non-trivial survey geometries. Let us start from the definition of the ``local" power spectrum, which sets the stage for the rest of calculations in this section.  It describes the power at position $\vec{x}_{c}$ contributed by Fourier modes $\vec{k}$ and is given by~\cite{2015PhRvD..92h3532S}: 
	\begin{equation}
		\label{localp}
		P_{g,n}^{\text{loc}}(\vec{k},\vec{x}_{c})= \underset{\vec{x}_{12}}\int e^{{-i\vec{k}\cdot\vec{x}_{12}}}\left\langle \begin{matrix} \delta_{g,n} \left( \vec{x}_{1}\right)\delta_{g,n} \left( \vec{x}_{2}\right) \end{matrix} \right\rangle
	\end{equation}
	where $\delta_{g}(\vec{x})$ is the observed galaxy density contrast, measured at comoving position $\vec{x}$. The index $n$ is a label that denotes the type of contribution, e.g. RSD, relativistic, etc. Note that in the most general case, for scales comparable to the survey size, Eq.~(\ref{localp}) must be generalized to its windowed form, the innermost integral in Eq.~(\ref{window2}). We will discuss this in section \ref{window-geometry}. 
	
\begin{figure}
		\centering
		\includegraphics[width=0.18\textwidth]{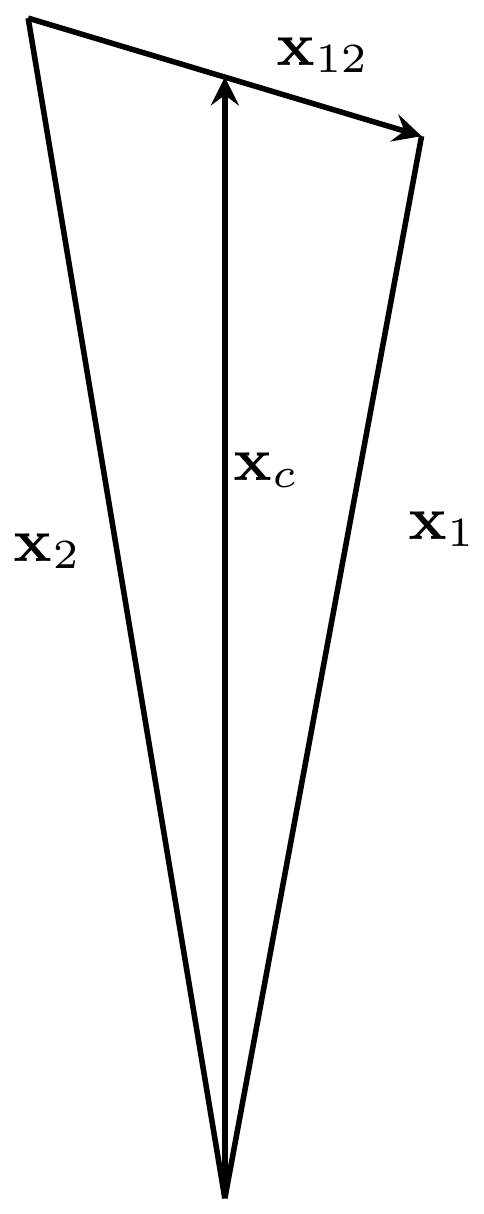}
		\caption{Relevant geometry for local power spectrum calculations. The vector $\vec{x}_{c}$ is the midpoint which we take to define the local line-of-sight and $\vec{x}_{12}$ = $\vec{x}_{1} - \vec{x}_{2}$. }
		\label{fig:midp}
	\end{figure}

	The relevant geometry under consideration is shown in Figure~\ref{fig:midp}. Since we are integrating over $\vec{x}_{12}$, the final local power spectrum is only a function of $\vec{x}_{c}$. We assume the time-dependence of fields implicitly, i.e. $\delta_{g,n}(\vec{x}) = \delta_{g,n}(\vec{x}, \tau)$, but do not write it for brevity.  

	The local power spectrum can be expanded in terms of Legendre polynomials as:
	\begin{equation}
		\label{p-expand}
		P_{g,n}^{loc}(\vec{k},\vec{x}_{c})= \sum_{\ell} \mathcal{P}^{(\ell)}_{g,n}(k, x_{c})\ \mathcal{L}_{\ell}( \hat{k}\cdot\hat{x}_{c})
	\end{equation}
	which reproduces plane-parallel multipoles in the limit $x_{c} \to \infty$  and constant $\hat{x}_{c}$.  In this equation, $\mathcal{P}^{(\ell)}_{g, n}(k, x_{c})$ can be considered  as the ``local" power spectrum multipole which is calculated as:
	\begin{equation}
		\label{estp1}
		\mathcal{P}_{g,n}^{(\ell)}(k,x_{c}) \equiv (2\ell + 1) \int \frac{d\Omega_{k}}{4\pi}\mathcal{L}_{\ell}(\hat{k}\cdot\hat{x}_{c})\, P_{g,n}^{loc}(\vec{k},\vec{x}_{c}) \,
	\end{equation}
    The full power spectrum multipoles are obtained by summing over all LOS vectors~\cite{2015PhRvD..92h3532S}:
	\begin{eqnarray}
		\label{estp2}
		P_{g,n}^{(\ell)}(k) \equiv  &&\int_{V_{s}}\frac{d^{3}\vec{x}_{c}}{V_{s}} \ \mathcal{P}^{(\ell)}_{g,n}(k, x_{c}) = \nonumber \\
		&&\int_{V_{s}} \frac{d\Omega_{s}}{V_{s}} \int_{V_{s}} x_{c}^2 d x_{c} \mathcal{P}^{(\ell)}_{g,n}(k, x_{c}),
	\end{eqnarray}
     which depends on the geometry of the survey volume $V_{s}$. 

\subsection{Galaxy density fluctuation with RSD and inverse-distance terms only}     
     
     We start from the simplest case, where we implement the mapping from real to redshift space, excluding corrections emanating from the effects of the perturbed background on light ray trajectories (observed angles and redshifts). Assuming uniform  selection function ($\bar{n}(\vec{x}) = \bar{n}(\vec{s}) $), here and throughout the paper, conservation of galaxy numbers in real and redshift spaces implies:
	\begin{equation}
		\label{conservationF}
		[1+\delta_{g,s}(\vec{s})]\,d^{3}\vec{s} = [1+\delta_{g}(\vec{x})]\,d^{3}\vec{x}
	\end{equation}
	Using the RSD map $\vec{s} = \vec{x} + \hat{x}\, ( \vec{v}_g\, .\, \hat{x} )/\mathcal{H}$ and Eq.~\eqref{conservationF}, one  finds
	
	\begin{equation}
		\label{density1}
		\delta_{g,s}(\vec{k}) =\underset{\vec{x}}{\int} e^{-i\vec{k}.\vec{x}} \{e^{-i \frac{1}{\mathcal{H}} (\vec{k}\cdot\hat{x})(\vec{v}_{g}(\vec{x})\cdot\hat{x})}[1+\delta_{g}(\vec{x})] -1 \}
	\end{equation}
	which can be used to obtain $\delta_{s}$ at any comoving position, 
	\begin{equation}
		\label{density2}
	\delta_{g,s}(\vec{x}) =\underset{\vec{x}', \vec{k}}{\int} e^{i\vec{k}.(\vec{x}-\vec{x}')} \{e^{-i\frac{1}{\mathcal{H}} (\vec{k}\cdot\hat{x}')(\vec{v}_{g}(\vec{x}')\cdot\hat{x}')}[1+\delta_{g}(\vec{x}')] -1 \}\,.
	\end{equation}
	The exponential factor inside the integral can be expanded to any desired order. Doing so, products involving Cartesian components of $\vec{k}$ will have the general form $k_{j}^{m}$ with $m$ being the power and $j$ = 1, 2, 3 representing $x$, $y$, $z$ components. These can be turned into derivatives with respect to $x_j$, $(- i\partial_j)^{m}$, which can then be pulled out of the integral. This results in:
\beqa
\label{exp22}
\delta_{g,s}& =& \delta_{g} + \nonumber \\ & & \sum_{n=1}^\infty {(-1)^n\over n!\, \mathcal{H}^n} \partial_{i_1}\ldots \partial_{i_n} \Big[ (1+\delta_g)\ \hx_{i_1}\ldots \hx_{i_n} (\vec{v} \cdot\hx)^n\Big] \nonumber \\ & & 
\eeqa
where all fields are evaluated at $\x$ and we have assumed that velocities are unbiased ($\vec{v}_g=\vec{v}$) with $\vec{v}$ being the dark matter velocity. This is valid at the large scales we are interested in. Now, when at least one  derivative acts on the unit vectors, it generates the {\em inverse-distance corrections} (absent for a fixed LOS) since 

\beq
\partial_i\, \hx_j = \frac{\delta_{ij}-\hx_i \hx_j}{|\x|},
\label{invDisGen}
\eeq
whereas when {\em all} derivatives act on the fields, they generate the usual plane-parallel result for $\delta_{g,s}(\x)$ where the fixed LOS is replaced by the true varying LOS $\hx$: expanding these type of terms around a fixed direction generates the strict plane-parallel results and their {\em wide-angle corrections}, as we discuss below.



\subsection{RSD, inverse-distance, relativistic  and wide-angle  corrections to the linear galaxy power spectrum}
\label{Linear}

At linear level, one only needs the first two terms in Eq.~\eqref{exp22} and should add to that relativistic terms that come from perturbed angles and redshifts of the received light due to background spacetime perturbations. The coefficient of the inverse-distance term should also be modified by magnification bias. Here we only consider metric perturbations caused by scalar modes in an otherwise flat metric. Therefore, for the perturbed flat FLRW metric, in conformal Newtonian gauge (CNG), we have:
	\beq
	ds^{2} =  a^2 \big[-(1 + 2\Phi)d\tau^2 + (1 - 2 \Psi)\delta_{i j}dx^i dx^j\big] \,.
	\eeq
 The linear, leading order relativistic calculation has received a fair amount of attention~\cite{J.Yoo_2009, 2009PhRvD..80h3514Y,2010PhRvD..82h3508Y, 2011PhRvD..84d3516C, 2011PhRvD..84f3505B, 2012PhRvD..85b3504J}. Here we follow in particular~\cite{2011PhRvD..84d3516C} (see their Eq.~37) and add oberver terms to that from \cite{Scaccabarozzi.et.al2018, Grimm.et.al.2020, Castorina-Dio-2021} and write:
\begin{widetext}
	\begin{eqnarray}
	\label{Relativistic}
	\delta_{g,L}(\vec{x}) &=&   {\delta}_{g} -\frac{1}{\mathcal{H}}\hat{x}_i \partial_i (\vec{v}\cdot\hat{x}) +  \left[\frac{2 - 5s}{\mathcal{H} x}  +5s - b_{e} + \frac{\dot{\mathcal{H}}}{\mathcal{H}^2} \right] \left[\Psi - \Psi_{o}+({\vec{v}_{o}} -  {\vec{v}})\cdot \hat{x} + \mathcal{H}_{0} V_{o} \right] + \frac{1}{\mathcal{H}} \dot{\Phi} +(5s-2) \Phi + \Psi   \nonumber \\
	&&  -\frac{2 - 5s}{ x} V_{o} + (2-5s ){\vec{v}_{o}} \cdot \hat{x} + \frac{(5s - 2)}{2} \int _{\tau_{0}}^{ \tau}d\tau' \frac{\tau - \tau'}{(\tau_{0} - \tau) (\tau_{0} -\tau')} \nabla^{2}_{\hat{x}'}( \Phi + \Psi) + \frac{2-5s}{x} \int _{\tau}^{ \tau_{0}}d\tau'  (\Phi + \Psi)  \nonumber \\
	&& + \left[\frac{2 - 5s}{\mathcal{H} x}  +5s - b_{e} + \frac{\dot{\mathcal{H}}}{\mathcal{H}^2} \right]  \int _{\tau}^{ \tau_{0}} d\tau' ( \dot{\Phi} + \dot{\Psi}) \,.
\end{eqnarray}
\end{widetext}
  All parameters here are in CNG . We will relate $\delta_{g}$ to the dark matter density contrast later. In this equation, terms with a subscript $o$ are evaluated at the observer's location and a dot refers to a partial derivative with respect to the conformal time. The remaining parameters in Eq.~\eqref{Relativistic} are as follows. The magnification bias is $s \equiv -(2/5) (\partial\, \text{ln}\, \bar{n}_{g}/{\partial \ln L} )|_{\bar{L}}$, with $\bar{n}_{g}$ being the background galaxy number density and ${\bar{L}}$ the threshold luminosity. The parameter $b_{e} \equiv \partial\, \text{ln} [\bar{n}_{g} a^3]/{\partial \text{ln} a}$  measures the evolution of the galaxy comoving number density (evolution bias) and the velocity potential, $V$, is related to the peculiar velocity via $\vec{v} = -\nabla{V}$. Also, the integrated terms are integrated along the LOS ($\hat{x}' = \hat{x}$)
  
  \subsubsection{Relativistic evolution of  fluctuations}
    In this section, we relate the Fourier modes of fluctuation fields appearing in Eq.~\eqref{Relativistic}, $\delta_{g}$, $\Phi$, $\vec{v}$, ...~to the dark matter density contrast, $\delta_{m}$, and its derivatives, and then find the evolution of $\delta_{m}$ in a relativistic context. While such calculations have been done before (see e.g. \cite{Yoo.Gong.2016}), the emphasis here is to find the evolution using a simple method that can be applied to more general situations.
	
	Denoting the perturbed density of cold dark matter  in the CNG by $\bar{\rho}_{m}\, (1 + \delta_{m})$, we aim to find the evolution of $\delta_{m} (\vec{k})$ starting at $\tau = \tau_{\star}$ long after the radiation-dominated era. Here we ignore contributions of  massless neutrinos and photons given the eras we are interested in. Also, since at the end we focus on power spectrum at large scales ($ k \lesssim 0.01 \text{h}/\text{Mpc}$)  we can ignore the contribution of massive neutrinos because these mostly impact smaller scales. We also neglect anisotropic stress which leads to $\Phi = \Psi$ from Einstein equations. Under these assumptions, at linear level, the continuity and Euler equations for the non-relativistic dark matter fluid lead to the following equations for fluctuations in Fourier space (in CNG):
 	\begin{eqnarray}
 	\label{EC1}
 	&& \dot{\delta}_{m}(\vec{k})  - 3\dot{\Phi}(\vec{k}) + \Theta_{v}(\vec{k})  =  0\\ \label{EC2}
 	&& \dot{\Theta}_{v}(\vec{k})  + \mathcal{H} \Theta_{v}(\vec{k}) =  k^2 \Phi(\vec{k}) 
    \end{eqnarray}	
	where $\Theta_{v}(x) = \nabla \cdot \vec{v}(x)$ with $v^{i} = dx^{i}/d\tau$  the peculiar velocity. The velocity potential at Fourier space can also be written as $V(\vec{k}) = \Theta_{v}(\vec{k}) / k^2$. Also all quantities have proper time dependence but we do not write it explicitly for  brevity. 
	
	Assuming the fluctuations are sourced by adiabatic (curvature)  perturbations, we have $\delta_{m}(\vec{k}) = \delta_{b}(\vec{k})$ where $b$ refers to baryons. We also assume that $\Theta_{v,m}(\vec{k}) = \Theta_{v,b}(\vec{k}) = \Theta_{v}(\vec{k})$ which is a good approximation at large scales. Then, Einstein equations for $G^{0}_{~0}$, $G^{0}_{~i}$ and $G^{i}_{~i}$ lead to the following relations for fluctuations in Fourier space (in units where $c = 1$): 
	\begin{widetext}
	\begin{eqnarray}
		\label{EIN1}
		&& \Phi(\vec{k}) + 3 \frac{\mathcal{H}}{k^2}\dot{\Phi}(\vec{k}) + 3 \frac{\mathcal{H}^{2}}{k^2} \Phi(\vec{k}) = -\frac{4 \pi G a^{2}}{k^2}(\bar{\rho}_m \delta_{m}(\vec{k}) +\bar{\rho}_{b} \delta_{b}(\vec{k}) ) = -\frac{3}{2} \frac{\mathcal{H}^2}{k^2}  \Omega_{M} \delta_{m}(\vec{k})  \\ \label{EIN2}
		&& \dot{\Phi}(\vec{k})  +  \mathcal{H} \Phi(\vec{k}) = 4 \pi G a^{2}(\bar{\rho}_m +\bar{\rho}_{b}) \frac{\Theta_{v}(\vec{k})}{k^2} = \frac{3}{2} \frac{\mathcal{H}^2}{k^2}  \Omega_{M} \Theta_{v}(\vec{k}) \\ \label{EIN3}
		&& \ddot{\Phi}(\vec{k}) + 3\mathcal{H} \dot{\Phi}(\vec{k})  +  (2 \dot{\mathcal{H} } + \mathcal{H}^2) {\Phi}(\vec{k})  = 0
	\end{eqnarray}	
	\end{widetext}
	where for the last equalities in Eq.~\eqref{EIN1} and Eq.~\eqref{EIN2} we have used Friedmann equations with $\Omega_{M} = \Omega_{m} + \Omega_{b}$ the total matter (dark + baryon) density parameter. Taking a time derivative of  the continuity equation, Eq.~\eqref{EC1}, and making use of a combination of Euler equation, Eq.~\eqref{EC2}, and equations above, one finds the following second order equation for the evolution of  dark matter density contrast:
	\begin{widetext}
	\begin{eqnarray}
		\label{delta-EV}
		\ddot{\delta}_{m}(\vec{k}) + 3 \mathcal{H} \dot{\Phi}(\vec{k}) + 6 \dot{\mathcal{H}} \Phi(\vec{k}) + \mathcal{H} \dot{\delta}_{m}(\vec{k}) -\frac{3}{2} \mathcal{H}^2 \Omega_{M} \delta_{m}(\vec{k} ) = 0 \,.
	\end{eqnarray}

	Now we need to express $ \dot{\Phi}(\vec{k})$ and $ {\Phi}(\vec{k})$ in terms of $ \delta_{m}(\vec{k})$, $ \dot{\delta}_{m}(\vec{k})$. To this end, we can take Eqs. \eqref{EC1}, \eqref{EIN1} and \eqref{EIN2} as a system of three equations, three unknows ($ {\Phi}(\vec{k})$ , $ \dot{\Phi}(\vec{k})$ , $\Theta(\vec{k})$) and solve for these in terms of other parameters. The result is: 

	\begin{eqnarray}
		\label{phi}
		&&{\Phi}(\vec{k}) = \frac{3 \mathcal{H}^2 \left(6  \mathcal{H}\, \Omega_{M} \dot{\delta}_{m}(\vec{k}) -  2 k^2  \Omega_{M} {\delta}_{m}(\vec{k}) + 9\mathcal{H}^2  \Omega_M^2 \delta_{m}(\vec{k})   \right)}{2(2 k^4 - 27 \mathcal{H}^4 \Omega_{M} - 9 \mathcal{H}^{2} k^{2} \Omega_{M})} \\ \label{phidot}
		&&\dot{{\Phi}}(\vec{k}) = \frac{ 3\left( -3 \mathcal{H}^4 \Omega_{M} \dot{\delta}_{m}(\vec{k})  -  \mathcal{H}^2 k^2 \Omega_{M} \dot{\delta}_{m}(\vec{k}) + \mathcal{H}^3 k^2  \Omega_M \delta_{m}(\vec{k})  \right)}{2 k^4 - 27 \mathcal{H}^4 \Omega_{M} - 9 \mathcal{H}^{2} k^{2} \Omega_{M}} \\ \label{thetav}
		&& \Theta_{v}(\vec{k}) = - \frac{ 2 k^4  \dot{\delta}_{m}(\vec{k}) - 9 \mathcal{H}^3 k^2  \Omega_{M} \delta_{m}(\vec{k})}{2 k^4 - 27 \mathcal{H}^4 \Omega_{M} - 9 \mathcal{H}^{2} k^{2} \Omega_{M}} \,.
	\end{eqnarray}
    After inserting these expressions into Eq.~\eqref{delta-EV}, one can factor out the common denominator (multiplied by $2 k^4$) to get: 
    \begin{eqnarray}
       \frac{2 k^4}{(2 k^4 - 27 \mathcal{H}^4 \Omega_{M} - 9 \mathcal{H}^{2} k^{2} \Omega_{M})} \,(\cdots) = 0 \,.
     \end{eqnarray} 
	For non-zero $k$, the only solution to this equation is $(\cdots ) = 0$. Therefore, the equation for the relativistic evolution of $\delta_{m}$ takes the following form:
		\begin{eqnarray}
			\label{DELEV}
			&& \ddot{\delta}_{m}(\vec{k}) + \dot{\delta}_{m}(\vec{k}) -\frac{3}{2} \mathcal{H}^2  \Omega_{M} \delta_{m}(\vec{k})  - \left( \frac{9 \mathcal{H}^2}{2 k^2} + \frac{27 \mathcal{H}^4}{2 k^4} \right)\Omega_{M} \ddot{\delta}_{m}(\vec{k}) - \left( \frac{9 \mathcal{H}^3}{ k^2}  + \frac{27 \mathcal{H}^5}{k^4} (1 - \frac{\dot{\mathcal{H}}}{\mathcal{H}^{2}}) \right) \Omega_{M} \dot{\delta}_{m}(\vec{k}) + \nonumber \\
			&& + \left(  \frac{9 \mathcal{H}^4}{2 k^2} (1 + \frac{3}{2} \Omega_M - 2 \frac{\dot{\mathcal{H}}}{\mathcal{H}^{2}}) +   \frac{81 \mathcal{H}^6}{4 k^4} \Omega_{M} (1 + 2 \frac{\dot{\mathcal{H}}}{\mathcal{H}^{2}} ) \right) \Omega_{M} {\delta}_{m}(\vec{k})   = 0 \,.
	   \end{eqnarray}
\end{widetext}
	
	The first three terms above constitute the well-known equation for the evolution of dark matter density fluctuation in the absence of relativistic corrections. The rest  are such corrections which are suppressed by powers of $\mathcal{H}/k$ and can be omitted for modes far below the (conformal) Hubble scale. The first piece has two solutions, $D_{+}(\tau) C_{\tau_{\star}}^{+}(\vec{k})$ (growing mode) and  $D_{-}(\tau) C_{\tau_{\star}}^{-}(\vec{k})$ (decaying mode) where $C_{\tau_{\star}}^{\pm}(\vec{k})$ are determined from initial conditions at $\tau = \tau_{\star}$ and:
	
	\begin{eqnarray}
		\label{DA}
		D_{+} (\tau)= \frac{5}{2} \Omega_{M} \mathcal{H}^3  \int _{0}^{a} \frac{d a'}{\mathcal{H}(a')^3}, \,\,\,\,\,\,\, D_{-} (\tau) = \frac{H(\tau)}{H_{0}}\,\,\,\,\,\,\,\,\,\,\,\
	\end{eqnarray}
	
	To find two solutions of Eq.~\eqref{DELEV} which includes relativistic corrections we can postulate that 
	\beqa
	\delta^{+}_{m}(\vec{k}) &=& D_{+}(\tau)\, \Big(1 + \frac{\mathcal{H}^2}{k^2} \xi (\tau)\Big)\ \mathcal{A}^{+}_{ \tau_{\star}}(\vec{k})\, , \\
	\delta^{-}_{m}(\vec{k}) &=& D_{-}(\tau)\, \Big(1 + \frac{\mathcal{H}^2}{k^2} \gamma (\tau)\Big)\ \mathcal{A}^{-}_{ \tau_{\star}}(\vec{k})\, ,
	\label{Ansatz}
	\eeqa
	which are inspired by the form of relativistic suppression factors, and plug them into Eq.~\eqref{DELEV} to find $\xi$ and $\gamma$. The quantities $\mathcal{A}^\pm_{ \tau_{\star}}(\vec{k})$ will be determined later.  After doing so and making use of 
		\begin{eqnarray}
		\label{aux1}
        &&\dot{D}_{+} =   \frac{\Omega_{M}}{2}(5 a - 3 D_{+})  \mathcal{H}\,,  \\ 
        && \dot{\mathcal{H}} = (1 - \frac{3}{2} \Omega_{M}) \mathcal{H}^2 \,,\\
        && \dot{\Omega}_{M} = 3 \mathcal{H}\,  \Omega_{M} (\Omega_{M} - 1),
	\end{eqnarray}
	the equation for growing and decaying modes take the following general forms:
	\begin{eqnarray}
		\label{xia}
		&&\frac{\mathcal{H}^2}{k^2} \Psi^{+}_{1} (\xi, \dot{\xi}, \ddot{\xi}) + \frac{\mathcal{H}^4}{k^4} \Psi^{+}_{2} (\xi, \dot{\xi}, \ddot{\xi})+\frac{\mathcal{H}^6}{k^6} \Psi^{+}_{3} (\xi, \dot{\xi}, \ddot{\xi}) = 0, \nonumber \\
		&& \frac{\mathcal{H}^2}{k^2} \Psi^{-}_{1} (\gamma, \dot{\gamma}, \ddot{\gamma}) + \frac{\mathcal{H}^4}{k^4} \Psi^{-}_{2} (\gamma, \dot{\gamma}, \ddot{\gamma})+\frac{\mathcal{H}^6}{k^6} \Psi^{-}_{3} (\gamma, \dot{\gamma}, \ddot{\gamma}) = 0 \nonumber
	\end{eqnarray}
   where the $\Psi^{+}$'s are also functions of  $D_{+}$, $\Omega_{M}$, $a$ and $\mathcal{H}$ and $\Psi^{-}$'s have the same functionality except for $D_{+}$. To find $\xi$ and $\gamma$ we can use a simple trick  inspired by the fact that any solution for $\xi$ and $\gamma$ needs to be valid at all scales in tandem with the fact that $\Psi$'s in the equations above  are suppressed at sub-Hubble scales by different powers of $\mathcal{H}/k$. This implies that one can find solutions for $\xi$ and $\gamma$ by requiring the $\Psi$'s to be separately zero for each of the equations above. For example, for the growing mode, requiring $\Psi^{+}_{m} = 0$ ($m = 1, 2, 3$),  leaves us with a system of three equations with three unknowns ($\xi$, $\dot{\xi}$, $\ddot{\xi}$) from which $\xi$ can be found. A similar procedure can be used for finding $\gamma$. One can also show that direct first and second order derivatives of $\xi$ and $\gamma$ are consistent with $\dot{\xi}$, $\ddot{\xi}$ and $\dot{\gamma}$, $\ddot{\gamma}$ respectively. The final results are:
	 \begin{eqnarray}
	 	\label{xisol}
	 	&&\xi(\tau) = \frac{3(5 a - 3 D_{+} ) \Omega_{M}}{2 D_{+}} = 3 \frac{d \ln D_{+}}{d \ln a} \equiv 3 f \\
	 	&&\gamma(\tau) = -\frac{9}{2} \Omega_{M}
	 \end{eqnarray}
	which show that unlike $\gamma$, $\xi$ is proportional to the growth rate parameter $f$, as one expects for the growing mode. Therefore, to sum the results up the full solution for the relativistic evolution of dark matter fluctuation is given by the simple expression:
		 \begin{eqnarray}
		\label{xisol-full}
     \delta_{m}(\vec{k}) &=& D_{+}(\tau)\, (1 + 3f\frac{\mathcal{H}^2}{k^2})\ \mathcal{A}^{+}_{ \tau_{\star}}(\vec{k}) \,\,\,\,+ \nonumber \\ 
     &&D_{-}(\tau)\, (1  - \frac{9}{2} \Omega_{M}\frac{\mathcal{H}^2}{k^2})\ \mathcal{A}^{-}_{ \tau_{\star}}(\vec{k}) \,.
	\end{eqnarray}
	
	Since we are interested in $\tau \gg \tau_{*}$ at which time the decaying mode can be neglected, we can set $\delta_{m} \approx \delta_{m,+}$ and define $\mathcal{A}^{+}_{ \tau_{\star}}(\vec{k})\equiv\mathcal{A}_{ \tau_{\star}}(\vec{k})$.  Finally, one can plug this result into Eqs.~\eqref{phi}-\eqref{thetav} to find the other parameters. After some algebra one gets:
	\begin{eqnarray}
		\label{deltafinal}
		&&\delta_{m}(\vec{k}) = D_{+} (1 + 3 f \frac{\mathcal{H}^2}{k^2}) \mathcal{A}_{ \tau_{\star}}(\vec{k}) \\ \label{fphi1}
		&&{\Phi}(\vec{k}) = -\frac{3}{2} \frac{\mathcal{H}^2}{k^2} D_{+} \Omega_{M} \mathcal{A}_{ \tau_{\star}}(\vec{k}) \\ \label{fphidot1}
		&& \dot{{\Phi}}(\vec{k}) = -\frac{3}{2} \frac{\mathcal{H}^3}{k^2} D_{+} \Omega_{M} (f -1) \mathcal{A}_{ \tau_{\star}}(\vec{k}) \\ \label{fthetav1}
		&& \Theta_{v}(\vec{k}) = - \mathcal{H} f  D_{+}  \mathcal{A}_{ \tau_{\star}}(\vec{k}) 
	\end{eqnarray}

    The relations above are for fluctuation modes in CNG at $\tau \gg \tau_{\star}$. However, comparison between the dark matter density contrast in CNG and its analogue in synchronous gauge gives us a better insight about $\mathcal{A}_{ \tau_{\star}}(\vec{k})$.  The density contrast and velocity divergence in CNG can be related to those in synchronous gauge at linear level via the standard gauge transformation (see e.g.~\cite{Ma_1995}):
    \begin{eqnarray}
    	\label{gauget1}
    	&&\tilde{\delta}_{m}(\vec{k}) = {\delta}_{m}(\vec{k}) -  \alpha \frac{\dot{\bar{\rho}}_{m}}{\bar{\rho}_{m}} \\ \label{gaugv}
    	&& \tilde{\Theta}_{v}(\vec{k}) = {\Theta}_{v}(\vec{k}) -  \alpha k^2
    \end{eqnarray}
where a tilde signifies synchronous gauge quantity. Now we can work in synchronous/comoving gauge (SCG) in which comoving coordinates are defined by free falling dark matter particles with $\tilde{v}^{i} = 0$. Therefore $\tilde{\Theta}_{v} = 0$ in SCG and from Eqs.~\eqref{gauget1}, \eqref{gaugv} one  finds:
\begin{eqnarray}
	\label{gauget2}
	\tilde{\delta}_{m}(\vec{k}) = {\delta}_{m}(\vec{k}) + 3 \frac{\mathcal{H}}{k^2} \Theta_{v}(\vec{k}) 
\end{eqnarray}
where we have used $\bar{\rho}_{m} \propto a^{-3}$. Now we can make use of Eqs.~\eqref{deltafinal} and \eqref{fthetav1} to obtain: 
\begin{eqnarray}
	\label{gauget3}
	\tilde{\delta}_{m}(\vec{k}) &=& D_{+} (1 + 3 f \frac{\mathcal{H}^2}{k^2}) \mathcal{A}_{ \tau_{\star}}(\vec{k})  - 3 \frac{\mathcal{H}}{k^2} ( \mathcal{H} f  D_{+}  \mathcal{A}_{ \tau_{\star}}(\vec{k}) ) \nonumber \\
	&&  = D_{+}  \mathcal{A}_{ \tau_{\star}}(\vec{k}) .
\end{eqnarray}
This result implies that one can replace $\mathcal{A}_{ \tau_{\star}}(\vec{k})$ in Eqs.~\eqref{deltafinal}-\eqref{fthetav1} by $\tilde{\delta}_{m}(\vec{k})/D_{+}$ to find:
	\begin{eqnarray}
	\label{deltafina2}
	&&\delta_{m}(\vec{k}) =  (1 + 3 f \frac{\mathcal{H}^2}{k^2})\tilde{\delta}_{m}(\vec{k}) \\ \label{fphi2}
	&&{\Phi}(\vec{k}) = -\frac{3}{2} \frac{\mathcal{H}^2}{k^2} \Omega_{M} \tilde{\delta}_{m}(\vec{k}) \\ \label{fphidot2}
	&& \dot{{\Phi}}(\vec{k}) = -\frac{3}{2} \frac{\mathcal{H}^3}{k^2} \Omega_{M} (f -1) \tilde{\delta}_{m}(\vec{k}) \\ \label{fthetav2}
	&& \Theta_{v}(\vec{k}) = - \mathcal{H} f  \tilde{\delta}_{m}(\vec{k}) \,.
\end{eqnarray}
This relates the quantities in CNG to the density fluctuation in SCG (with linear power spectrum $\simeq k^{n_s}$ as $k\to0$, where $n_s$ is the scalar spectral index) which will make calculations much easier.

The final step for calculating the observed galaxy power spectrum from Eq.~\eqref{Relativistic} is relating $\delta_{g}$ to $\delta_{m}$ via an appropriate bias relation.  In a relativistic context, galaxy density fluctuation is most easily related to the dark matter fluctuation in the SCG~\cite{2011PhRvD..84d3516C, 2012PhRvD..85b3504J}, via $\tilde{\delta}_{g} = b_{1} \tilde{\delta}_{m}$ at linear level. This bias relation is also gauge invariant \cite{2012PhRvD..85b3504J}, allowing one to find the corresponding bias relation in other gauges. The gauge transformation for the galaxy density fluctuation and velocity modes are similar to Eqs.~\eqref{gauget1} and \eqref{gaugv} except ${\dot{\bar{\rho}}_{g}}/{\bar{\rho}_{g}} = (-3 + b_{e}) \mathcal{H}$. Assuming zero velocity bias at large scales ($\Theta_{v, g} = \Theta_{v}, \tilde{\Theta}_{v, g} = 0 $) one finds  $\tilde{\delta}_{g}(\vec{k}) = {\delta}_{g}(\vec{k}) + (3-b_{e}) \mathcal{H} \Theta_{v}(\vec{k})/k^2$.  Now, using  $\tilde{\delta}_{g}(\vec{k}) = b_{1} \tilde{\delta}_{m}(\vec{k})$ and Eq.~\eqref{fthetav2} one obtains:
\begin{eqnarray}
	\label{rbias}
		{\delta}_{g}(\vec{k})  = b_{1} \tilde{{\delta}}_{m}(\vec{k}) + (3- b_{e}) \frac{\mathcal{H}^{2}}{k^2} f \tilde{\delta}_{m}(\vec{k})
\end{eqnarray}

\subsubsection{The full linear galaxy density contrast}
Using equations \eqref{deltafina2}-\eqref{rbias}, one can rewrite the linear galaxy density fluctuation in Eq.~\eqref{Relativistic} by expressing its constituents in terms of  inverse Fourier integrals. We have:
\begin{eqnarray}
	\label{deltagl}
	{\delta}_{g,L}(\vec{x})  =  \delta^{\text{RSD}}{(\vec{x})} + \delta^{\text{NI}}(\vec{x}) + \delta^{\text{I}}(\vec{x}) 
\end{eqnarray}
where we have ignored observer terms. It has been shown in \cite{Grimm.et.al.2020, Castorina-Dio-2021} that the inclusion of observer terms renders the variance of the observed galaxy density fluctuations free of IR-divergences and is also necessary for a gauge invariant expression of the observed galaxy density contrast. It is also shown that when all relativistic terms including observer contributions are included, relativistic effects in cosmological observable statistics become infrared insensitive at linear and nonlinear levels \cite{Mitsou.et.al.2023}. However, for modes well within survey volume, observer effects on Fourier space correlation functions should be negligible because these are evaluated at a single position of the observer. For modes comparable to the survey size when window function is included (see section \ref{window-geometry}), the convolution with window function and the mixing between window and correlation function multipoles can make observer terms relevant. For example, contributions like observer's velocity induce a kinematic dipole in the measured galaxy density contrast which can affect the measured galaxy power spectrum multipoles in realistic situations (see e.g. \cite{Elkhashab.et.al2021}). However, for the scales we are interested in, we can omit these contributions.  Therefore, we will only focus on other terms. In the equation above, $\delta^{\text{NI}}(\vec{x})$ is comprised of all relativistic and inverse-distance terms that are evaluated at the source position.  This also includes the second term in Eq.~\eqref{rbias}. The $\delta^{\text{I}}(\vec{x})$ terms are integrated relativistic contributions. The total linear galaxy density fluctuation in Eq.~\eqref{deltagl} can be written in the instructive form:

\begin{eqnarray}
	\label{rps}
	&&\delta_{g,L}(\vec{x})=\underset{\vec{k}'  }{\int} e^{i \vec{k}'\cdot\vec{x}} (\text{F}^{\text{RSD}}+\text{F}^{\text{NI}})(\vec{k}', \vec{x})\, \tilde{\delta}_{m0}(\vec{k'})  \nonumber \\ 
	&& + \,\,\,\, \underset{\vec{k}'  }{\int} \int_{0}^{x} dx'  e ^{i \vec{k}'\cdot \vec{x}' }\text{F}^{\text{I}}(\vec{k}', \vec{x},\vec{x}')  \tilde{\delta}_{m0}(\vec{k'}) 
\end{eqnarray}
 where $\text{F}$ functions  are:
 
\begin{widetext}
	\begin{eqnarray}
		\label{dg1}
		 &&\text{F}^{\text{RSD}}(\vec{k}', \vec{x}) =  \left( b_{1} + f(\hat{k}'\cdot \hat{x})^2 \right ) \tilde{\mathcal{D}}_{+} (\tau_{x})\\ \label{dg2}
		 &&\text{F}^{\text{NI}}(\vec{k}', \vec{x})   = \biggl \{  (3 - b_{e}) f \frac{\mathcal{H}^2}{k'^2} + \frac{3}{2} \frac{\mathcal{H}^2}{k'^2} \Omega_{M} \left[\frac{5s-2}{\mathcal{H} x} - 10 s +b_e  +\frac{3}{2} \Omega_{M} - f +1\right] \nonumber  \\ \label{dg4}
		 &&\hspace{1.9cm} - f  \frac{i (\hat{k}'\cdot \hat{x})\mathcal{H}}{k'}  \left[ 5s  - b_{e}  + 1 - \frac{3}{2}\Omega_{M} - \frac{5s - 2}{\mathcal{H} x} \right]  \biggl \}\tilde{\mathcal{D}}_{+} (\tau_{x})  \\ \label{integratedF}
		  &&\text{F}^{\text{I}}(\vec{k}', \vec{x},\vec{x}') = 3 H_{0}^2 \Omega_{M0} (1 + z_{x'})\left\{ {(5s - 2)} \frac{(x - x')x'}{2 x}  \left[ 1 - (\hat{k}'\cdot\hat{x}')^2 + 2i \frac{(\hat{k}'\cdot \hat{x}') }{k' x'} \right] \right. \nonumber \\  
		  && \hspace{1.9cm}\left.   + \left[ \frac{2-5s}{\mathcal{H} x} + 5 s -b_e +1  -\frac{3}{2} \Omega_{M} \right]  \frac{(1 -f_{x'}) \mathcal{H}(x') }{k'^2}  \,\, + \,\, \frac{5s - 2}{k'^2 x} \,\, \right \}  \tilde{\mathcal{D}}_{+} (\tau_{x'}) 
	\end{eqnarray}
\end{widetext}
where $\tilde{\mathcal{D}}_{+}(\tau_{x}) \equiv \tilde{D}_{+}(\tau_{x}) / \tilde{D}_{+}(\tau_{0})$ represents the normalized growth factor of the SCG density fluctuation and $\tilde{D}_{+}$ can be obtained from Eq.\eqref{DA}. For these results we have used $\nabla^{2}_{\hat{x}'} \phi = x'^2 \vec{\nabla}^2 \phi - \partial_{x'} (x'^2 \partial_{x'} \phi)$  and $\mathcal{H}^{2} \Omega_{M} =  H_{0}^2 \Omega_{M0}(1+z)$. Also for calculating integrated contributions in Eq.~\eqref{Relativistic} proper time differences are converted to comoving distances, e.g. $\tau_{0} - \tau' = x'$. Now we can construct the corresponding local power spectrum defined in Eq.~\eqref{localp}. This becomes:
\begin{widetext}
		\beq
			\label{lin-local}
		P^{\text{loc}}_{g, L}(\vec{k}, \vec{x}_{c}) = 
			 \underset{\vec{x}_{12} }{\int} e^{-i\vec{k \cdot\vec{x}_{12} }}\ \langle\delta_{g,L}( \vec{x}_c + \vec{x}_{12}/2) \,	\delta_{g,L}(\vec{x}_c - \vec{x}_{12}/2) \rangle	\eeq

\end{widetext}
	where we have used that $\vec{x}_{1,2}=\vec{x}_c \pm \vec{x}_{12}/2$, following the geometry of Fig.~\ref{fig:midp}. In the following we will also use $\langle \tilde{\delta}_{m0}(\vec{k}_{1}) \tilde{\delta}_{m 0}(\vec{k}_{2}) \rangle = \delta_{D}(\vec{k}_{1} + \vec{k}_{2}) \tilde{P}_{m0}(k)$ with $\tilde{P}_{m0}(k)$ being the linear dark matter power spectrum in the SCG at the corresponding redshift.

\subsubsection{Wide-angle corrections}
We now illustrate, with an example, how wide-angle effects can be easily calculated using perturbative expansions. Let's consider, for instance, the RSD-RSD contribution to the local power spectrum in Eq.~\eqref{lin-local} which, from Eqs \eqref{rps} and \eqref{dg1}, takes the form:
\begin{eqnarray}
\underset{\vec{x}_{12},\vec{k}' }{\int} && e^{-i(\vec{k} - \vec{k}' )\cdot\vec{x}_{12} }\, [b_{1} + f\,  (\hat{k}'\cdot \hat{x}_{1})^2] \, [b_{1} + f \, (\hat{k}'\cdot \hat{x}_{2})^2]\nonumber \\ & & \times\ \tilde{P}_{m}({k}').	 \label{example1} 
\end{eqnarray}
Wide-angle corrections arise when one relaxes the assumption that  $\hat{x}_{1}$ and $\hat{x}_{2}$ are equal to $\hat{x}_{c}$. Here we ignore the variation of $f$ itself and assume it is evaluated at $x_{c}$. Similarly, $\tilde{P}_{m}({k}')$ is evaluated at $x_{c}$. In general, calculation of wide-angle corrections can be complicated but for sufficiently small ratio of $\vec{x}_{12}/x_{c}$ one can compute them perturbatively. 

Perturbative methods for calculation of leading order wide-angle corrections to two-point correlation functions have been used in~\cite{2016JCAP...01..048R,  2018MNRAS.476.4403C} by choosing $\vec{x}_{12}/x_{c}$ or other suitable choices (like the bisector instead of the midpoint in the denominator) for quantifying magnitudes of corrections. Here our approach for perturbative calculations is relatively different as we find analytic relations in Cartesian Fourier space that can be easily extended to nonlinear regime and also higher-order statistics like bispectrum.

Let's again consider the geometry of Fig.~\ref{fig:midp} as a sketch of galaxy pair positions and our choice of LOS. We define the perturbation parameter vector to be:
\begin{equation}
	\label{p1}
	\boldsymbol \epsilon \equiv \frac{\vec{x}_{12}}{x_{c}}\,
\end{equation}
with $\vec{x}_{12}=\vec{x}_{1}-\vec{x}_{2}$. We can assume that pairs are within spherical bins with mean radius at redshift $\bar{z}$ and depth $\Delta z$. For most surveys $\Delta z / \bar{z} <$ 1. Therefore, the sample volume can be written as:
\begin{equation}
	\label{VP}
	V_{s} = \frac{4 \pi}{3} f_{\rm sky} \left[ (\bar{r} + \frac{\Delta r}{2})^3 - (\bar{r} -  \frac{\Delta r}{2})^3 \right ]
\end{equation}
where $f_{\rm sky} \equiv \Omega_{s}/\Omega_{\text{tot}}$ is the fraction of the sky covered by the  survey ($\Omega_{\text{tot}} \simeq$ 41253 deg$^2$ ) and $\bar{r}$ and $\Delta r$ correspond to the comoving distances at $\bar{z}$ and $\Delta z$. We assume for most galaxy pairs within the bin, the ratio of Eq. \eqref{p1} is small. It is also worth noting that, looking at Eq. \eqref{window2}, the results that will be obtained in the following can be windowed by any survey-specific window function (see section \ref{window-geometry}). Also, There can be pairs for which \eqref{p1} is larger than one. However, their corresponding separations are most of the time much larger than $k^{-1}$ for the scales we are interested in. Therefore, these have negligible contributions and one can still use the entire survey volume. 

To evaluate integrals such as Eq.~\eqref{example1}, one can expand $\hat{x}_{1}$, $\hat{x}_{2}$ and any scalar function made out of them in terms of $\vec{\epsilon}$ to any desired order. For example:

\begin{eqnarray}
	\label{pexp}
	&&\hat{x}_{1}  = \frac{\vec{x}_{c} + \vec{x}_{12}/2}{|\vec{x}_{c} + \vec{x}_{12}/2|} = \hat{x}_{c} +\frac{1}{2} \boldsymbol {\epsilon} + ... + \mathcal{O}(\epsilon^{3})\,, 
	\nonumber \\
	&& \frac{1}{x_{1}} = \frac{1}{x_{c}}( 1 -\frac{1}{2} \boldsymbol{\epsilon} \cdot \hat{x}_{c} + ... )
\end{eqnarray}
and similar expansions for $\hat{x}_{2}$, $1/x_{2}$. Here we keep terms up to second order to compute leading order contributions. But in general, These help us separate wide-angle corrections in a perturbative manner to any desired order. 

After these expansions, we can decompose $\boldsymbol \epsilon$ into its Cartesian components, with the ``z"-direction defined to be along the LOS unit vector, $\hat{x}_{c}$. Doing so, Eq.~\eqref{example1} can be expressed as a sum of integrals that have the following general form:
\begin{eqnarray}
	\label{o2}
	\mathcal{I}_{m,n,l}^{L} = \underset{\vec{x}_{12},\vec{k}'  }{\int} e^{i(\vec{k}'-\vec{k})\cdot\vec{x}_{12} }  \mathcal{C}_{m,n,l}(\vec{k}',\hat{x}_{c}, x_{c}) (\boldsymbol \epsilon_{x})^{m} (\boldsymbol \epsilon_{y})^{n} (\boldsymbol \epsilon_{z})^{l} \nonumber \\
\end{eqnarray}
where $\mathcal{C}$ contains all corresponding terms apart from perturbation factors and ($m, n, l $) can be $0$, $1$ and $2$ such that $(m + n + l) \leq 2$ (for terms up to second order). Now, according to Eq.~\eqref{p1}, all $(\boldsymbol \epsilon_{j})^{a}$ terms turn into the $a$-th derivative of the exponential factor with respect to $\vec{k}_{j}$ and can be pulled out of the integral. This results in  no $\vec{x}_{12}$ factor being left inside and the integral over $\vec{x}_{12}$ becomes delta function, $\delta_{D}(\vec{k}' - \vec{k})$, which results in:
\begin{eqnarray}
	\label{o3}
	\mathcal{I}_{m,n,l}^{L} = \left( \frac{i^{(m+n+l)}}{x_{c}^{m}\,x_{c}^{n}\,x_{c}^{l}} \right) \left(\frac{\partial^{m}}{\partial{k_{x}^{m}}}\,\frac{\partial^{n}}{\partial{k_{y}^{n}}}\,\frac{\partial^{l}}{\partial{k_{z}^{l}}}\right) \,\,\mathcal{C}_{m,n,l}(\vec{k},\hat{x}_{c}, x_{c}) \nonumber \\
\end{eqnarray}
and we sum over all such terms:
\begin{eqnarray}
	\label{so3}
	\mathcal{I}^{L}(k, \hat{k}.\hat{x}_{c}, x_{c}) = \sum_{(m + n + l) \leq 2} \mathcal{I}_{m,n,l}^{L} (k, \hat{k}.\hat{x}_{c}, x_{c})
\end{eqnarray}
This result depends only on $\vec{k}$ and $\vec{x}_{c}$, as expected. Also with our symmetric choice of LOS there is no imaginary term and, therefore, no $1/(kx_{c})$ contribution. The leading order wide-angle corrections to the power spectrum are therefore suppressed by ${1}/(kx_{c})^{2}$ or $(1/(kx_{c}))(\mathcal{H}/k)$ where the latter corresponds to  mixing with relativistic corrections. 
\subsubsection{Non-integrated relativistic corrections}
The procedure explained above for wide-angle effects involves computing the mixing between wide-angle and relativistic contributions. Therefore, to find non-integrated relativistic terms one can evaluate the corresponding contributions to $P^{\text{loc}}_{g, L}$ with plane-parallel assumption.

Finally, it is worth mentioning that in previous parts, we assumed that parameters like $f$, $\tilde{\mathcal{D}}_{+}$, $\Omega_{M}$, etc. are evaluated at $x_{c}$. As an example, we assumed $f(x_{1}) f(x_{2}) \approx f(x_{c})^2$. While one would expect that this is a good approximation for surveys that radially narrow, it is strightforward to obtain corrections to this assumption by performing an expansion around $x_{c}$. This will result in terms that are suppressed by $\mathcal{H}^2/k^2$ where $\mathcal{H}$ is evaluated at $x_{c}$. Then, the result is integrated over, as in Eq. \eqref{estp2}, to yield the final multipoles. 
\subsubsection{Integrated relativistic corrections}
To find the leading order correction from integrated contributions to the local power spectrum, one should start from Eq.~\eqref{lin-local} and, considering Eq.~\eqref{rps}, cross correlate  Eq.~\eqref{integratedF} with the RSD term, Eq.~\eqref{dg1}. This results in:
\begin{widetext}
	\begin{eqnarray}
	\underset{\vec{x}_{12},\vec{k}' }{\int} && e^{-i\vec{k \cdot\vec{x}_{12} }} \left ( \int_{0}^{x_{1}} dx' e^{i \vec{k}'\cdot (\vec{x}' - \vec{x}_{2})}\, \text{F}^{\text{I}}(k', x_{1},x', \hat{k}'\cdot\hat{x}') \,\text{F}^{\text{RSD}}(-\hat{k}'\cdot\hat{x}_{2}) +\int_{0}^{x_{2}} dx' e^{i \vec{k}'\cdot (\vec{x}_{1} - \vec{x}')}\, \text{F}^{\text{I}}(k', x_{2},x', -\hat{k}'\cdot\hat{x}')\, \text{F}^{\text{RSD}}(\hat{k}'\cdot\hat{x}_{1})  \right )  \nonumber \\ & & \times \ \tilde{P}_{m0}(k')
	\end{eqnarray}
Now we can make use of the fact that  $\hat{x}' = \hat{x}_{1} = \vec{x}_{1}/x_{1}$ in the first integral and $\hat{x}' = \hat{x}_{2} = \vec{x}_{2}/x_{2}$ in the second one (at linear level in fluctuations) and also perform the coordinate transformation $y = x' x_{c}/x_{1}$ and  $y = x' x_{c}/x_{2}$ in the first and second integrals respectively. Then, using that $\vec{x}_{1,2}=\vec{x}_c \pm \vec{x}_{12}/2$ one obtains:
	\begin{eqnarray}
	\underset{\vec{x}_{12},\vec{k}' }{\int} e^{-i(\vec{k} - \frac{x_c + y}{2 x_c} \vec{k}') \cdot\vec{x}_{12} }&& \left ( \int_{0}^{x_{c}} dy \,(\frac{x_{1}}{x_{c}}) \,e^{i \vec{k}'\cdot (\frac{y}{x_c} - 1)\cdot\vec{x}_{c}}\, \text{F}^{\text{I}}(k',x_{1}, \frac{x_{1}}{x_{c}} y, \hat{k}'\cdot\hat{x}_{1})\, \text{F}^{\text{RSD}}(-\hat{k}'\cdot\hat{x}_{2})\,\,  +  \right. 
	 \nonumber \\
	 &&\left. \int_{0}^{x_{c}} dy \,(\frac{x_{2}}{x_{c}}) \,e^{-i \vec{k}'\cdot (\frac{y}{x_c} - 1)\cdot\vec{x}_{c}}\, \text{F}^{\text{I}}(k',x_{2}, \frac{x_{2}}{x_{c}}y, -\hat{k}'\cdot\hat{x}_{2})\, \text{F}^{\text{RSD}}(\hat{k}'\cdot\hat{x}_{1})   \right )\tilde{P}_{m0}(k')\,.
\end{eqnarray}
Now, to find the leading order contribution, we can ignore mixing with wide-angle corrections and set $\hat{x}_{1} = \hat{x}_{2} = \hat{x}_{c}$ in the equation above. In addition $x_{1,2}/x_{c}$ can also be set to one as any correction will produce subleading terms. The final result then becomes:
 
 	\begin{eqnarray}
 		\label{int2}
   \int_{0}^{x_{c}} dy \,(\frac{2 x_c}{x_c + y})^3 &&\left( \,e^{i 2(\frac{y - x_c}{y + x_c})\vec{k}\cdot\vec{x}_{c}}\, \text{F}^{\text{I}}(\frac{2 x_{c}k}{x_{c}+y}, x_{c}, y, \hat{k}\cdot\hat{x}_{c})\, \text{F}^{\text{RSD}}(-\hat{k}\cdot\hat{x}_{c})  + \right. \nonumber \\
 	&&\left. e^{-i 2(\frac{y - x_c}{y + x_c})\vec{k}\cdot\vec{x}_{c}}\, \text{F}^{\text{I}}(\frac{2 x_{c}k}{x_{c}+y}, x_{c} ,y, -\hat{k}\cdot\hat{x}_{c})\, \text{F}^{\text{RSD}}(\hat{k}\cdot\hat{x}_{c}) \right) \tilde{P}_{m0}\Big(\frac{2 x_{c}k}{x_{c}+y}\Big) 
 \end{eqnarray}
\end{widetext}
which is a simple 1D integral for numerical integration as will be discussed in the following. 

\subsection*{Final Results}
 The final results for the local linear power spectrum with non-integrated relativistic (including inverse-distance) and wide-angle corrections take the following form:
	\begin{widetext}
		\begin{eqnarray}
			\label{final-pr}
			P_{g, L}^{\text{loc, NI}}(k, \eta, x_{c}) &=& P_{g, L}^{\text{pp}}(k, \eta) + \frac{1}{(k x_{c})^2}P_{g}^{\text{R1}}(k, \eta)+\frac{\mathcal{H}^2}{k^2}P_{g}^{\text{R2}}(k, \eta) + \frac{1}{(k x_{c})} \frac{\mathcal{H}}{k}P_{g}^{\text{R12}}(k, \eta) \, + \nonumber \\
			&& + \frac{1}{(k x_{c})^2}P_{g}^{\text{w1}}(k, \eta)+ \frac{1}{(k x_{c})} \frac{\mathcal{H}}{k}P_{g}^{\text{w2}}(k, \eta) + \mathcal{O}(\frac{1}{(kx_{c})^4}) + ...
		\end{eqnarray}
	\end{widetext}
   \begin{figure*}
	\centering
	\includegraphics[width=0.8\textwidth]{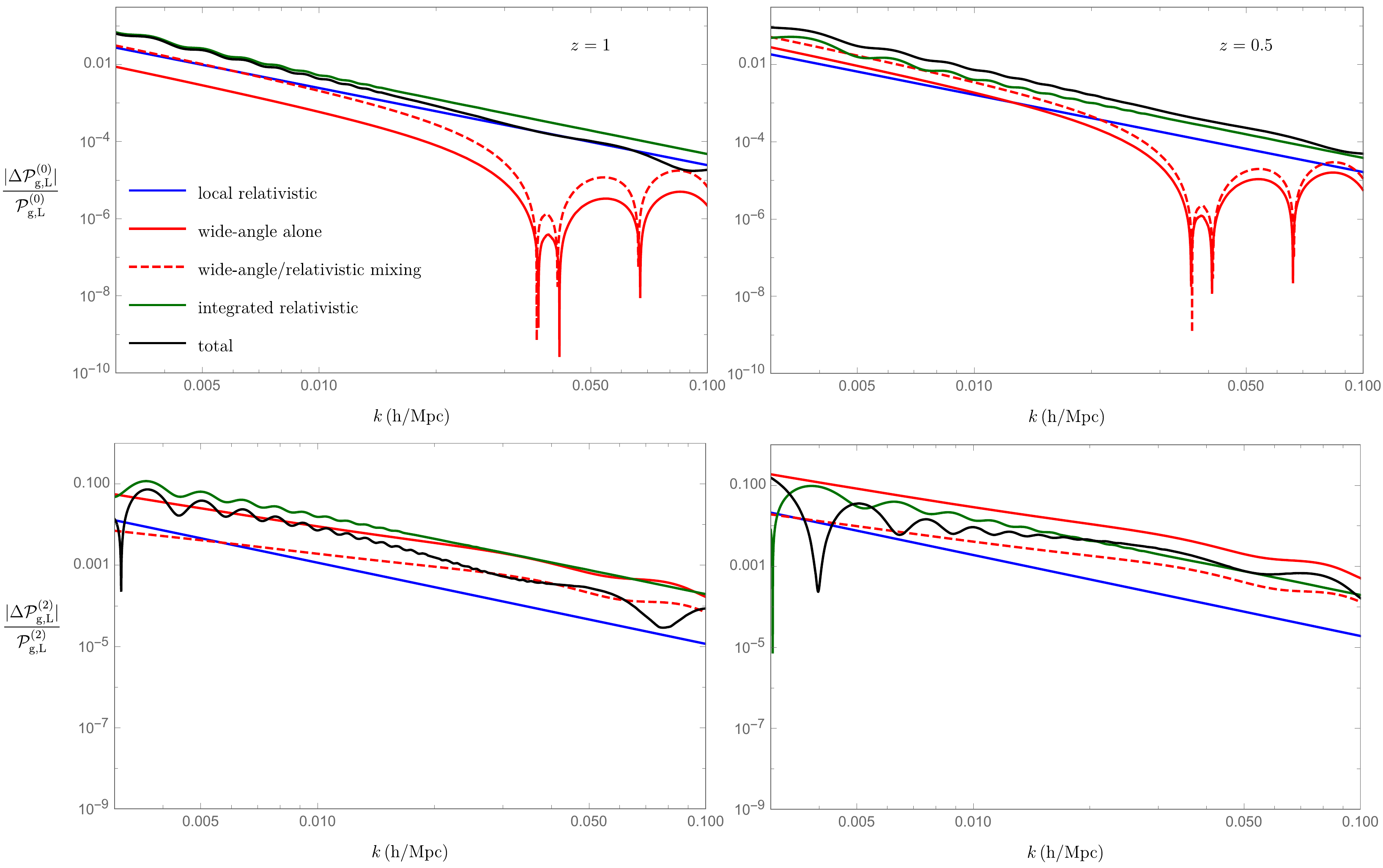}
	\caption{Absolute values of relative corrections to monopole and quadrupole of the plane-parallel linear local power spectrum from leading  local (non-integrated) relativistic (corresponding to functions $R_{1}$, $R_{2}$ and $R_{12}$ defined above), wide-angle alone (function $w1$ which also includes the mixing with inverse-distance corrections modified by magnification bias), wide-angle/relativistic mixing (function $w2$) and integrated relativistic corrections (based on the integral of \eqref{int2}) The black line shows the sum of all these contributions. The linear bias is $b_{1} = 1.5$, and magnification and evolution biases are set to $s = 1$ and $b_{e} = 0$ respectively.}
	\label{fig:Lensing}
\end{figure*}
	where again $\eta = \hat{k} \cdot \hat{x}_{c}$. The first term is just the local plane-parallel approximation result: 
	\begin{equation}
		\label{pp}
	P_{g, L}^{\text{pp}}(k, \eta) = (b_{1} + f \eta^2)^2 \, \tilde{P}_{L}(k)\,. 
    \end{equation}
	The next three terms are non-integrated relativistic corrections which include inverse-distance (R$_{1}$) modified by magnification bias and other local (non-integrated) relativistic (R$_{2}$) contributions and the mixing between them (R$_{12}$). We have: 
	\begin{equation}
		\label{rr1}
		P_{g}^{\text{R}_{1}}(k, \eta) = f^2 \eta ^2 (2-5 s)^2\, \tilde{P}_{L}(k) \,
	\end{equation}
and

		\begin{widetext}
		\begin{eqnarray}
			\label{rr2}&&P_{g}^{\text{R}_{2}}(k, \eta) =  \frac{1}{4}\biggl [ -4 b_{1} f (-6 + 2 b_{e} + 3 \Omega_{M}) + 6 b_{1} \Omega_{M} (2 + 2 b_{e} - 20 s + 3 \Omega_{M}) + f \eta^2 \biggl(6 \Omega_{M} (2 + 2 b_{e}-20 s  + 3 \Omega_{M})    \\  
			&& \hspace{1.9cm}   + f (28 + 4 b_{e}^2 + 40 s + 100 s^2 - 4 b_{e}(4 + 10 s - 3 \Omega_{M})  - 24 \Omega_{M} - 60 \,s \, \Omega_{M}+ 9\, \Omega_{M}^2 ) \biggl ) \biggl] \tilde{P}_{L}(k)  \nonumber\\
			\label{rr3}&&P_{g}^{\text{R}_{12}}(k, \eta) = (-2+5 s) \left(3\, b_{1} \Omega_{M} +3 f \eta ^2 \Omega_{M} +f^2 \eta ^2 (-2+2 b_{e}-10 s+3 \Omega_{M} )\right)
			\tilde{P}_{L}(k) \,.
		\end{eqnarray} 
\end{widetext}
	 \begin{figure*}
	\centering
	\includegraphics[width=0.9\textwidth]{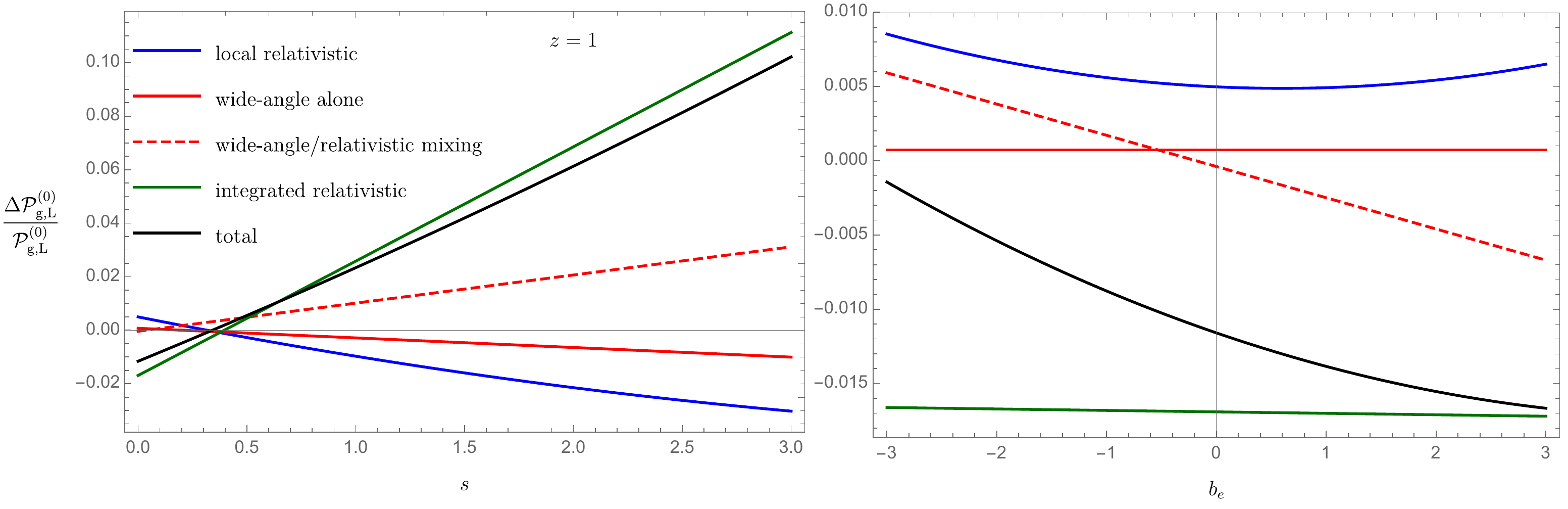}
	\caption{Similar plot as above (not absolute value) for monopole but at a fixed scale, $k = 0.005$ $h$/Mpc and varying magnification bias with $b_{e} = 0$ (left) and evolution bias with $s = 0$ (right)  at redshift $z = 1$ }
	\label{fig:SBE}
\end{figure*}  
	 \begin{figure*}
	\centering
	\includegraphics[width=0.7\textwidth]{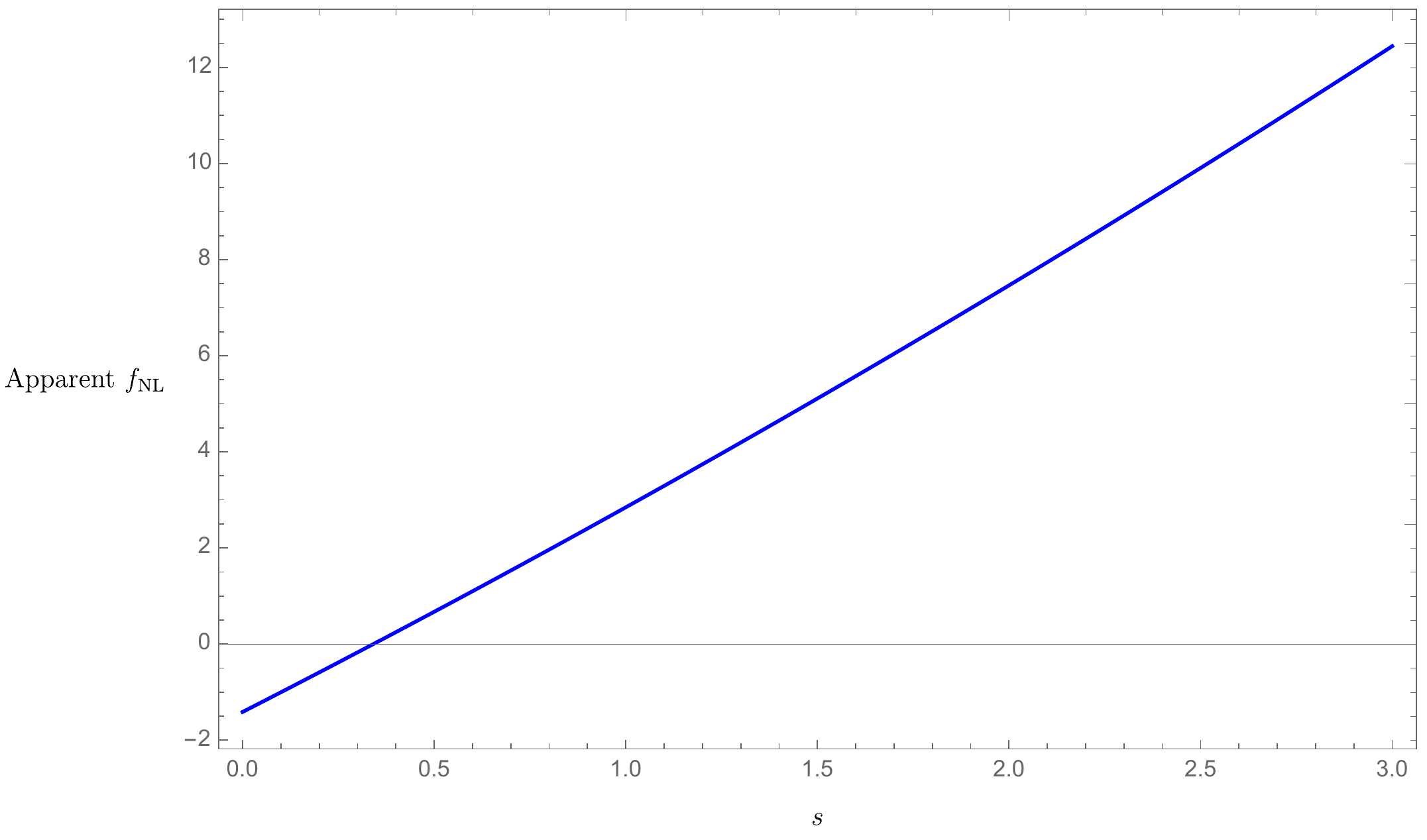}
	\caption{Apparent $f_{\text{NL}}$ induced by deviations of the plane-parallel approximation for linear galaxy bias $b_{1} = 1.5$ as a function of magnification bias at $k = 0.005$ h/Mpc and assuming $b_{e} = 0$.}
	\label{fig:figurefnl}
\end{figure*}  
	The last two terms in Eq.~\eqref{final-pr} are the leading order wide-angle corrections which have mixing with inverse-distance and other non-integrated relativistic terms. These two contributions vanish in the limit $\vec{x}_{1} = \vec{x}_{2}$. We find:
		\begin{widetext}
		\begin{eqnarray}
			\label{final-wa1}
			P_{g}^{\text{w1}}(k, \eta) &=&  f \left(b_{1} \left(-16 \eta ^4+17 \eta ^2+\left(5-20\, \eta ^2\right) s-2\right)+f \eta ^2
			\left(-24\, \eta ^4+26\, \eta ^2+5 s-7\right)\right) \tilde{P}_{L}(k) \nonumber \\
			&& + f  \left(b_{1} \left(10\, \eta ^4-11\, \eta ^2+5 \left(2 \eta ^2-1\right) s+2\right)+f \eta
			^2 \left(9\, \eta ^4-8\, \eta ^2+5 s\right)\right)k \tilde{P}_{L}'(k) \\
			&&-\frac{1}{2} f \left(\eta ^2-1\right) \left(4 b_{1} \eta ^2-b_{1}+2 f \eta ^4+f \eta ^2\right)
			k^2\tilde{P}_{L}''(k) \nonumber \,.
		\end{eqnarray}
	and 
	\begin{eqnarray}
		\label{final-wa2}
	    P_{g}^{\text{w2}}(k, \eta) &=& -f \eta ^2  \left(b_{1}-2 f \eta ^2+f\right) (2 b_{e} -10 s+3\, \Omega_{M} -2) \tilde{P}_{L}(k)  
		+ \frac{1}{2} f \left(\eta ^2-1\right) \left(b_{1}-f \eta ^2\right) (2 b_{e}-10 s+3 \Omega_{M} -2) k \tilde{P}_{L}'(k) \nonumber
	\end{eqnarray}
	\end{widetext}
	with $\tilde{P}' = d\tilde{P}/dk$, etc. A similar procedure is followed for integrated relativistic correction. In fact, the local power spectrum for these contributions can be found by inserting  \eqref {int2} into the integral of Eq. \eqref{local-m} and doing the integral over $\eta  = \hat{k}\cdot\hat{x}_{c}$ first. This results in a straightforward one-dimensional integral over $y$. 
	
	 From these results, one can insert them into Eq.~\eqref{estp1} to find the local multipoles:
	\begin{equation}
		\label{local-m}
		\mathcal{P}^{(\ell)}_{g, L}(k, x_{c}) \equiv \frac{(2\ell + 1)}{2} \int_{-1}^{1} d\eta\, \mathcal{L}_{\ell}(\eta)P_{g,L}(k,\eta,x_{c}) \,
	\end{equation}
    which can be used  in Eq.~\eqref{estp2} to obtain the power spectrum multipoles:
     \begin{equation}
    	\label{local-v}
    	P_{g}^{(\ell)}(k) \equiv  \int_{V_s}\frac{d^{3}\vec{x}_{c}}{V_s} \, \mathcal{P}^{(\ell)}_{g, L}(k, x_{c}),
    \end{equation}
     where the integration is done within the volume of the galaxy sample, $V_s$, see Eq.~\eqref{VP}. 

Let us now illustrate these corrections and estimate their relative importance. Here we show plots of corrections to local power spectrum and do not specify a survey bin. However, it is very straightforward to find the final power spectrum multipoles by performing the integration in Eq. \eqref{local-v} since the local power spectrum multipoles are functions of magnitude of $\vec{x}_{c}$ only. In addition, for bins with narrow depth, the results are close.   Figure~\ref{fig:Lensing} shows our results  at redshift ${z} = 1$ and linear bias $b_{1} = 1.5$. For the cosmology, we assumed a flat $\Lambda$CDM model with $\Omega_{M} = 0.32$, $\sigma_{8} = 0.828$, $n_{s} = 0.968$ and negligible radiation. For this plot,  we also assumed $s =1$ and $b_{e}  = 0$ (i.e $n_{g} \propto 1/a^3$). In general, the relative importance of different contributions depends on redshift and total corrections scale almost like $1/k^2$ at these $k's$. Figure~\ref{fig:SBE} shows the variation of monopole corrections at a fixed scale $k = 0.005$ $h$/Mpc. We can see here that there is much more sensitivity to magnification bias as compared with evolution bias. 
  
 \subsection*{Implications for local $f_{\text{NL}}$ measurements}  
 It is well known that local primordial non-Gaussianity  (PNG) induces a scale-dependent bias with the following form  \cite{Dalal.et.al.2007, Slosar.et.al.2008}:
 \begin{eqnarray}
 	\label{fnl-bias}
   \Delta b(k) = 3 f_{\text{NL}} \delta_{c}(b_{1} - 1)\frac{\Omega_{M} \mathcal{H}^{2} a}{\tilde{D}_{+}(z) k^{2} T(k) }
\end{eqnarray}
which modifies the (plane-parallel) linear galaxy power spectrum as $P^{\text{pp, NG}}_{g}(k,\eta) \to (b_{1} + \Delta b(k) + f \eta ^{2}) \tilde{P}_{L}(k)$. At large scales ($k \lesssim 0.01$ h/Mpc ) where $T(k) \to 1$,  the $1/k^{2}$ scaling of $\Delta b$ resembles that of wide-angle and relativistic contributions calculated already. This implies that such corrections can be mistaken with a true $f_{\text{NL}}$ term if are not taken into account in theoretical modeling. Comparing power spectrum monopoles, for example, we can find such apparent $f_{\text{NL}}$. Figure~\ref{fig:figurefnl} shows such a term as a function of magnification bias. It is estimated by comparing the PNG correction to the linear monopole power spectrum with the similar corrections from the total wide-angle and relativistic (integrated and non-integrated) contributions. It is smaller than current limits but not negligible compared to upcoming survey forecasts~\cite{ZhaPulMan2111}.

\subsection{The One-loop Power Spectrum}  
\subsubsection{Beyond Linear Theory}
	\label{1-loop} 
	We now show how to apply the perturbative approach followed in linear case to include nonlinear effects. Here we keep terms up to third order in SPT (1-loop). For this case we neglect relativistic $\mathcal{H}/k$ contributions and consider inverse distance and wide-angle corrections only. This is because their inclusion is an unnecessary complication for our purpose of comparing the wide-angle corrections with other local effects as inverse-distance terms and other local relativistic terms are typically of the same order (see e.g. \cite{Hwang.et.al.2015, Bertacca.et.al.2015} for an investigation of relativistic effects in nonlinear regimes). Therefore, hereafter we remove the "tilde" on top of SCG quantities for simplicity and, for example, use $\delta$ instead of $\tilde{\delta}$ .
	
	Equation~(\ref{exp22}) can be recast by writing $\delta_{g}$ and $\vec{v}$ in terms of their Fourier representations, and assuming negligible vorticity the velocity field can be rewritten in terms of its scalar mode, the divergence $\Theta_{v} =\boldsymbol{\nabla}\cdot\vec{v} \equiv -\mathcal{H} f  \boldsymbol{\nabla}\cdot\vec{u} = -\mathcal{H} f  \Theta$ which carries the same amount of information. Therefore, Equation~(\ref{exp22}) can be rewritten as:
	
	\begin{widetext}
		\begin{eqnarray}
			\label{exp3}
			\delta_{g,s}(\vec{x})&=&\delta_{g}(\vec{x}) + f \underset{\vec{k}  }{\int}e^{i \vec{k}\cdot\vec{x}} D_{\theta}(\vec{k},\vec{x})\,\Theta(\vec{k}) 
			+f \underset{\vec{k}_{1}, \vec{k}_{2} }{\int} e^{i\vec{k}_{12}\cdot\vec{x}} D_{\theta \delta}(\vec{k}_{1}, \vec{k}_{2},\vec{x})\, \Theta(\k_{1}) \delta_{g}(\vec{k}_{2}) + 
			\nonumber \\
			&& f^{2} \underset{\vec{k}_{1}, \vec{k}_{2}  }{\int} e^{i\vec{k}_{12}\cdot\vec{x}} D_{\theta \theta}(\vec{k}_{1}, \vec{k}_{2},\vec{x})\,\Theta(\vec{k}_{1}) \Theta(\vec{k}_{2}) + 
			f^{2} \underset{\vec{k}_{1}, \vec{k}_{2}, \vec{k}_{3}  }{\int} e^{i\vec{k}_{123}\cdot\vec{x}} D_{\theta \theta \delta}(\vec{k}_{1}, \vec{k}_{2},\vec{k}_{3},\vec{x})\,\Theta(\vec{k}_{1})\Theta(\vec{k}_{2})\delta_{g}(\vec{k}_{3}) + 
			\nonumber \\
			&& f^{3}  \underset{\vec{k}_{1}, \vec{k}_{2}, \vec{k}_{3} }{\int} e^{i \vec{k}_{123}\cdot\vec{x}} D_{\theta \theta \theta}(\vec{k}_{1}, \vec{k}_{2},\vec{k}_{3},\vec{x})\,\Theta(\vec{k}_{1})\Theta(\vec{k}_{2})\Theta(\vec{k}_{3}) + \ldots
		\end{eqnarray}
	\end{widetext}
	where the $D$ kernels collect terms proportional to $\delta_g$ and powers of $\Theta$ as denoted by their subindex, and we have displayed only the necessary terms for a 1-loop power spectrum calculation. These kernels are given by,
	
	\begin{widetext}
		\begin{eqnarray}
			\label{D-functions}
			&&D_{\theta}(\vec{k},\vec{x}) = \frac{(\vec{k}\cdot\hat{x})^{2}}{k^{2}} - \frac{2 i}{x}\,\frac{\vec{k}\cdot\hat{x}}{k^{2}}
			\,,
			\nonumber \\
			&&D_{\theta\delta}(\vec{k}_{1},\vec{k}_{2},\vec{x}) = \frac{(\vec{k}_{12}\cdot\hat{x})(\vec{k}_{1}\cdot\hat{x})}{k_{1}^{2}} - \frac{2 i}{x}\,\frac{\vec{k}_{1}\cdot\hat{x}}{k_{1}^{2}}
			\,,
			\\
			&&D_{\theta \theta}(\vec{k}_{1},\vec{k}_{2},\vec{x}) =\frac{(\vec{k}_{1}\cdot\hat{x})(\vec{k}_{2}\cdot\hat{x})}{k_{1}^{2} k_{2}^{2}}\,\left(\frac{1}{2}(\vec{k}_{12}\cdot\hat{x})^{2} - \frac{2i}{x} (\vec{k}_{12}\cdot\hat{x}) - \frac{1}{x^{2}} \right )\,,
			\nonumber \\
			&&D_{\theta \theta \delta}(\vec{k}_{1},\vec{k}_{2},\vec{k}_{3},\vec{x}) =\frac{(\vec{k}_{1}\cdot\hat{x})(\vec{k}_{2}\cdot\hat{x})}{k_{1}^{2} k_{2}^{2}}\,\left(\frac{1}{2}(\vec{k}_{123}\cdot\hat{x})^{2} - \frac{2i}{x} (\vec{k}_{123}\cdot\hat{x}) - \frac{1}{x^{2}} \right )\,,
			\nonumber \\
			&&D_{\theta \theta \theta}(\vec{k}_{1},\vec{k}_{2},\vec{k}_{3},\vec{x}) =\frac{(\vec{k}_{1}\cdot\hat{x})(\vec{k}_{2}\cdot\hat{x})(\vec{k}_{3}\cdot\hat{x})}{k_{1}^{2} k_{2}^{2} k_{3}^{2}}\,\left(\frac{1}{6}(\vec{k}_{123}\cdot\hat{x})^{3} - \frac{i}{x} (\vec{k}_{123}\cdot\hat{x})^{2} - \frac{1}{x^{2}}(\vec{k}_{123}\cdot\hat{x}) \right ). \nonumber
		\end{eqnarray}
	\end{widetext} 
	where $\vec{k}_{12...} = \vec{k}_{1} + \vec{k}_{2} + ... $\,. The first term in each of these kernels corresponds to the plane-parallel result where the fixed LOS is  replaced by $\hx$, the second and further terms correspond to inverse-distance corrections. 
	
	These expressions include all RSD contributions needed for the calculation of the tree-level bispectrum and one-loop galaxy power spectrum.  These results can also be used to derive the RSD perturbatiom theory (PT) kernels, after using the PT expansion for $\delta$ and $\Theta$. We discuss them in Appendix~\ref{Kernels}.
	
	We start from Eq.~\eqref{exp3} and plugging it into Eq.~\eqref{localp} we find 
		\begin{widetext}
		\begin{eqnarray}
			\label{final-local}
			&P_{g,s}^{loc}(\vec{k},\vec{x}_{c})&= \underset{\vec{x}_{12},\vec{k}_{1} }{\int} e^{i(\vec{k}_{1}-\vec{k})\cdot\vec{x}_{12} } \biggl \{ P_{gg}(k_{1})  +  f \bigg( D_{\theta}(-\vec{k}_{1},\vec{x}_{2}) + D_{\theta}(\vec{k}_{1},\vec{x}_{1}) \bigg) P_{g\theta}(k_{1}) + f^{2} D_{\theta}(\vec{k}_{1},\vec{x}_{1}) D_{\theta}(-\vec{k}_{1},\vec{x}_{2}) P_{\theta \theta}(k_{1}) \biggl \} 
			\nonumber \\
			&&+\underset{\{\vec{x}_{12},\vec{k}_{1},\vec{k}_{2}\}}{\int}e^{i(\vec{k}_{12}-\vec{k})\cdot\vec{x}_{12} } \biggl\{ f^{2} b_{1}^{2} D_{\theta\delta}(\vec{k}_{1},\vec{k}_{2},\vec{x}_{1}) \bigg ( D_{\theta\delta}(-\vec{k}_{1},-\vec{k}_{2},\vec{x}_{2})+  D_{\theta\delta}(-\vec{k}_{2},-\vec{k}_{1},\vec{x}_{2}) \bigg)  P_{L}(k_{1}) P_{L}(k_{2})  
			\nonumber \\
			&&   + f^{4} D_{\theta\theta}(\vec{k}_{1},\vec{k}_{2},\vec{x}_{1}) \bigg( D_{\theta\theta}(-\vec{k}_{1},-\vec{k}_{2},\vec{x}_{2}) + D_{\theta\theta}(-\vec{k}_{2},-\vec{k}_{1},\vec{x}_{2}) \bigg) P_{L}(k_{1}) P_{L}(k_{2}) \biggl \} \,+ \, 
			\\ 
			&&+\biggl[\biggl( \underset{\vec{x}_{12},\vec{k}_{1},\vec{k}_{2}  }{\int} e^{i(\vec{k}_{1}-\vec{k})\cdot\vec{x}_{12} } \biggl \{ f D_{\theta \delta}(\vec{k}_{2},-\vec{k}_{12},\vec{x}_{2}) B_{g\theta g}(\vec{k}_{1},\vec{k}_{2},-\vec{k}_{12}) + f^{2} D_{\theta \theta}(\vec{k}_{2},-\vec{k}_{12},\vec{x}_{2}) B_{g\theta \theta}(\vec{k}_{1},\vec{k}_{2},-\vec{k}_{12}) 
			\nonumber \\ 
			&& + f^{2} D_{\theta}(\vec{k}_{1},\vec{x}_{1}) D_{\theta \delta}(\vec{k}_{2},-\vec{k}_{12},\vec{x}_{2})B_{\theta \theta g}(\vec{k}_{1},\vec{k}_{2},-\vec{k}_{12}) + f^{3} D_{\theta}(\vec{k}_{1},\vec{x}_{1}) D_{\theta \theta}(\vec{k}_{2},-\vec{k}_{12},\vec{x}_{2})B_{\theta \theta \theta}(\vec{k}_{1},\vec{k}_{2},-\vec{k}_{12}) 
			\nonumber \\
			&&+ f^{2} b_{1}^{2} \bigg( D_{\theta \theta \delta}(-\vec{k}_{1},\vec{k}_{2},-\vec{k}_{2},\vec{x}_{2}) + D_{\theta \theta \delta}(\vec{k}_{2},-\vec{k}_{1},-\vec{k}_{2},\vec{x}_{2}) + D_{\theta \theta \delta}(\vec{k}_{2},-\vec{k}_{2},-\vec{k}_{1},\vec{x}_{2}) \bigg)  P_{L}(k_{1}) P_{L}(k_{2})
			\nonumber \\
			&&+ f^{3}b_{1} \bigg( D_{\theta \theta \theta}(-\vec{k}_{1},\vec{k}_{2},-\vec{k}_{2},\vec{x}_{2}) + D_{\theta \theta \theta}(\vec{k}_{2},-\vec{k}_{1},-\vec{k}_{2},\vec{x}_{2}) + D_{\theta \theta \theta}(\vec{k}_{2},-\vec{k}_{2},-\vec{k}_{1},\vec{x}_{2}) \bigg) P_{L}(k_{1}) P_{L}(k_{2})\, 
			\nonumber \\
			&&+  f^{3}b_{1} D_{\theta}(\vec{k}_{1},\vec{x}_{1})\bigg( D_{\theta \theta \delta}(-\vec{k}_{1},\vec{k}_{2},-\vec{k}_{2},\vec{x}_{2}) + D_{\theta \theta \delta}(\vec{k}_{2},-\vec{k}_{1},-\vec{k}_{2},\vec{x}_{2}) + D_{\theta \theta \delta}(\vec{k}_{2},-\vec{k}_{2},-\vec{k}_{1},\vec{x}_{2}) \bigg) P_{L}(k_{1}) P_{L}(k_{2})
			\nonumber \\
			&&+ f^{4}D_{\theta}(\vec{k}_{1},\vec{x}_{1})\bigg( D_{\theta \theta \theta}(-\vec{k}_{1},\vec{k}_{2},-\vec{k}_{2},\vec{x}_{2}) + D_{\theta \theta \theta}(\vec{k}_{2},-\vec{k}_{1},-\vec{k}_{2},\vec{x}_{2}) + D_{\theta \theta \theta}(\vec{k}_{2},-\vec{k}_{2},-\vec{k}_{1},\vec{x}_{2}) \bigg) P_{L}(k_{1}) P_{L}(k_{2})\biggl\}\,
			\nonumber \\
			&&+ \underset{\vec{x}_{12},\vec{k}_{1},\vec{k}_{2}  }{\int} e^{i(\vec{k}_{12}-\vec{k})\cdot\vec{x}_{12} }\biggl\{f^{3}b_{1}D_{\theta \delta}(\vec{k}_{1},\vec{k}_{2},\vec{x}_{1}) \bigg( D_{\theta\theta}(-\vec{k}_{1},-\vec{k}_{2},\vec{x}_{2}) + D_{\theta\theta}(-\vec{k}_{2},-\vec{k}_{1},\vec{x}_{2}) \bigg) P_{L}(k_{1}) P_{L}(k_{2}) \biggl\}\,\biggl) +  \nonumber \\
			&&\hspace{40mm}+\hspace{10mm}(\vec{x}_{1}\leftrightarrow\vec{x}_{2} )\,\, \biggl ] \nonumber\,.
		\end{eqnarray}
	where we have used that four-point functions can be treated in the Gaussian limit due to our one-loop approximation, the $D$-functions are defined as in Eq.~\eqref{D-functions} and, as usual, 
	\beq
	\left<\delta_{g}(\vec{k}_{1})\,\delta_{g}(\vec{k}_{2})\right>= \delta_{D}(\vec{k}_{1}+\vec{k}_{2})\, P_{gg}(k_{1}), \ \ \ \ \ 
	\left<\delta_{g}(\vec{k}_{1})\,\Theta(\vec{k}_{2})\right>=\delta_{D}(\vec{k}_{1}+\vec{k}_{2})\, P_{g \theta}(k_{1}), \ \ \  \ \ 
	\left<\Theta(\vec{k}_{1})\,\Theta(\vec{k}_{2})\right>=\delta_{D}(\vec{k}_{1}+\vec{k}_{2})\, P_{\theta\theta}(k_{1}) 
	\eeq
	\end{widetext}
	\beq
	\left<\delta_{g}(\vec{k}_{1})\delta_{g}(\vec{k}_{2})\Theta(\vec{k}_{3})\right> = \delta_{D}(\vec{k}_{1} + \vec{k}_{2} + \vec{k}_{3})\ B_{gg\theta}(\vec{k}_{1}, \vec{k}_{2},\vec{k}_{3})
	\eeq
and similarly for other bispectra terms such as $B_{g\theta g}$, etc. To perform a one-loop calculation of the galaxy power spectrum in redshift space we need biased bispectra at tree-level and biased power spectra at one-loop. Therefore we  include local bias terms up to second order ($b_{1}$, $b_{2}$) with non-local bias contributions from tidal fields at second and third order ($\gamma_{2}$ and $\gamma_{21}$ in the parametrization of~\cite{2012PhRvD..85h3509C,2019PhRvD..99l3514E, 2020arXiv200609729E}). The bias expansion then reads,
	\begin{eqnarray}
		\label{bias}
		&\delta_{g}(\vec{x}) =& b_{1} \delta(\vec{x}) + \frac{b_{2}}{2} \delta^{2}(\vec{x}) + \gamma_{2}\, \mathcal{G}_{2}(\Phi_{v}|\vec{x})  \\
		&&+ \,\gamma_{21}\, \mathcal{G}_{2}(\varphi_{1},\varphi_{2}\,|\vec{x})\nonumber + ... 
	\end{eqnarray}
	where ``..." refers to higher order terms or those that enforce $\langle\delta_{g}\rangle = 0$. The latter don't change power spectrum as long as $k \neq 0$. In this equation $ \mathcal{G}_{2}(\Phi_{v}|\vec{x})$ is the second-order Galileon arising from tidal effects defined as:
	\begin{eqnarray}
		\label{bias2}
		\mathcal{G}_{2}(\Phi_{v}|\vec{x}) \equiv  ({\nabla}_{i j } \Phi_{v})^2 - ({\nabla}^2 \Phi_{v})^2
	\end{eqnarray}
	with ${\nabla}^2 \Phi_{v} = \Theta $. In addition, $\mathcal{G}_{2}(\varphi_{1},\varphi_{2}\,|\vec{x})$ is a nonlocal third-order bias contribution that depends on the Lagrangian linear and quadratic potentials $\varphi_{1,2}$~\cite{2019PhRvD..99l3514E}
	\begin{eqnarray}
		\label{bias3}
		\mathcal{G}_{2}(\varphi_{1}, \varphi_{2}|\vec{x}) \equiv  {\nabla}_{i j } \varphi_{1}  {\nabla}_{i j } \varphi_{2} - {\nabla}^2 \varphi_{1} {\nabla}^2 \varphi_{2}
	\end{eqnarray}
	where ${\nabla}^2 \varphi_{1} = - \delta$ and ${\nabla}^2 \varphi_{2} = -\mathcal{G}_{2}(\varphi_{1})$. 
	
	The galaxy-galaxy and matter-galaxy power spectra in this bias basis has been calculated in \cite{2019PhRvD..99l3514E, 2020arXiv200609729E}. Using their formulas, one can also find galaxy-velocity power spectrum. We have:
	\begin{widetext}
		\begin{eqnarray}
			\label{bias-Pgg}
			&P_{\text{gg}}(k) =& b_{1}^{2} P_{mm}(k) + 4 \,b_{1} P_{L}(k) \underset{\vec{q}}{\int} \left [ 2 \gamma_{2} G_{2}(\vec{k},-\vec{q}) + \gamma_{21} K(\vec{k}, \vec{q}) \right] K(\vec{k} - \vec{q},\vec{q})P_{L}(|\vec{k} - \vec{q}|)\nonumber  \\
			&& +2\,b_{1} \underset{\vec{q}}{\int}\left[b_{2} + 2 \gamma_{2} K(\vec{k} - \vec{q},\vec{q}) \right] F_{2}(\vec{k} - \vec{q},\vec{q})P_{L}(|\vec{k} - \vec{q}|) P_{L}(q)  \\
			&& + \frac{1}{2}\underset{\vec{q}}{\int} \left[b_{2}^{2} + 2 b_{2} \gamma_{2} K(\vec{k} - \vec{q},\vec{q}) + 4 \gamma_{2}^{2} K(\vec{k} - \vec{q},\vec{q})^{2}\right] P_{L}(|\vec{k}-\vec{q}|)P_{L}(q)\,, \nonumber
		\end{eqnarray}
		and
		\begin{eqnarray}
			\label{bias-Pgt}
			&P_{\text{g$\theta$}}(k) =& b_{1} P_{m \theta}(k) + 2 P_{L}(k) \underset{\vec{q}}{\int} \left [ 2 \gamma_{2} G_{2}(\vec{k},-\vec{q}) + \gamma_{21} K(\vec{k}, \vec{q}) \right] K(\vec{k} - \vec{q}, \vec{q})P_{L}(|\vec{k} - \vec{q}|) \nonumber \\
			&&+ \underset{\vec{q}}{\int}\left[b_{2} + 2 \gamma_{2} K(\vec{k} - \vec{q},\vec{q}) \right] G_{2}(\vec{k} - \vec{q},\vec{q})P_{L}(|\vec{k} - \vec{q}|) P_{L}(q)\,.
		\end{eqnarray}
	In these equations $P_{mm}$ and $P_{m \theta}$ are the one-loop matter-matter and matter-velocity power spectra. These, along with $P_{\theta \theta}$, have the following well-known expressions~\cite{2002PhR...367....1B}:
		\begin{eqnarray}
			\label{pmm}
			P_{mm}(k) =P_{\text{L}}(k)+ 2 \underset{\{ \vec{q}\}}{\int} F_{2}(\vec{k} - \vec{q}, \vec{q})^{2} P_{\text{L}}(|\vec{k} - \vec{q}|) P_{\text{L}}(q) + 6\, P_{\text{L}}(k)  \underset{\{ \vec{q}\}}{\int} F_{3} (\vec{k}, -\vec{q}, \vec{q}) P_{\text{L}}(q) \nonumber \,,
		\end{eqnarray}	
		\begin{eqnarray}
			\label{pmt}
			P_{m \theta}(k) =P_{\text{L}}(k) + 2 \underset{\{ \vec{q}\}}{\int} F_{2}(\vec{k} - \vec{q}, \vec{q}) G_{2}(\vec{k} - \vec{q}, \vec{q}) P_{\text{L}}(|\vec{k} - \vec{q}|) P_{\text{L}}(q) + 3\, P_{\text{L}}(k)  \underset{\{ \vec{q}\}}{\int}( F_{3} (\vec{k}, -\vec{q}, \vec{q}) + G_{3} (\vec{k}, -\vec{q}, \vec{q})) P_{\text{L}}(q) \nonumber \,,
		\end{eqnarray}	
		\begin{eqnarray}
			\label{ptt}
			P_{\theta \theta}(k) =P_{\text{L}}(k) + 2 \underset{\{ \vec{q}\}}{\int} G_{2}(\vec{k} - \vec{q}, \vec{q})^{2} P_{\text{L}}(|\vec{k} - \vec{q}|) P_{\text{L}}(q) + 6\, P_{\text{L}}(k)  \underset{\{ \vec{q}\}}{\int} G_{3} (\vec{k}, -\vec{q}, \vec{q}) P_{\text{L}}(q) \,
		\end{eqnarray}
where $F_{n}$ and $G_{n}$ represent SPT kernels~\cite{2002PhR...367....1B}, e.g. ($x_{12}\equiv \widehat{\vec{q}}_1\cdot \widehat{\vec{q}}_2$)
\beqa
&&F_{2}(\vec{q}_{1},\vec{q}_{2}) = {5\over 7} + {x_{12}\over 2}\Big({q_{1}\over q_{2}} + {q_{2}\over q_{1}}\Big) + {2\over 7} \, x_{12}^2, \ \ \ \ \ 
G_{2}(\vec{q}_{1},\vec{q}_{2}) = {3\over 7} + {x_{12}\over 2}\Big({q_{1}\over q_{2}} + {q_{2}\over q_{1}}\Big) + {4\over 7} \, x_{12}^2, \ \ \ \ \ 
K(\vec{q}_{1},\vec{q}_{2}) =   x_{12}^2-1. \nonumber \\ & & 
\label{F2G2K}
\eeqa
For  the bispectra  appearing in Eq.~\eqref{final-local} we have,
	\begin{eqnarray}
		\label{bggt}
		B_{g \theta g}(\vec{k}_{1}, \vec{k}_{2}, -\vec{k}_{12})\, &=& 
		 \Gamma^{(2)}_{g}(\vec{k}_{1}, \vec{k}_{2}) \Gamma^{(1)}_{g}(k_{1})  \Gamma^{(1)}_{\theta}(k_{2}) P_{\text{L}}(k_{1})P_{\text{L}}(k_{2}) 
		 + \Gamma^{(2)}_{\theta}(\vec{k}_{1}, -\vec{k}_{12}) \Gamma^{(1)}_{g}(k_{1})  \Gamma^{(1)}_{g}( k_{12}) P_{\text{L}}(k_{1})P_{\text{L}}(k_{12}) \nonumber \\&& 
		 + \Gamma^{(2)}_{g}(\vec{k_{2}}, -\vec{k_{12}}) \Gamma^{(1)}_{g}(k_{2})  \Gamma^{(1)}_{\theta}(k_{12}) P_{\text{L}}(k_{2})P_{\text{L}}(k_{12}) \nonumber 
	\end{eqnarray}
	and similar expressions for other combinations $B_{g \theta \theta}$, $B_{\theta \theta g}$ and $B_{\theta \theta \theta}$, with $\Gamma$-functions up to the needed order given by~\cite{2019PhRvD..99l3514E, 2020arXiv200609729E},
	\beq
	\Gamma^{(1)}_g(k) = b_1, \ \ \ \ \  \Gamma^{(1)}_\theta(k) = 1, \ \ \ \ \	
	\Gamma^{(2)}_{g}(\vec{k}_{1}, \vec{k}_{2}) = 2\,b_{1} F_{2}(\vec{k}_{1}, \vec{k}_{2}) + b_{2} + 2\,\gamma_{2} K(\vec{k}_{1}, \vec{k}_{2}),  \ \ \ \ \ 
	\Gamma^{(2)}_{\theta}(\vec{k}_{1}, \vec{k}_{2}) = 2\, G_{2}(\vec{k}_{1}, \vec{k}_{2})
	\eeq
	 Now we turn to a discussion of the impact of going beyond the plane-parallel approximation from these one-loop expressions. 
	\end{widetext}
	
	\subsubsection{Inverse-distance and wide-angle effects}
	We proceed in the same way as done in linear theory. Considering the geometry of Fig.~\ref{fig:midp} again, the inverse-distance corrections come from $1/x_{n}$ terms in Eq.~\eqref{final-local} by setting $x_{1} = x_{2}= x_{c}$ and $\hat{x}_{1} = \hat{x}_{2} = \hat{x}_{c}$. Then, since there will be no explicit dependence 
	on $\vec{x}_{1}$ and $\vec{x}_{2}$ in the $D$-functions, the integral over $\vec{x}_{12}$ turns into a delta function and we are left with just one loop integral which can be easily evaluated numerically. At the end, these terms produce $1/(kx_{c})^2$ corrections but without wide-angle contributions. 	For wide-angle corrections, before applying the perturbative method to evaluate the integrals in Eq.~\eqref{final-local}, it is instructive to highlight a few steps one can take to reduce the number of parameters involved and to make the expressions more amenable to perturbative expansion. As an example, let's consider the second and forth integrals in Eq.~\eqref{final-local} (others are simpler) which take the following general form:
	\begin{eqnarray}
		\label{samp1}
		\mathcal{I}_{2, 4} = \underset{\vec{x}_{12},\vec{k}_{1},\vec{k}_{2} }{\int} e^{i(\vec{k}_{12}-\vec{k})\cdot\vec{x}_{12} }\ \mathcal{C}( \vec{k}_{1}, \vec{k}_{2},\vec{x}_{1},\vec{x}_{2})\,
	\end{eqnarray}
	where $\mathcal{C}$ contains all parameters and functions in the integrand. This integral, via coordinate transformations $\vec{k}_{1} + \vec{k}_{2} = \vec{k}_{R}$ and $\vec{k}_{1} - \vec{k}_{2} = \vec{k}_{r}$, can be re-written as:
	\begin{equation}
		\label{samp2}
		\mathcal{I}_{2, 4} = \frac{1}{8}\underset{\vec{x}_{12},\vec{k}_{R},\vec{k}_{r}}{\int} e^{i(\vec{k}_{R}-\vec{k})\cdot\vec{x}_{12} }\ \mathcal{C}( \vec{k}_{R}, \vec{k}_{r},\vec{x}_{1},\vec{x}_{2})\,.
	\end{equation}
	Inside $\mathcal{C}$, there are combinations of products such as $\hat{k}_{R}\cdot\hat{k}_{r}$, $\hat{k}_{R}\cdot\hat{x}_{1}$, $\hat{k}_{r}\cdot\hat{x}_{1}$, etc. We first take the integral over $\vec{k}_{r}$. For this purpose, we choose the coordinate system whose $``z"$ direction is defined by $\hat{k}_{R}$, i.e. $\hat{z}' \equiv \hat{k}_{R}$ and $\hat{x}'$, $\hat{y}'$ are defined accordingly on its normal plane. As an example, let us consider the following integral as one of the most general form of the terms that appear in Eq.~\eqref{samp2}:
	\begin{eqnarray}
		\label{samp3}\underset{\vec{x}_{12},\vec{k}_{R},\vec{k}_{r} }{\int}&& e^{i(\vec{k}_{R}-\vec{k})\cdot\vec{x}_{12} } (\hat{k}_{r}\cdot\hat{x}_{1})^{2} (\hat{k}_{r}\cdot\hat{x}_{2})^{2} \times \nonumber \\
		&& \times\, \mathcal{F}( \vec{k}_{R}\cdot\hat{x}_{1}, \vec{k}_{R}\cdot\hat{x}_{2}, \hat{k}_{r}\cdot\hat{k}_{R}, k_{r}, k_{R})\, 
	\end{eqnarray}
	with $\mathcal{F}$ being a function. The first expression can be decomposed in the above-mentioned coordinate system as: $(\hat{k}_{r}\cdot\hat{x}_{1})^{2} (\hat{k}_{r}\cdot\hat{x}_{2})^{2} = ( \hat{k}_{r i'} \hat{x}_{1 i'} )^2  ( \hat{k}_{r j'} \hat{x}_{2 j'})^2$ with reapeted indices are being summed over and $i'$, $j'$ denoting (${x}'$, ${y}'$, ${z}'$) each. Now we can write the components in terms of the corresponding spherical coordinates: $\hat{k}_{r x'} = \sin(\theta_{r}) \cos(\phi_{r})$, $\hat{k}_{r y'} = \sin(\theta_{r}) \sin(\phi_{r})$, $\hat{k}_{r z'} =  \cos(\theta_{r})$ and similarly,  $\hat{x}_{1x'} = \sin(\theta_{1}) \cos(\phi_{1})$, $\hat{x}_{1 y'} = \sin(\theta_{1}) \sin(\phi_{1})$, $\hat{x}_{1 z'} =  \cos(\theta_{1})$, etc. Also let's define $ \hat{k}_{r}\cdot\hat{k}_{R} = \cos(\theta_{r}) \equiv \mu_{r}$. Plugging all these into Eq.~\eqref{samp3}, we first take the integral over $\phi_{r}$ (note that arguments of $\mathcal{F}$ are all independent of $\phi_{r}$). The result, after some algebra, is:
	\begin{widetext}
		\begin{eqnarray}
			\label{samp4}
			\underset{\vec{x}_{12},\vec{k}_{R} }{\int}&&e^{i(\vec{k}_{R}-\vec{k})\cdot\vec{x}_{12} }\int dk_{r} d\mu_{r}  k_{r}^{2} (\frac{\pi}{4} )  \bigg [ \left(1 + 2 (\hat{x}_{1}\cdot\hat{x}_{2})^{2}\right)(\mu_{r}^{2} - 1)^2 - 4 \left(1 - 6 \mu_{r}^2 + 5 \mu_{r}^4 \right)(\hat{x}_{1}\cdot\hat{x}_{2}) (\hat{k}_{R} \cdot \hat{x}_{1}) (\hat{k}_{R} \cdot \hat{x}_{2}) + \nonumber \\
			&& +\, \left(-1 + 6 \mu_{r}^2 - 5 \mu_{r}^4 \right) \left( (\hat{k}_{R} \cdot\hat{x}_{1})^{2} + (\hat{k}_{R} \cdot \hat{x}_{2})^{2} \right) + (3 - 30 \mu_{r}^{2} + 35 \mu_{r}^{4})\, (\hat{k}_{R} \cdot\hat{x}_{1})^{2} (\hat{k}_{R} \cdot \hat{x}_{2})^{2} \bigg ] \times \\
			&& \times \mathcal{F}( \vec{k}_{R}\cdot \hat{x}_{1}, \vec{k}_{R}. \hat{x}_{2}, \mu_{r}, k_{r}, k_{R}) \nonumber
		\end{eqnarray}
	\end{widetext}
	Now, considering the 2D integral inside, we are left with two remaining integrals (over $\vec{k}_{R}$ and $\vec{x}_{12}$) and a dependence on $\hat{x}_{1}.\hat{x}_{2}$ that can be added to the $\mathcal{F}$ function. The same is true for the first and third integrals of Eq.~\eqref{final-local} and they are even simpler because there is no $\hat{x}_{1}.\hat{x}_{2}$ term produced in the calculations. For the first one the analogue of the 2D integral is the loop integral inside power spectra. For the third one, $\vec{k}_{2}$ plays the role of $\vec{k}_{r}$ and the steps are the same. 
	
	The two-dimensional  loop integrals are the only objects that need to be evaluated numerically which is not difficult given their simple structure. The rest of the calculations are analytic. Therefore, the most general outcome for each of the four integrals of the local power spectrum Eq.~\eqref{final-local} takes the form:
	
	\begin{eqnarray}
		\label{o1}
		&&\mathcal{I}_{\alpha} = \\ 
		&&\underset{\vec{x}_{12},\vec{k}'  }{\int} e^{i(\vec{k}'-\vec{k})\cdot\vec{x}_{12} }\  \mathcal{C}_{\alpha} (\hat{k}'\cdot\hat{x}_{1},\, \hat{k}'\cdot\hat{x}_{2},\, \hat{x}_{1}\cdot\hat{x}_{2},\, k',\, \frac{1}{x_{1}},\, \frac{1}{x_{2}} ) 
		\nonumber 
	\end{eqnarray}
	 with $\alpha = 1, 2, 3, 4$. This is where we can employ the perturbative expansions to find wide-angle corrections to any desired order. Replacing the unit vectors and $1/x$ factors with the expansions in Eq.~\eqref{pexp}, we can follow the same steps as in linear case (Eqs \ref{o2} -- \ref{so3}) and find all leading order wide-angle contributions. Although the procedure is straightforward, the number of terms involved is quite large and therefore we do not write them down explicitly here.
		
	\begin{figure*}
		\centering
		\includegraphics[width=1\textwidth]{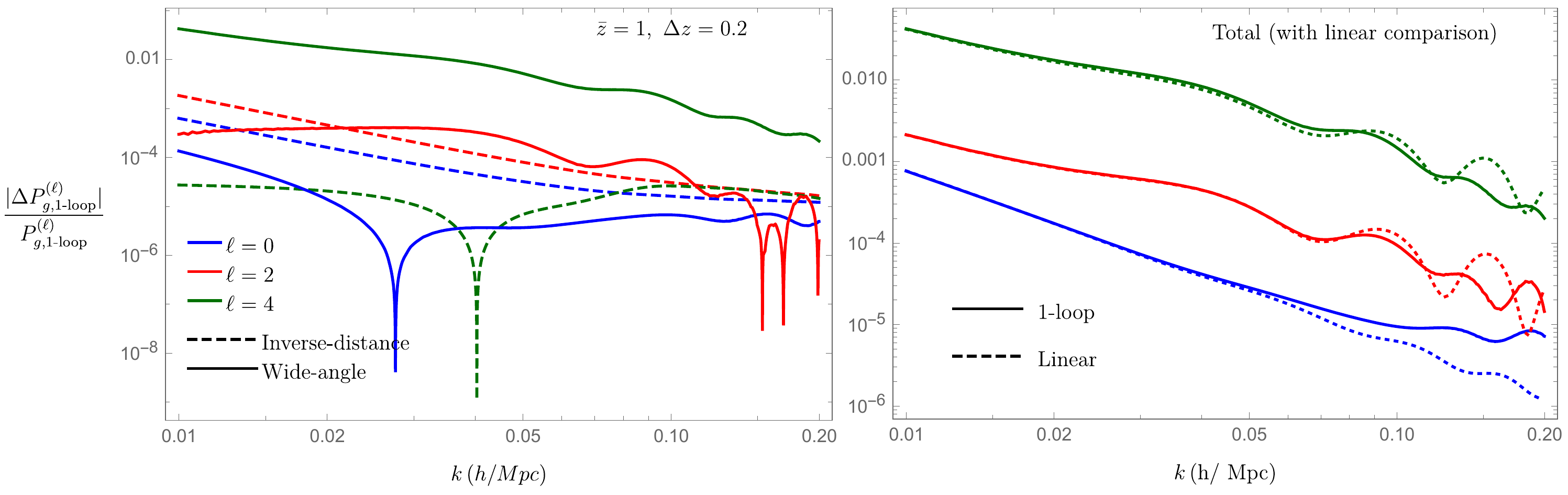}
		\caption{Non-linear beyond plane-parallel corrections to multipoles of 1-loop galaxy power spectrum, for a Euclid-like galaxy sample. The left figure separates wide-angle and inverse-distance corrections, both relative to nonlinear plane-parallel multipoles. The bias parameters are $b_{1} = 1.46$ and $b_{2}$, $\gamma_{2}$ and $\gamma_{21}$ which are obtained from Eqs.~(\ref{biasn1}-\ref{biasn3}). The figure on the right combines both corrections and shows the total result. Also on that panel, the total nonlinear corrections are compared with the corresponding linear ratio.  }
		\label{fig:figure3}
	\end{figure*}
	The final result for the one-loop local galaxy power spectrum beyond the plane-parallel limit takes the following form:
	\begin{eqnarray}
		\label{final-local3}
		&P_{g,s}^{\text{loc}}(\vec{k},\vec{x}_{c}) &= P_{\text{g,1-loop}}^{pp}(k,\eta) + \frac{1}{(k x_{c})^2} P_{\text{g,1-loop}}^{\text{id}}(k,\eta)\nonumber \\
		&&+ \frac{1}{(k x_{c})^2} P_{\text{g,1-loop}}^{\text{wa}}(k,\eta) + \mathcal{O}(\frac{1}{(k x_{c})^4})
	\end{eqnarray}
	where as before $\eta \equiv \hat{k} \cdot \hat{x}_{c}$. $P_{\text{g,1-loop}}^{pp}(k,\eta) $ represents the plane-parallel result of Eq.~\eqref{final-local} which is obtained by neglecting all inverse-distance terms in the D-functions (see Eq.~\ref{D-functions}) and also setting $\hat{x}_{1} = \hat{x}_{2} = \hat{x}_{c}$. The following two terms are inverse-distance and wide-angle corrections respectively, which are obtained following the steps explained  previously. For bias parameters we use local-Lagrangian values for $\gamma_2,\gamma_{21}$ and peak-background split fit for quadratic bias~\cite{2012PhRvD..85h3509C,2012PhRvD..86h3540B,2016JCAP...02..018L,2019PhRvD..99l3514E,2020arXiv200609729E}:
	\begin{align}
		\label{biasn1}
		b_{2} & = 0.412 - 2.143 b_{1} + 0.929 b_{1}^{2} +  0.008 b_{1}^{3} + \frac{4}{3} \gamma_{2} \,, \\ \label{biasn2}
		\gamma_{2} &= - \frac{2}{7} (b_{1} - 1)\,, \\ \label{biasn3}
		\gamma_{21} &= \frac{2}{21}(b_{1} - 1) + \frac{6}{7} \gamma_{2}\,.
	\end{align}

	Inserting Eq.~\eqref{final-local3} into Eq.~\eqref{estp1}, one can find the local multipoles and then, by plugging the result into Eq.~\eqref{estp2}, find the final multipoles. For the plane-parallel one-loop case, this yields even parity multipoles up to $\ell = 8$ and wide-angle corrections increase it to $\ell = 10$. However, given the difficulty in detecting high multipoles, we only plot the first three, using Eq. \eqref{VP} for a Euclid like survey bin \cite{Euclid2020.Blanchard.et.al} with $\bar{z} = 1$, $\Delta z = 0.2$ and $b_{1} = 1.46$. It is worth reminding that for loop calculations, we didn't include magnification bias. 
	
    Figure \ref{fig:figure3} shows relative plots of such corrections for this survey bin. The figure on the right shows the total corrections which are also compared with their linear counterparts. Although loop effects start becoming important after $k \gtrsim 0.1 $ $h$/Mpc, they don't make any  enhancement in the magnitudes of corrections beyond the plane-parallel approximation that could be important in practice. The most significant enhancement is seen in the monopole, but this happens when the corrections are already well below detectable levels for the foreseeable future.  In any case, the perturbative procedure developed here can be a fast and easy way for assessing the importance of such effects for different galaxy samples, if desired.

	\subsubsection{Velocity dispersion and cosmological distortions}
	
	Velocities of galaxies within virialized halos can be significantly larger than peculiar velocities predicted from linear theory. This causes apparent elongation of galaxy clusters along the line-of-sight known as the Fingers of God (FoG) effect. To see how this changes in the presence of corrections to the plane-parallel approximation, it is enough to consider a simple approach to velocity dispersion effects, where the pairwise velocity distribution is Gaussian and scale-independent characterized by variance $\sigma_{p}$. In this case,  the linear redshift space power spectrum gets modulated by a damping factor~\cite{1994MNRAS.267.1020P, 1995MNRAS.275..515C, 1996MNRAS.280L..19P}. Neglecting wide-angle and relativistic effects for now, this reads
	\begin{eqnarray}
		\label{fog}
		&&P^{\text{loc}}_{g, L}(\vec{k}, \vec{x}_{c}) = \\ 
		&& (b_{1} + f\, (\hat{k}\cdot\hat{x}_{c})^{2})^{2} \exp\left[- \frac{(k \sigma_{p} (\hat{k}\cdot\hat{x}_{c}))^{2}}{2}\right] P_{L}(k) \nonumber
	\end{eqnarray}
	\begin{figure}[t!]
	\centering
	\includegraphics[width=0.5\textwidth]{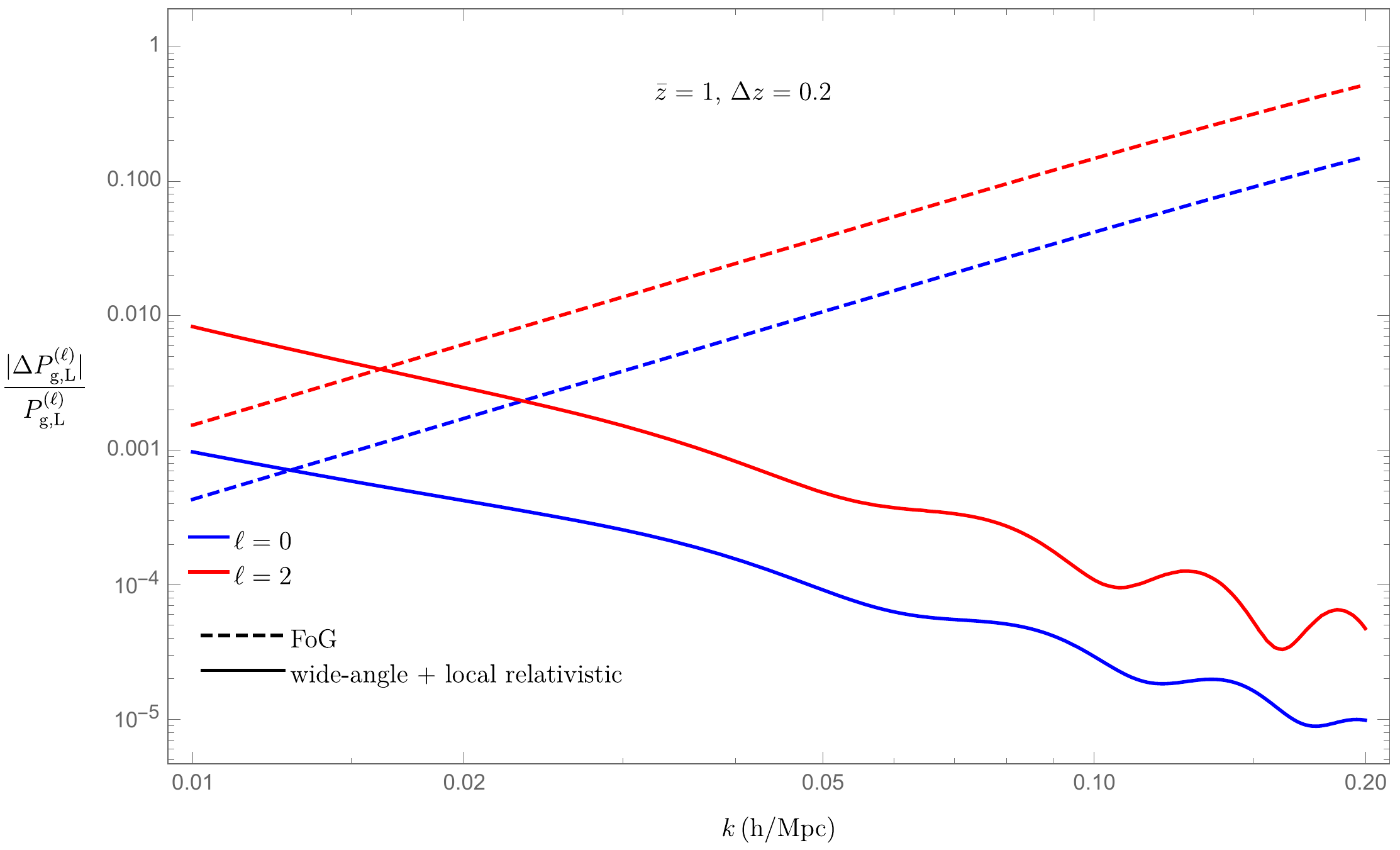}
	\caption{Comparison between corrections to plane-parallel monopole and quadrupole power and the corresponding corrections due to velocity dispersion effects. Here we have included local (non-integrated) relativistic corrections only, for the sake of illustration.}
	\label{fig:figurefog}
\end{figure}
	Similar to previous cases, the final multipoles for this modulated local power spectrum can be found. Figure~\ref{fig:figurefog} shows a comparison between FoG corrections to the monopole and quadrupole power against the corrections from linear wide-angle and local (non-integrated) relativistic terms we found earlier. We have used $\sigma_{p} \approx$ 4.5 Mpc/$h$ for $\bar{z} = $1 from~\cite{2019MNRAS.483.2078Y}, $b_{e} = 0$, $s=1$ and other parameters are as presented in the 1-loop case. This figure implies that one can ignore the mixing between wide-angle/relativistic corrections with FoG corrections at large scales shown as these are all sub-leading. In fact, wide-angle and relativistic corrections are suppressed by $1/(kx_{c})^2$ and $(\mathcal{H}/k)^{2}$ while leading FoG corrections are suppressed by $\sim(k \sigma_p)^2$. Therefore, the mixing are suppressed by $\sim\sigma_p/x_{c}$ and $\sim\mathcal{H}\sigma_p$ which are subleading to wide-angle/relativistic terms at these scales. Therefore, one can separately modulate the plane-parallel power spectrum as in Eq.~\eqref{fog} and then add wide-angle and relativistic contributions. In other words:
	\begin{eqnarray}
		\label{fog1}
		&&P^{\text{loc}}_{g, L}(\vec{k}, \vec{x}_{c}) = \\ 
		&& P^{\text{PP, FoG}}_{g, L}(\vec{k}, \vec{x}_{c}) + \text{wide-angle}+ \text{relativistic}\,. \nonumber
	\end{eqnarray}
Therefore, in practice, one can compute the nonlinear redshift-space power spectrum (including loop and FoG effects) in the plane-parallel approximation and  then simply add linear wide-angle and relativistic corrections on top.	

    Finally, let us briefly comment on the impact of going beyond the plane-parallel approximation in cosmological distortions. The    Alcock-Paczynski (AP) parameters ($\alpha_{\perp}$, $\alpha_{\parallel}$) \cite{1979Natur.281..358A}  scale the actual modes in perpendicular and parallel directions to account for the anisotropic distortion caused by the assumption of the wrong cosmology when mapping angles and redshifts to comoving positions or wavevectors. In other words, one replaces $k_{\parallel} \to k_{\parallel} / \alpha_{\parallel}$ (or $x_{\parallel} \to x_{\parallel}  \alpha_{\parallel}$) and $k_{\perp} \to k_{\perp} / \alpha_{\perp}$ (or $x_{\perp} \to x_{\perp}  \alpha_{\perp}$) and constrains $\alpha$'s from data (using the true cosmological model leads to both parameters being measured to be $1$). This parametrization can be easily applied to the previous calculations of wide-angle effects based on local power spectrum definitions. For example, at the linear level, one can start from Eq.~\eqref{final-pr} and rescale $k$, $x_{c}$ and $\eta \equiv \hat{k}.\hat{x}_{c}$ as:
    \begin{eqnarray}
	\label{AP1}
	&& k \to k\, \left (\frac{\eta^{2}}{\alpha_{\parallel}^{2}} + \frac{(1 - \eta^{2})}{\alpha_{\perp}^{2}} \right )^{1/2}\,, \\ \nonumber
	&& x_{c} \to \alpha_{\parallel}\, x_{c}\,, 	\\
	&& \eta \to \frac{\eta}{\alpha_{\parallel}}  \left(\frac{\eta^{2}}{\alpha_{\parallel}^{2}} + \frac{(1 - \eta^{2})}{\alpha_{\perp}^{2}} \right )^{-1/2} \,. \nonumber
\end{eqnarray}
    In this way, AP effects can be included in $k$-space in the presence of wide-angle corrections at both linear and nonlinear regimes. Such effects have been investigated for  wide-angle two-point correlation functions of galaxies at the linear level in~\cite{2021arXiv210308126S}.
    
  \subsubsection{Wide-angle and relativistic corrections in the presence of nonilinearities}
    
We now summarize the results of previous parts and make a general conclusion. Figures~\ref{fig:figure3} and \ref{fig:figurefog} imply that at large scales the presence of nonlinear effects will not change the relative size of the wide-angle corrections. The same thing can be said about relativistic effects. This means that in practice we can separately calculate corrections from non-linear terms and wide-angle and relativistic contributions and calculate the latter in linear theory. In other words we have:
\begin{eqnarray}
	\label{nlnarw}
			&&P^{\text{loc}}_{g, L}(\vec{k}, \vec{x}_{c}) = \\ 
	&& P^{\text{PP, L}}_{g, L}(\vec{k}, \vec{x}_{c}) + \text{non-linear corrections} + \nonumber \\ &&\text{linear wide-angle}+\text{linear relativistic} + \nonumber \\ && 
	\text{subleading cross-terms}\,. \nonumber
\end{eqnarray}
which significantly simplifies the calculations. 

 \subsection{Inclusion of window effects}
 \label{window-geometry}
  
 We finish the power spectrum section by discussing how the results of previous parts can be generalized to situations where window effects become important. Here by window function we also mean the function that determines the shape of the survey volume. Beginning from Eq. \eqref{window2}, one can write the product of window functions as 
 	\begin{equation}
	\label{ww}
	W(\vec{x}_{1}) W(\vec{x}_{2}) = \underset{\vec{x}_{12}'}\int \delta_{D}(\vec{x}_{12}' - \vec{x}_{12}) 	W(\vec{x}_{c}+\frac{\vec{x}_{12}'}{2}) W(\vec{x}_{c}-\frac{\vec{x}_{12}'}{2})
\end{equation} 
which leads to the convolution:
  \begin{widetext}
 	\begin{equation}
 		\label{window3}
 		P_{g}^{\ell}(k)= \frac{(2 \pi)^3}{V_{s}}(2 \ell+1)\int \frac{d \Omega_{k}}{4 \pi}\underset{\vec{x}_{c}}\int \mathcal{L}_{\ell}(\hat{k}\cdot\hat{x}_{c}) \int\frac{d^3q}{(2 \pi)^3}\underset{\vec{x}_{12}}\int e^{{-i(\vec{k}- \vec{q})\cdot\vec{x}_{12}}} \left\langle \begin{matrix} \delta_{g} \left( \vec{x}_{1}\right)\delta_{g} \left( \vec{x}_{2}\right) \end{matrix}  \right\rangle \underset{\vec{x}_{12}'}\int  e^{{-i \vec{q}\cdot\vec{x}_{12}'}} W(\vec{x}_{c}+\frac{\vec{x}_{12}'}{2}) W(\vec{x}_{c}-\frac{\vec{x}_{12}'}{2})
 	\end{equation}
 \end{widetext}
 where the integration over $d^3 q$ comes from the definition of delta function for $\delta_{D}(\vec{x}_{12}' - \vec{x}_{12})$ . The integral in front of $d^3 q$ is the local power spectrum according to Eq. \eqref{localp} which is exactly what we calculated in previous parts. One can use those results and replace $\vec{k}$ with $\vec{k} - \vec{q}$. More straightforwardly, one can make a change of variables: $\vec{k} - \vec{q} \equiv \vec{p}$ to write:
   \begin{widetext}
 	\begin{equation}
 		\label{window4}
 		P_{g}^{\ell}(k)= \frac{(2 \pi)^3}{V_{s}}(2 \ell+1)\int \frac{d \Omega_{k}}{4 \pi}\underset{\vec{x}_{c}}\int \mathcal{L}_{\ell}(\hat{k}\cdot\hat{x}_{c}) \int\frac{d^3p}{(2 \pi)^3} P_{g}^{\text{loc}}(\vec{p},\vec{x}_{c})\underset{\vec{x}_{12}'}\int  e^{{-i (\vec{k} - \vec{p})\cdot\vec{x}_{12}'}} W(\vec{x}_{c}+\frac{\vec{x}_{12}'}{2}) W(\vec{x}_{c}-\frac{\vec{x}_{12}'}{2})\,.
 	\end{equation}
 \end{widetext}
This result implies that in order to incorporate window effects which also include the effects of non-trivial geometry all one needs to do is to find the convolution between the results of previous parts (for $P_{g}^{\text{loc}}(\vec{p},\vec{x}_{c})$)  and window products. It is clear from this integral that for modes that are well below the  survey size (i.e. $k^{-1}$ much smaller than survey size), the integral over window functions can be approximated with a delta function that reproduces the previous results (the window function always enforces the integral over $x_{c}$ to be limited to the survey volume, as we also assumed). 
 
\section{The Redshift-Space Bispectrum}
	\label{bispectrum}
\begin{figure}
	\centering
	\includegraphics[width=0.27\textwidth]{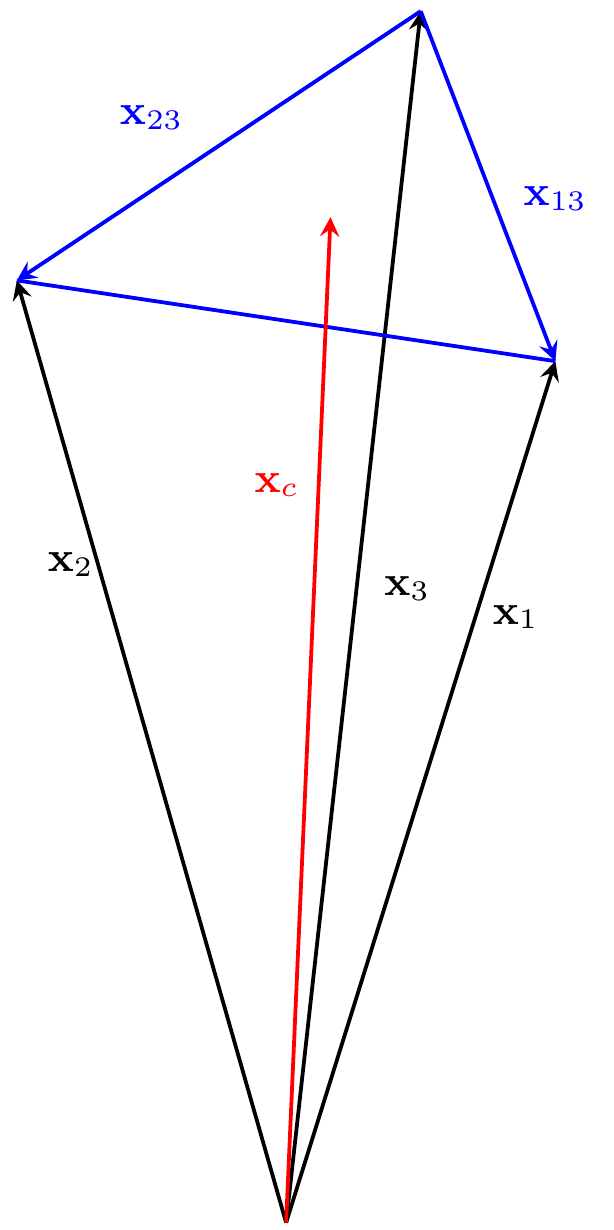}
	\caption{Relevant geometry for local bispectrum calculations. Here $\vec{x}_{c}$ is the centroid of the triangle which we take to be te line-of-sight. Also, $\vec{x}_{n m} = \vec{x}_{n} - \vec{x}_{m}$.}
	\label{fig:midpb}
\end{figure}

	Similar to the power spectrum case, we first start with the definition of the local bispectrum \cite{2015PhRvD..92h3532S}:
	\begin{eqnarray}
		\label{bispectrum1}
		B_{g,s}^{loc}(\vec{k}_{1},\vec{k}_{2},\vec{x}_{c})&=& \int \frac{d^{3}\vec{x}_{13}} {\left( 2\pi \right) ^{3}}\frac{d^{3}\vec{x}_{23}} {\left( 2\pi \right)^{3}}\,e^{-i\vec{k}_{1}\cdot\vec{x}_{13}}e^{-i\vec{k}_{2}\cdot\vec{x}_{23}}\times
		\nonumber \\
		&&\left\langle\delta_{g,s} \left( \vec{x}_{1}\right)\delta_{g,s} \left( \vec{x}_{2}\right) \delta_{g,s} \left( \vec{x}_{3}\right) \right\rangle
	\end{eqnarray}
	for which we are considering triangle configurations in $k$-space satisfying  $\vec{k}_{3} + \vec{k}_{1} + \vec{k}_{2} = 0$. For the local bispectrum, $\vec{x}_{c} = (\vec{x}_{1} + \vec{x}_{2} + \vec{x}_{3})/3$ is the centroid of the corresponding real-space triangles made by the points at which vectors $\vec{x}_{1}$, $\vec{x}_{2}$ and $\vec{x}_{3}$ end. The geometry of the problem is drawn in Figure~\ref{fig:midpb}. We choose $\vec{x}_{c}$ to define our LOS for the local bispectrum. This has the advantage of being fully symmetric with respect to point exchanges.
	To describe $k$-space triangle configurations one needs five parameters. Using the parametrization in \cite{1999ApJ...517..531S}, three of these can be $k_{1}$, $k_{2}$ and the cosine of angle between their vectors ($\hat{k}_{1}\cdot\hat{k}_{2} \equiv \cos\theta \equiv \mu$).  The other two determine the orientation of the triangle with respect to the observer. One of them can be defined, without loss of generality, as the angle between the centroid (LOS) vector and  the $k_{1}$ side ($\cos\theta_{1} \equiv \hat{k}_{1}\cdot\hat{x}_{c}$). The other one, $\phi_{12}$, can be defined in the following way: in the plane whose normal is determined by $\vec{k}_{1}$, $\phi_{12}$ is the angle between projections of $\vec{k}_{2}$ and $\vec{x}_{c}$ onto that plane. This implies  $\hat{k}_{2} \cdot \hat{x}_{c} = \mu \cos \theta_{1}+ \sqrt{1 - \mu^2}\, \sin\theta_{1}\, \cos\phi_{12}$. 

	The local bispectrum can then be expanded in terms of spherical harmonics defined by orientation angles:
	\begin{equation}
		\label{bi-expand}
		B_{g,s}^{loc}(\vec{k}_{1},\vec{k}_{2},\vec{x}_{c})= \sum_{\ell, m}\, \mathcal{B}_{g, s}^{(\ell,m)}({k}_{1},{k}_{2},\mu, {x}_{c})\ Y_{\ell m }(\theta_{1},\phi_{12})\,, \nonumber
	\end{equation}
	and, similar to the power spectrum, ``local  multipoles" of the bispectrum can be obtained from:
	\begin{eqnarray}
		\label{estB1}
		&&\mathcal{B}_{g,s}^{(\ell,m)}({k}_{1},{k}_{2},\mu, x_{c})  =
		\nonumber \\
		&&\frac{(2\ell + 1)}{4\pi} \int d\Omega_{T}\, B_{g,s}^{loc}(\vec{k}_{1},\vec{k}_{2},\vec{x}_{c})\ Y_{\ell m}^{*}(\theta_{1},\phi_{12}) \,.
	\end{eqnarray}
    where $d\Omega_{T} \equiv d\cos\theta_{1} \, d\phi_{12}$. Ultimately, the ``final" multipoles can be calculated by integrating over LOS vectors within the survey volume:
	\begin{eqnarray}
		\label{estB2}
		&&B_{g,s}^{(\ell,m)}({k}_{1},{k}_{2},\mu) =  \int_{V_{s}}\frac{d^{3}\vec{x}_{c}}{V_{s}}\ \mathcal{B}_{g,s}^{(\ell,m)}({k}_{1},{k}_{2},\mu, x_{c})\,. \nonumber \\
		&&
	\end{eqnarray}
	
	The plane-parallel multipoles of the bispectrum can be obtained taking the limit $\vec{x}_{1} = \vec{x}_{2} = \vec{x}_{c}$ and $\vec{x}_{c} \to \infty$. In the following subsections we go beyond this limit and investigate contributions from relativistic and wide-angle terms. 
	
\subsection{Odd-Parity (Imaginary) Multipoles from Relativistic and Wide-angle contributions}
	\label{b-imaginary}
	Inclusion of relativistic corrections to the galaxy number density contrast can lead to imaginary odd-parity multipoles \cite{2019MNRAS.486L.101C, 2019arXiv190605198J,2020JCAP...03..065M} that are totally absent in the plane-parallel limit. In this section we show that, in addition to relativistic terms, wide-angle corrections alone produce odd-parity multipoles too. We calculate these imaginary terms from different physical origins and compare them. 
	
	For the tree-level galaxy bispectrum, only terms up to second order in $\delta$ or $\vec{u}$ are needed. Also for relativistic contributions, we only need to keep terms that produce odd-parity multipoles in which we are interested in this part. We use the results in \cite{2019JCAP...04..050D} (see also \cite{2014PhRvD..90b3513Y, 2014JCAP...09..037B, 2020JCAP...09..058D} ) for relevant second-order relativistic terms. Combining Eq.~\eqref{exp22} with these terms leads to:
	\begin{widetext}
		\begin{eqnarray}
			\label{gal-a}
			\delta_{g,s}(\vec{x}_{1}) &=&  \delta_{g}(\vec{x}_{1}) +\biggl [ f\,\hat{x}_{1j}\, \partial_{x_{1j}}\mathlarger{[}  (\vec{u}_{g}(\vec{x}_{1})\cdot \hat{x}_{1}) (1 + \delta_{g}(\vec{x}_{1}))\mathlarger{]}  + \frac{f^{2}}{2}\,\hat{x}_{1k}\,\hat{x}_{1j} \,\partial_{x_{1j}} \partial_{x_{1k}}\, \mathlarger{[} (\vec{u}_{g}(\vec{x}_{1})\cdot\hat{x}_{1})^2 \mathlarger{]}\biggl] \nonumber \\
			&&  +\,\frac{1}{x_{1}}\biggl[ f\,(2 -5s)(\vec{u}_{g}(\vec{x}_{1}) \cdot\hat{x}_{1}) \mathlarger{[}1 + \delta_{g}(\vec{x}_{1})\mathlarger{]} + (2 -\frac{5}{2}s)f^2 \,\hat{x}_{1j}\, \partial_{x_{1j}}\mathlarger{[}  (\vec{u}_{g}(\vec{x}_{1})\cdot\hat{x}_{1})^2\mathlarger{]}\biggl]  \\
			&&+\, \mathcal{H}\, \biggl[ \mathcal{A}_{1} \,f\, \vec{u}_{g}(\vec{x}_{1})\cdot\hat{x} + \mathcal{A}_{2}\,f^{2}\, \hat{x}_{1j}\, \partial_{x_{1j}}\mathlarger{[}  (\vec{u}_{g}(\vec{x}_{1})\cdot\hat{x}_{1})^2\mathlarger{]} + \mathcal{A}_{1}\,f\,\vec{u}_{g}(\vec{x}_{1})\cdot\hat{x}_{1} \, \delta_{g}(\vec{x}_{1}) - \frac{f}{\mathcal{H}} \vec{u}_{g}(\vec{x}_{1})\cdot\hat{x}_{1} \dot{\delta}_{g}(\vec{x}_{1})\nonumber \\	
			&&   - 2 \, f^{2}\,  \vec{u}_{g j} \partial_{{x}_{1 j}}  (\vec{u}_{g}(\vec{x}_{1})\cdot\hat{x}_{1}) + \frac{f}{\mathcal{H}^{2}}\, \phi \,\, \hat{x}_{1 j} \hat{x}_{1 k}\, \partial_{x_{1j}} \partial_{x_{1k}}\, (\vec{u}_{g}(\vec{x}_{1})\cdot\hat{x}_{1}) + \frac{1}{\mathcal{H}^{2}}\, \phi\,\, \hat{x}_{1 j} \partial_{x_{1 j}} \delta_{g}(\vec{x}_{1})  \nonumber \\
			&& - \frac{f}{\mathcal{H}^2} (\vec{u}_{g}(\vec{x}_{1})\cdot\hat{x}_{1})\, \hat{x}_{1 j} \hat{x}_{1 k}\, \partial_{x_{1j}} \partial_{x_{1k}} \phi \biggl ] \,. \nonumber 
		\end{eqnarray}
	\end{widetext}
	where $\delta_{g}$'s here are in SCG (as the difference with the CNG parameter is subleading for odd-parity multipoles) but we have dropped the "tilde" for simplicity. Also here $\mathcal{A}_{1} \equiv (5s+1 -\frac{3}{2}\Omega_{M} - b_{e})$ and $\mathcal{A}_{2} \equiv (3 - \frac{9}{4} \Omega_{M} + \frac{5}{2}s -  b_{e} )$. Also we have used $\dot{\mathcal{H}}/\mathcal{H}^2= 1 - (3/2)\, \Omega_{M}$.
	
	There are three brackets after $\delta_{g}(\vec{x}_{1})$ in this equation. The first one is the RSD contribution. The second one is inverse-distance term which is part of Eq.~\eqref{exp22} but its coefficient has been modified to include magnification bias \cite{2019JCAP...04..050D}. Since the leading suppression factor in the imaginary bispectrum will be $1/(k x_{c})$, we neglect $1/x_{n}^2$ terms in Eq.~\eqref{exp22}. They will be relevant for real (even parity) multipoles which we consider later. The last bracket is the ``pure" relativistic contribution which leads to a $\mathcal{H}/k_{n}$ suppression in the bispectrum. Again, usually the last two brackets combined are called relativistic terms.
	
	Similar to the power spectrum case, we assume here that the velocity field is unbiased, i.e. $\vec{u}_{g}=\vec{u}$ and for density field biasing we use Eq.~\eqref{bias} up to quadratic order needed for the tree-level bispectrum, i.e. we omit the $\gamma_{21}$ contribution. This leads to the following Fourier space relation:
	\begin{widetext}
	\begin{eqnarray}
		\label{gal-b}
		\delta_{g}(\vec{k}) =b_{1}\, \delta(\vec{k})\, +\frac{b_{2}}{2} \underset{\vec{x},\vec{q}_{1},\vec{q}_{2}}{\int} e^{i (\vec{q}_{1} + \vec{q}_{2} - \vec{k})\cdot\vec{x}} \,\delta(\vec{q}_{1}) \delta(\vec{q}_{2})
		+\, \gamma_{2} \underset{\vec{x},\vec{q}_{1},\vec{q}_{2}}{\int}e^{i (\vec{q}_{1} + \vec{q}_{2} - \vec{k})\cdot\vec{x}} \, K(\vec{q}_{1}, \vec{q}_{2})\,\delta(\vec{q}_{1}) \delta(\vec{q}_{2})
	\end{eqnarray}
	where $K$ is given in Eq.~(\ref{F2G2K}). Since we keep terms up to second order in $\delta$ or $\Theta$, most of the Fourier fields from $\delta_{g}$'s and $\vec{u}_{g}$'s in Eq.~\eqref{gal-a} can be treated as linear. For a few other terms which appear at first order we can write:
	\begin{eqnarray}
		\label{kernel1}
		\delta(\vec{k}) = \delta_{L}(\vec{k}) \, + \underset{\vec{q}_{1},\vec{q}_{2}}{\int}\delta_{D}(\vec{k} - \vec{q}_{1}-\vec{q}_{2}) \, F_{2}(\vec{q}_{1},\vec{q}_{2})    \delta_{L}(\vec{q}_{1})  \delta_{L}(\vec{q}_{2})\,
	\end{eqnarray}
and
	\begin{eqnarray}
		\label{kernel2}
		\Theta(\vec{k}) = \delta_{L}(\vec{k}) \, + \underset{\vec{q}_{1},\vec{q}_{2}}{\int}\delta_{D}(\vec{k} - \vec{q}_{1}-\vec{q}_{2}) \, G_{2}(\vec{q}_{1},\vec{q}_{2})  \delta_{L}(\vec{q}_{1})  \delta_{L}(\vec{q}_{2}) \,.
	\end{eqnarray}
	where the second-order SPT kernels $F_{2},G_{2}$ are given in Eq.~(\ref{F2G2K}). Note that in Eq.~\eqref{gal-a} we also use $\dot{\delta} = f \mathcal{H} \delta$ and $\phi(\vec{k}) = -(3/2) \Omega_{M} (\mathcal{H}^2/k^2) \delta(\vec{k})$ from Eq. \eqref{fphi2} (again, tilde sign for SCG parameter is dropped for simplicity). Taking all these into account and sorting terms out, the final result for galaxy number density contrast up to second order in $\delta_{L}$ becomes:
		\begin{eqnarray}
			\label{deltab}
			\delta_{g,s}(\vec{x}_{1}) = \underset{\vec{q}_{1}}{\int}  e^{i \vec{q}_{1}\cdot\vec{x}_{1}} R_{1}^{\text{total}}(\vec{q}_{1}, \vec{x}_{1} )\, \delta_{L}(\vec{q}_{1})\,+ \underset{\vec{q}_{1}, \vec{q}_{2}}{\int} e^{i \vec{q}_{12}\cdot\vec{x}_{1}}   R_{2}^{\text{total}}(\vec{q}_{1},\vec{q}_{2}, \vec{x}_{1} )\,\delta_{L}(\vec{q}_{1})\, \delta_{L}(\vec{q}_{2})
		\end{eqnarray}
	where $\vec{q}_{12} = \vec{q}_{1} + \vec{q}_{2}$.  We have $R_{1}^{\text{total}}(\vec{q}_{1}, \vec{x}_{n} ) \equiv R_{1}^{\text{RSD}}(\vec{q}_{1}, \vec{x}_{n} )+ R_{1}^{\text{ID}}(\vec{q}_{1}, \vec{x}_{n} ) + R_{1}^{\text{rel}}(\vec{q}_{1}, \vec{x}_{n} )$ with $n = 1, 2$ and similarly for $R_{2}^{\text{total}}(\vec{q}_{1}, \vec{q}_{2}, \vec{x}_{n} )$. Here we have separated inverse-distance (denoted `ID') terms from ``pure" relativistic ones (denoted `rel') just to keep track of different suppression factors. For these functions we have:
		\begin{eqnarray}
			\label{RSD1}
			&&\hspace{-4.5cm} R_{1}^{\text{RSD}}(\vec{q}_{1}, \vec{x}_{1}) = b_{1} + f  \frac{(\vec{q}_{1}\cdot\hat{x}_{1})^{2}}{q_{1}^{2}}\,, \\
			&&\hspace{-4.5cm} R_{2}^{\text{RSD}}(\vec{q}_{1},\vec{q}_{2}, \vec{x}_{1} ) = b_{1}\, F_{2}(\vec{q}_{1},\vec{q}_{2}) + \frac{b_{2}}{2} +  \gamma_{2} \, K(\vec{q}_{1},\vec{q}_{2}) +  f\, \frac{(\vec{q}_{12}\cdot\hat{x}_{1})^{2}}{q_{12}^{2}} G_{2}(\vec{q}_{1},\vec{q}_{2}) \nonumber \\ 
			&&\label{RSD2} \hspace{-1.5cm}+  f \,b_{1} \frac{(\vec{q}_{12}\cdot\hat{x}_{1})(\vec{q}_{1}\cdot\hat{x}_{1})}{q_{1}^{2}}+ f^{2} \,\frac{(\vec{q}_{1}\cdot\hat{x}_{1})(\vec{q}_{2}\cdot\hat{x}_{1})}{2 q_{1}^{2}q_{2}^{2}} (\vec{q}_{12}\cdot\hat{x}_{1})^{2} \,,
		\end{eqnarray}
		\begin{eqnarray}
			\label{ID1}
			&&\hspace{-4cm}R_{1}^{\text{ID}}(\vec{q}_{1}, \vec{x}_{1}) =  - f(2 - 5s) \frac{ i}{x_{1}}\,\frac{\vec{q}_{1}\cdot\hat{x}_{1}}{q_{1}^{2}} \,,\\
			&&\label{ID2} \hspace{-4cm}R_{2}^{\text{ID}}(\vec{q}_{1},\vec{q}_{2}, \vec{x}_{1} ) =  - \frac{ i}{x_{1}}\,\left[f(2-5s)\frac{\vec{q}_{12}\cdot\hat{x}_{1}}{q_{12}^{2}} G_{2}(\vec{q}_{1},\vec{q}_{2}) +  f(2-5s)\, b_{1} \,\frac{\vec{q}_{1}\cdot\hat{x}_{1}}{q_{1}^{2}}  \, \right. \\
			&&+ \left. f^{2}(2 - \frac{5}{2}s) \frac{(\vec{q}_{1}\cdot\hat{x}_{1})(\vec{q}_{2}\cdot\hat{x}_{1})}{q_{1}^{2}q_{2}^{2}} (\vec{q}_{12}\cdot\hat{x}_{1}) \right]\,
		\end{eqnarray}
		\begin{eqnarray}
			\label{REL1}
			&&R_{1}^{\text{rel}}(\vec{q}_{1}, \hat{x}_{1}) =  - \frac{\mathcal{H}}{q_{1}}\,i f\, \mathcal{A}_{1} \frac{\vec{q}_{1}\cdot\hat{x}_{1}}{q_{1}}\,,\\
			&&R_{2}^{\text{rel}}(\vec{q}_{1},\vec{q}_{2}, \hat{x}_{1} ) = \mathcal{H} \left[ - i f \mathcal{A}_{1} \frac{\vec{q}_{12}\cdot\hat{x}_{1}}{q_{12}^2}\, G_{2}(\vec{q}_{1}, \vec{q}_{2}) - i \mathcal{A}_{2} f^{2} \frac{(\vec{q}_{1}\cdot\hat{x}_{1}) \,(\vec{q}_{2}\cdot\hat{x}_{1})(\vec{q}_{12}\cdot\hat{x}_{1})}{q_{1}^{2}\, q_{2}^{2}} - i b_{1} f\, (\mathcal{A}_{1} - f) \frac{\vec{q}_{1}\cdot\hat{x}_{1}}{q_{1}^2}    \right. \nonumber \\
			&&\label{REL2} \hspace{2.6cm}\left.+ 2\, i f^{2} \frac{\hat{q}_{1}\cdot \hat{q}_{2}}{q_{1} q_{2}} (\vec{q}_{1} \cdot \hat{x}_{1})\, - \frac{3}{2} i \,f\, \Omega_{m} \frac{(\vec{q}_{1} \cdot \hat{x}_{1})^{3}}{q_{1}^{2} \, q_{2}^{2}} - \frac{3}{2} i\,b_{1}\, \Omega_{m} \frac{\vec{q}_{2}\cdot \hat{x}_{1}}{q_{1}^{2}}   +  \frac{3}{2} i \,f\, \Omega_{m} \frac{(\vec{q}_{1}\cdot \hat{x}_{1})^{2}   \vec{q}_{2} \cdot \hat{x}_{1}}{q_{1}^{2} q_{1}^{2}}     \right]
		\end{eqnarray}
 Now we can calculate the local bispectrum by plugging Eq.~\eqref{deltab} into Eq.~\eqref{bispectrum1}. The result is:
		\begin{eqnarray}
			\label{localbis}
			B_{g, s}^{\text{loc}}(\vec{k}_{1},\vec{k}_{2},\vec{x}_{c}) &=&  \underset{\vec{q}_{1}, \vec{q}_{2}, \vec{x}_{13},\vec{x}_{23}}{\int} e^{i(\vec{q}_{1} - \vec{k}_{1})\cdot\vec{x}_{13}} e^{i(\vec{q}_{2} - \vec{k}_{2})\cdot\vec{x}_{23}} R_{1}^{\text{total}}(\vec{q}_{1}, \vec{x}_{1})\,R_{1}^{\text{total}}(\vec{q}_{2}, \vec{x}_{2})\biggl[ R_{2}^{\text{total}}(-\vec{q}_{1},-\vec{q}_{2}, \vec{x}_{3} ) \,\, + \nonumber \\
			&&+\,\,R_{2}^{\text{total}}(-\vec{q}_{2},-\vec{q}_{1}, \vec{x}_{3}) \biggl ] P_{L}(q_{1}) P_{L}(q_{2})\,\,\, +\,\,\,\, \begin{pmatrix}
				\vec{k}_{1} \\
				\vec{k}_{2} \to \vec{k}_{3} 
			\end{pmatrix}\,\,\,\, + \,\,\,\, \begin{pmatrix}
				\vec{k}_{1} \to \vec{k}_{2} \\
				\vec{k}_{3}  
			\end{pmatrix} \,,
		\end{eqnarray}
	which assumes closed triangle configurations, i.e. $\vec{k}_{3} = -\vec{k}_{1} - \vec{k}_{2}$.
	
	Our goal is to find odd-parity ``local"  multipoles (all imaginary) by inserting this equation into Eq.~\eqref{estB1}. In fact, there is no plane-parallel contributions (which come from $R^{\text{RSD}}$ terms only) for these multipoles. The leading order imaginary terms are produced by inverse-distance, pure relativistic and wide-angle contributions. Before calculating them we find the plane-parallel local bispectrum, which is all real, by inserting $R_{1,2}^{\text{RSD}}$ into the equation above and taking the limit $\hat{x}_{1} = \hat{x}_{2} = \hat{x}_{c}$. It can be easily seen that the result is:
		\begin{eqnarray}
			\label{ppb}
			B_{g }^{\text{local,pp}}(\vec{k}_{1},\vec{k}_{2}, \vec{x}_{c}) &=&  R_{1}^{\text{RSD}}(\vec{k}_{1}, \vec{x}_{c})\,R_{1}^{\text{RSD}}(\vec{k}_{2}, \vec{x}_{c})\biggl[ R_{2}^{\text{RSD}}(-\vec{k}_{1},-\vec{k}_{2}, \vec{x}_{c} ) \,\, +\,\, R_{2}^{\text{RSD}}(-\vec{k}_{2},-\vec{k}_{1}, \vec{x}_{c}) \biggl ]P_{L}(k_{1}) P_{L}(k_{2}) \nonumber \\
			&&+ \begin{pmatrix}
				\vec{k}_{1} \\
				\vec{k}_{2} \to \vec{k}_{3} 
			\end{pmatrix}\,\,\,\, + \,\,\,\, \begin{pmatrix}
				\vec{k}_{1} \to \vec{k}_{2} \\
				\vec{k}_{3}  
			\end{pmatrix} \,
		\end{eqnarray}
	\end{widetext}
	which can then be used in Eq.~\eqref{estB1} to yield the local plane-parallel multipoles:
	
	\begin{eqnarray}
		\label{plane-b}
		&&\mathcal{B}_{g, \text{pp}}^{(\ell, m)}(k_{1}, k_{2}, \mu, x_{c}) =  \\
		&&  \frac{(2\ell + 1)}{4\pi} \int d\Omega_{T}\, B_{g }^{\text{local,pp}}(\vec{k}_{1},\vec{k}_{2}, \vec{x}_{c})\, Y_{\ell m}^{*}(\theta_{1},\phi_{12}) \nonumber \,
	\end{eqnarray}
	and the final multipoles are found by integrating this over LOS vectors within the survey volume via Eq.~\eqref{estB2}. The only non-zero multipoles for the plane-parallel case are those of even $\ell$.
	
	Let us now explain how each of the aforementioned odd-parity contributions are obtained.
	
	\subsection*{ Odd-parity Multipoles from Inverse-distance and Relativistic Terms}
	For inverse-distance terms we mix RSD functions, $R_{1,2}^{\text{RSD}}$ from Eqs.~(\ref{RSD1}-\ref{RSD2}),  with $R_{1,2}^{\text{ID}}$  from Eqs.~(\ref{ID1}-\ref{ID2}) in Eq.~\eqref{localbis} and separate the imaginary part. For relativistic terms we do the same but with $R_{1,2}^{\text{rel}}$ from Eqs.~(\ref{REL1}-\ref{REL2}). Also, since these contributions are separate from wide-angle ones (the leading order suppression factors of all three contributions are of the same order, i.e.~it is $1/(kx_{c})$ for ID and wide-angle terms and $\mathcal{H}/k$ for relativistic terms), they do not mix with each other and we can simply set $\hat{x}_{n} = \hat{x}_{c}$ and $1/{x}_{n} = 1/{x}_{c}$ for inverse-distance and relativistic contributions. Therefore we have:
	\begin{widetext}
		\begin{eqnarray}
			\label{localid}
			&B_{g}^{\text{local,ID/rel}}(\vec{k}_{1},\vec{k}_{2},\vec{x}_{c}) &\,=  \biggl \{R_{1}^{\text{RSD}}(\vec{k}_{1}, \vec{x}_{c})\,R_{1}^{\text{RSD}}(\vec{k}_{2}, \vec{x}_{c})\biggl[ R_{2}^{\text{ID/rel}}(-\vec{k}_{1},-\vec{k}_{2}, \vec{x}_{c} ) \,\, +\,\,R_{2}^{\text{ID/rel}}(-\vec{k}_{2},-\vec{k}_{1}, \vec{x}_{c}) \biggl ] + \nonumber \\
			&& + \left(R_{1}^{\text{ID/rel}}(\vec{k}_{1}, \vec{x}_{c})\,R_{1}^{\text{RSD}}(\vec{k}_{2}, \vec{x}_{c}) + R_{1}^{\text{RSD}}(\vec{k}_{1}, \vec{x}_{c})\,R_{1}^{\text{ID/rel}}(\vec{k}_{2}, \vec{x}_{c})\right)\biggl[ R_{2}^{\text{RSD}}(-\vec{k}_{1},-\vec{k}_{2}, \vec{x}_{c} ) + \nonumber \\
			&& + \,\,R_{2}^{\text{RSD}}(-\vec{k}_{2},-\vec{k}_{1}, \vec{x}_{c}) \biggl ] \, \biggl \} P_{L}(k_{1}) P_{L}(k_{2}) \,\,+\,\, \begin{pmatrix}
				\vec{k}_{1} \\
				\vec{k}_{2} \to \vec{k}_{3} 
			\end{pmatrix}\,\,\,\, + \,\,\,\, \begin{pmatrix}
				\vec{k}_{1} \to \vec{k}_{2} \\
				\vec{k}_{3}  
			\end{pmatrix}  
		\end{eqnarray}
	\end{widetext}
	where we use $k_{3} = |\vec{k}_{1} + \vec{k}_{2}|$ for cyclic permutations. Again, suppression factors at leading order are $1/(k_{n} x_{c})$ and $\mathcal{H}/k_{n}$ for inverse distance and relativistic terms respectively. This result depends on $k_{1}$, $k_{2}$, $\mu$, $\hat{k}_{1}\cdot \hat{x}_{c} = \cos\theta_{1}$ and $\hat{k}_{2} \cdot \hat{x}_{c} = \mu \cos \theta_{1}+ \sqrt{1 - \mu^2}\, \sin\theta_{1}\, \cos\phi_{12}$. Therefore, to find the corresponding local imaginary multipoles, we insert this expression into Eq.~\eqref{estB1} to integrate over angular dependencies:
		\begin{eqnarray}
		\label{estBri}
		&&\mathcal{B}_{\text{Im}, \text{ID/rel}}^{(\ell,m)}({k}_{1},{k}_{2},\mu, x_{c}) \\
		&&\frac{(2\ell + 1)}{4\pi} \int d\Omega_{T}\, B_{g}^{\text{local,ID/rel}}(\vec{k}_{1},\vec{k}_{2},\vec{x}_{c})\, Y_{\ell m}^{*}(\theta_{1},\phi_{12})  \nonumber
	\end{eqnarray}
	and use this into Eq.~\eqref{estB2} to integrate over $x_{c}$ and find the final multipoles. We will later plot them together and call this combination as ``relativistic terms" to be consistent with the literature.   
	\subsection*{ Odd-parity Wide-angle Multipoles}
	Leading order wide-angle terms that produce imaginary odd-parity multipoles are also suppressed by $1/(kx)$ factors and, as  mentioned above, there is no mixing between these and inverse-distance/relativistic terms at leading order. Therefore, to find the leading wide-angle contributions we 
	only need to put the $R_{1,2}^{\text{RSD}}$ functions into Eq.~\eqref{localbis} and take the imaginary part (as real parts are all even-parity). We have,
	\begin{widetext}
		\begin{eqnarray}
			\label{wab1}
			B_{g}^{\text{local,wa}}(\vec{k}_{1},\vec{k}_{2},\vec{x}_{c}) &=&\Im \bigg\{ \underset{\vec{q}_{1}, \vec{q}_{2}, \vec{x}_{13},\vec{x}_{23}}{\int} e^{i(\vec{q}_{1} - \vec{k}_{1})\cdot \vec{x}_{13}} e^{i(\vec{q}_{2} - \vec{k}_{2})\cdot\vec{x}_{23}} R_{1}^{\text{RSD}}(\vec{q}_{1}, \vec{x}_{1})\,R_{1}^{\text{RSD}}(\vec{q}_{2}, \vec{x}_{2})\biggl[ R_{2}^{\text{RSD}}(-\vec{q}_{1},-\vec{q}_{2}, \vec{x}_{3} ) \,\, + \nonumber \\
			&&+\,\,R_{2}^{\text{RSD}}(-\vec{q}_{2},-\vec{q}_{1}, \vec{x}_{3}) \biggl ] P_{L}(q_{1}) P_{L}(q_{2})\bigg\}\,\,\, +\,\,\,\, \begin{pmatrix}
				\vec{k}_{1} \\
				\vec{k}_{2} \to \vec{k}_{3} 
			\end{pmatrix}\,\,\,\, + \,\,\,\, \begin{pmatrix}
				\vec{k}_{1} \to \vec{k}_{2} \\
				\vec{k}_{3}  
			\end{pmatrix} \,,
		\end{eqnarray}
	\end{widetext}
	Now, to extract wide-angle terms we use a method inspired by the geometry of the problem in Fig.~\ref{fig:midpb}, which is the natural extension of what we did in the power spectrum case. Let's define the perturbation vectors as:
	\begin{equation}
		\label{p23}
		\boldsymbol{\epsilon}_{1} \equiv \frac{\vec{x}_{13}}{x_{c}}\,,\,\,\,\,\,\,\,\,\,\,\,\,\,\,\,\,\boldsymbol{\epsilon}_{2} \equiv \frac{\vec{x}_{23}}{x_{c}}\,.
	\end{equation}
	There can be pairs for which these quantities are larger than one. However, their corresponding separations are most of the time much larger than $k_{1}^{-1}$ and $k_{2}^{-1}$ for the scales we are interested in. Therefore, these have negligible contributions and one can still use the entire survey volume. 
	
	Now, using $\vec{x}_{1} = \vec{x}_{c} +2\,\vec{x}_{13}/3 - {\vec{x}_{23}}/3$,\, $\vec{x}_{2} = \vec{x}_{c} +2\,\vec{x}_{23}/3 - {\vec{x}_{13}}/3$ and $\vec{x}_{3} = \vec{x}_{c} -\,\vec{x}_{13}/3- {\vec{x}_{23}}/3$, we can expand $\hat{x}_{n}$'s up to any desired order in $\epsilon$'s. For example we have:
	\begin{eqnarray}
		\label{p23exp}
		&&\hat{x}_{1} = \frac{\vec{x}_{c} +2\,\vec{x}_{13}/3 - {\vec{x}_{23}}/3 }{|\vec{x}_{c} +2\,\vec{x}_{13}/3 - {\vec{x}_{23}}/3 |} = \hat{x}_{c} + \frac{2}{3} \vec{\epsilon}_{1} \,+ ... \,,
		\\  
		&& \label{p23expp} \frac{1}{x_{1}} =  \frac{1}{|\vec{x}_{c} +2\,\vec{x}_{13}/3 - {\vec{x}_{23}}/3 |} = \frac{1}{x_{c}} ( 1 - \frac{2}{3}\boldsymbol{\epsilon}_{1}.\hat{x}_{c} +  \nonumber \\
		&& + \frac{1}{3}\boldsymbol{\epsilon}_{2}.\hat{x}_{c} + ... )
	\end{eqnarray}
	and similar expansions for $\hat{x}_{2}$, $\hat{x}_{3}$, $1/x_{2}$ and $1/x_{3}$. For the odd-parity case, we neglect $ \mathcal{O}(\epsilon_{1,2}^{2})$ terms and higher. Again, plane-parallel contributions, Eq.~\eqref{ppb}, are reproduced in the limit ${\epsilon}_{1, 2} \to 0$ or $\hat{x}_{1} \simeq \hat{x}_{2} \simeq \hat{x}_{3} \simeq \hat{x}_{c}$ and $1/x_{c} \to \infty$. 
	
	By applying these expansions to unit vectors in the $R_{1,2}^{\text{RSD}}$ functions, defined in Eqs.~(\ref{RSD1}-\ref{RSD2}) we can separate all terms of order $\epsilon_{1}$ and $\epsilon_{2}$ in Eq.~\eqref{wab1} and find the leading order wide-angle contributions. Such terms have the  following general form:
	
	\begin{eqnarray}
		\label{o2b}
		&&\mathcal{I}^{wa}_{(a_{1},...,a_{6})}(\vec{k}_{1}, \vec{k}_{2}, \vec{x}_{c}) =  \\
		&& \underset{\vec{q}_{1}, \vec{q}_{2}, \vec{x}_{13},\vec{x}_{23}}{\int} e^{i(\vec{q}_{1} - \vec{k}_{1})\cdot\vec{x}_{13}} e^{i(\vec{q}_{2} - \vec{k}_{2})\cdot\vec{x}_{23}}  \mathcal{C}_{a_{1},...,a_{6}}(\vec{q}_{1}, \vec{q}_{2}, \hat{x}_{c}) \times \nonumber \\
		&&\times (\boldsymbol \epsilon_{1x})^{a_{1}} (\boldsymbol \epsilon_{1y})^{a_{2}} (\boldsymbol \epsilon_{1z})^{a_{3}}(\boldsymbol \epsilon_{2x})^{a_{4}} (\boldsymbol \epsilon_{2y})^{a_{5}} (\boldsymbol \epsilon_{2z})^{a_{6}}  + \text{cyclic} \nonumber \,
	\end{eqnarray}
	where $\mathcal{C}$ contains all objects except perturbation parameters and $a_{1} ... a_{6}$ can be $0$ or $1$ with $\sum_{j = 1}^{6} a_{j} = 1$. 
	
	Now, from Eq.~\eqref{p23}, we can pull $\epsilon_{1j}$ and $\epsilon_{2 j}$ factors out of integral by replacing $\vec{x}_{13j}$ and $\vec{x}_{23j}$ with derivatives $i\, \partial / \partial k_{1 j}$ and $i\, \partial / \partial k_{2 j}$ which act on the entire integral. Doing so, the integrals over $\vec{x}_{13}$ and $\vec{x}_{23}$ turn into delta functions and that simply replaces $\vec{q}$'s with $\vec{k}$'s. In this way the above expression becomes:
	\begin{eqnarray}
		\label{o3b}
		&&\mathcal{I}^{wa}_{(a_{1} , ... , a_{6})}(\vec{k}_{1}, \vec{k}_{2}, \vec{x}_{c}) = \nonumber \\
		&&\left( \frac{i^{(a_{1} + ... + a_{6})}}{x_{c}^{a_{1}} ... x_{c}^{a_{6}}} \right) \left(\frac{\partial^{a_{1}}}{\partial{k_{1x}^{a_{1}}}}\,...\, \frac{\partial^{a_{6}}}{\partial{k_{2z}^{a_{6}}}}\right) \,\,\mathcal{C}_{a_{1}, ... , a_{6}}(\vec{k}_{1},\vec{k}_{2}, \hat{x}_{c}) \nonumber \\ 
		&&\,\,\, + \,\,\, \text{cyclic} \nonumber \,
	\end{eqnarray}
	where we use $\vec{k}_{3} = - \vec{k}_{1} - \vec{k}_{2}$ for cyclic permutations. Given the condition on $a_{j}$'s, this result produces imaginary terms that are suppressed by factors of $1/(k_{1}x_{c})$, $1/(k_{2}x_{c})$ and $1/(k_{12}x_{c})$. The final wide-angle local bispectrum is the sum of all such terms:
	\begin{eqnarray}
		\label{o4b}
		B_{g,\text{Im}}^{\text{loc,wa}}(\vec{k}_{1},\vec{k}_{2},\vec{x}_{c}) = \sum_{\{a_{1}, ..., a_{6}\}} \mathcal{I}^{wa}_{(a_{1},  ... , a_{6})}(\vec{k}_{1}, \vec{k}_{2}, \vec{x}_{c}) \hspace{10mm}
	\end{eqnarray}
	Again, this expression depends on $k_{1}$, $k_{2}$, $\mu$, $\hat{k}_{1}\cdot \hat{x}_{c} = \cos\theta_{1}$ and $\hat{k}_{2} \cdot \hat{x}_{c} = \mu \cos \theta_{1}+ \sqrt{1 - \mu^2}\, \sin\theta_{1}\, \cos\phi_{12}$. Finally, to get odd-parity multipoles, we repeat the same steps of using this result in Eq.~\eqref{estB1} (which integrates over $\cos\theta_{1}$ and $\phi_{12}$) to find the corresponding local multipoles $\mathcal{B}_{\text{Im, wa}}^{(\ell, m)}(k_{1}, k_{2}, \mu, x_{c})$, and then insert these into Eq.~\eqref{estB2} to find the final multipoles. 
   \begin{figure*}
		\centering
		\includegraphics[width=1\textwidth]{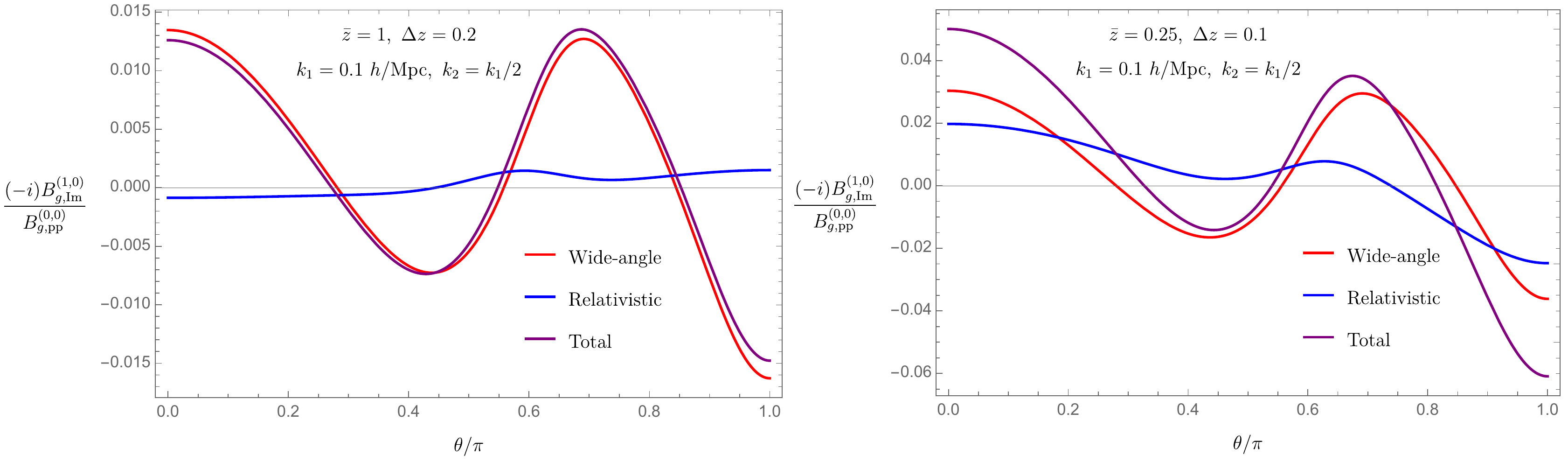}
		\caption{Examples for imaginary dipole moments of galaxy bispectrum from different origins. Relativistic terms also include inverse-distance contributions. These are plotted for different triangle configurations as function of the angle between $\vec{k}_{1}$ and $\vec{k}_{2}$ ($\theta \equiv \cos^{-1} \mu$). Bias parameters are $b_{1} = 1.46$ for $\bar{z} = 1$ bin and $b_{1} = 1.5$ for the other one with $b_{2}$, $\gamma_{2}$ calculated from Eqs.~\eqref{biasn1} and \eqref{biasn2} for each bin. Also for relativistic coefficients, we set $b_{e} = 0$ and $s = 1$.}
		\label{fig:figure4}
	\end{figure*}
	\subsection*{Dipole Moment, Signal-to-Noise and Measurements}
	From the previous two subsections we get the final local odd-parity multipoles as:
	\begin{widetext}
		\begin{eqnarray}
			\label{finalbispectrum2}
			&&\mathcal{B}_{g, \text{Im}}^{(\ell, m)}(k_{1}, k_{2}, \mu, x_{c}) =  \mathcal{B}_{\text{Im, ID}}^{(\ell, m)}(k_{1}, k_{2}, \mu, x_{c}) +  \mathcal{B}_{\text{Im, rel}}^{(\ell, m)}(k_{1}, k_{2}, \mu) + \mathcal{B}_{\text{Im, wa}}^{(\ell, m)}(k_{1}, k_{2}, \mu, x_{c}) .\,\,\,\,
		\end{eqnarray}
	\end{widetext}
	with different contributions separated as labeled. We can then find the final multipoles using
		\begin{eqnarray}
		\label{estB22}
		&&B_{g,\text{Im}}^{(\ell,m)}({k}_{1},{k}_{2},\mu) =  \int_{V_{\text{bin}}}\frac{d^{3}\vec{x}_{c}}{V_{\text{bin}}} \ \mathcal{B}_{g, \text{Im}}^{(\ell, m)}(k_{1}, k_{2}, \mu, x_{c})\nonumber \\
		&&
	\end{eqnarray}
	and separate each contribution. 

	Figure~\ref{fig:figure4} shows examples for the relative size of the dipole moment ($\ell=1$) from wide-angle contributions ($B_{\text{Im, wa}}^{(\ell, 0)}$) and inverse-distance + pure relativistic (or relativistic in general) terms ($ B_{\text{Im, ID}}^{(\ell, 0)} +  B_{\text{Im, rel}}^{(\ell, 0)}$) with respect to the plane-parallel monopole moment ($B_{g, \text{pp}}^{(0,0)}$). These quantities are plotted against $\theta/\pi$ which is the angle between two sides of the triangle, $k_{1}$ and $k_{2}$. The plots in Fig.~\ref{fig:figure4} correspond to two redshift bins, one being a Euclid-like sample bin with $\bar{z} = 1$, $\Delta z =0.2$ and $V_s = 7.94 \,\text{Gpc}^{3} h^{-3}$ \cite{Euclid2020.Blanchard.et.al} and the other one being a DESI BGS (Bright Galaxy Survey)-like sample bin with $\bar{z} = 0.25$, $\Delta z =0.1$ and $V_{s} = 0.58\, \text{Gpc}^{3} h^{-3}$ \cite{2016arXiv161100036D}.
	
	From these plots we clearly see that that wide-angle contributions have important effects on odd-parity signals and should not be neglected if one intends to constrain relativistic effects from imaginary signals. 
	
	Let us now estimate signal-to-nose ratio for the bispectrum dipole as a function of relativistic parameters $b_{e}$ and $s$. For noise we use the Gaussian cosmic variance of the bispectrum dipole, assuming $m = 0$ for simplicity. This can be a reasonable approximation for the scales we are considering. We have
		\begin{eqnarray}
			\label{v1}
			&(\sigma^{(1, 0)}_{B})^{2}& = \langle \widehat{B}_{g}^{(1,0)}(k_{1}, k_{2}, \mu)^{2} \rangle -  \langle \widehat{B}_{g}^{(1,0)}(k_{1}, k_{2}, \mu) \rangle ^{2} \nonumber \\
			&&
		\end{eqnarray}
	and we can use the following estimator (we only need leading contributions to dipole variance so we can neglect relativistic and wide-angle corrections here):
	\begin{widetext}
		\begin{eqnarray}
			\label{v2}
			\widehat{B}_{g}^{(1,0)}(k_{1}, k_{2}, \mu) = \frac{3}{V_{123}} \prod_{i =1}^{3} \int_{k_{i}} d^{3}q_{i} \delta_{D}(\vec{q}_{123}) \mathcal{L}_{1} (\hat{q}_{1}.\hat{x}_{c}) \delta_{g,s} \left( \vec{q}_{1}\right)\delta_{g,s} \left( \vec{q}_{2}\right) \delta_{g,s} \left( \vec{q}_{3}\right)
		\end{eqnarray}
	\end{widetext}
    where we have set: $\int (d\Omega_{T}/ 4\pi) F(\vec{q}_{n}) = (1/V_{123}) \prod_{i = 1}^{3} \int_{k_{n}} d^{3}\vec{q}_{n} \delta_{D}(\vec{q}_{123}) F(\vec{q}_{i})$ with $V_{123} = \int_{k_{n}} d^{3}\vec{q}_{n} \delta_{D}(\vec{q}_{123}) \simeq 8 \pi^{2} k_{1} k_{2} k_{3} \Delta k^{3}$. Again, since we are only interested in leading contributions to the variance, we can put $\delta_{g,s}(\vec{q}) = (b_{1} + f (\hat{q} \cdot \hat{x}_{c})^{2})\,\delta_{\text{L}}(q)$. Using  Eq.~\eqref{v2} in Eq.~\eqref{v1}, after some straightforward calculations and taking shot noise into account, one can write:
	 
	\begin{widetext}
		\begin{eqnarray}
			\label{varbi}
			(\sigma^{(1, 0)}_{B})^{2} =   - s_{123} \frac{ V_{f}}{V_{123}} \frac{9}{4 \pi} \int_{-1}^{1} d \eta_{1}\int_{0}^{2 \pi} d \phi_{12}\, \mathcal{L}_{1}(\eta_{1})^{2} \prod_{a = 1}^{3} \left [ P_{g, pp} (k_{a}, \eta_{a}) + \frac{1}{(2 \pi)^{3} \bar{n}_{g}} \right ]
		\end{eqnarray}
	\end{widetext}
	which leads to an imaginary error for dipole, as expected. Here $s_{123} = 6, 2, 1$ for equilateral, isosceles and other triangles respectively, $V_{f}$ is the volume of the fundamental cell in Fourier space, i.e. $V_f\equiv (2 \pi)^{3}/V_s$ and $P_{g, pp}(k, \eta) = (b_{1} + f \eta^{2})^{2} P_{L}(k) $. Here $\eta_{1} = \hat{k}_{1} \cdot \hat{x}_{c} = \cos\theta_{1}$, $\eta_{2} = \hat{k}_{2} \cdot \hat{x}_{c} = \eta_{1} \mu + \sqrt{1 - \eta_{1}^2} \sqrt{1 - \mu^{2}} \cos\phi_{12}$ and  $\eta_{3} = \hat{k}_{3} \cdot \hat{x}_{c}$ which can be expressed in terms of the previous two expressions using $\vec{k}_{3} = - \vec{k}_{1} - \vec{k}_{2}$.
	
\begin{figure*}
    	\centering
    	\includegraphics[width=1\textwidth]{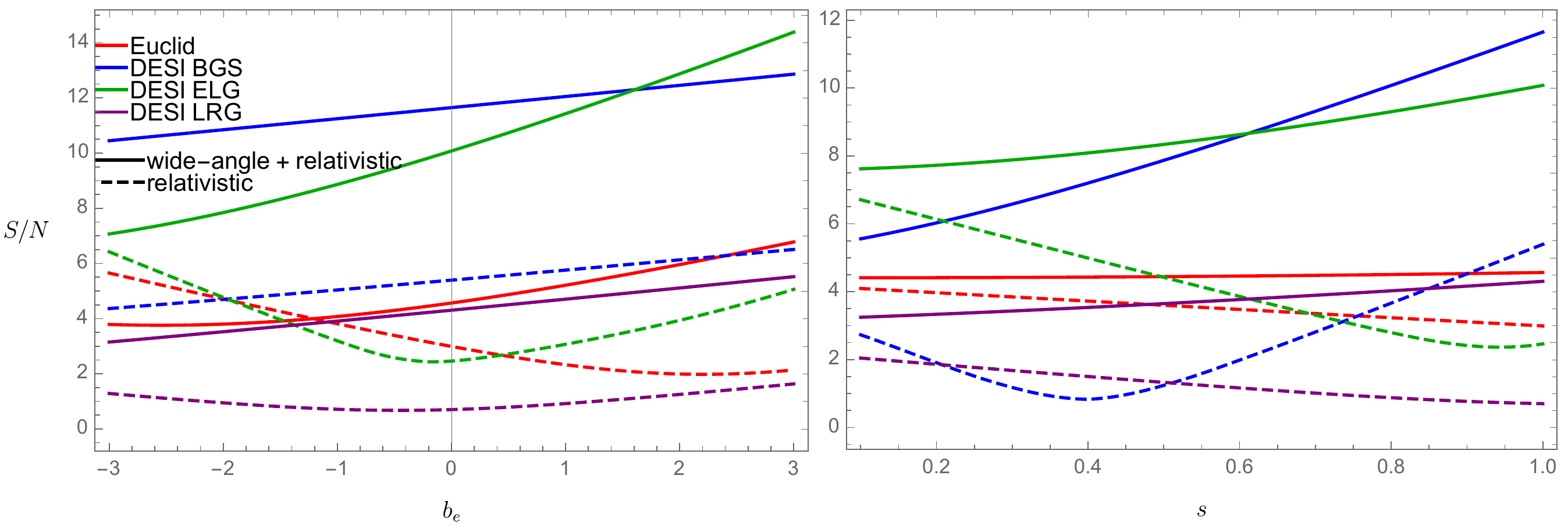}
    	\caption{Estimates of signal-to-noise ratio for bispectrum dipole moments in different surveys as a function of evolution bias ($b_{e}$) and magnification bias ($s$). For the former (left plots) we set $s = 1$ and for the latter (right) we used $b_{e} = 0$. All forecast parameters are taken from \cite{Euclid2020.Blanchard.et.al,  2016arXiv161100036D} for DESI and Euclid surveys respectively. Solid lines show results when all contributions to dipole are included. Dashed lines show results when wide-angle terms are neglected. }
    	\label{fig:sn}
    \end{figure*}

	We calculate signal-to-noise S/N for a given redshift bin using:
	\begin{eqnarray}
		\label{sn1}
		&&(S/N)^{2} = \sum_{\{k_{1}, k_{2},k_{3}\}} \frac{B_{g,\text{Im}}^{(\ell,m)}({k}_{1},{k}_{2},\mu)B_{g,\text{Im}}^{*(\ell,m)}({k}_{1},{k}_{2},\mu)}{(\sigma^{(1, 0)}_{B})^{2}} \nonumber \\
		&&
	\end{eqnarray}
    where we sum over all triangles assuming $k_{1} \leq k_{2} \leq k_{3}$, from a minimum wavenumber $k_{\text{min}} = 0.007 \,h$/Mpc in steps of $\Delta k = k_{f} = (2 \pi )/V_s^{1/3}$. Figure \ref{fig:sn} shows S/N for different surveys as a function of evolution bias, $b_{e}$, (with $s=1$) and magnification bias, $s$, with ($b_{e} = 0$). For Euclid results we used forecast parameters from \cite{Euclid2020.Blanchard.et.al} (Table 3) and considered the whole range (0.9 $< z <$ 1.8) as one bin centered at $\bar{z} = 1.35$. For DESI we have used forecast data in \cite{2016arXiv161100036D} (Tables 2.3 and 2.5). We take $0 < z < 0.5$ for BGS (Bright Galaxy survey) and $0.6 < z < 1.2$ for LRG (Luminous Red Galaxies) as single bins. For ELG (Emission Line Galaxies), we break the redshift range into two bins:  $0.6 < z < 1.2$, $1.2<z<1.7$ and add up their corresponding S/N. The signal-to-noise for relativistic-only dipole (dashed line) is quite consistent with the results in \cite{2020JCAP...03..065M}, re-emphasizing its detectability in upcoming surveys. However, we are also showing here that wide-angle effects are even more important in many cases (as can be seen from the solid lines) and should be taken into account for any likelihood analysis of relativistic parameters from bispectrum dipole data.  
    
	Let us finish this section commenting on the measurement of odd-parity signals accounting for wide-angle effects that we just calculated. Using Eqs.~(\ref{bispectrum1}-\ref{estB2}), and changing variables ($\vec{x}_{13}$, $\vec{x}_{23}$, $\vec{x}_{c}$) back to ($\vec{x}_{1}$,$\vec{x}_{2}$,$\vec{x}_{3}$) we can rewrite the multipole formula as (we take $m = 0$  for simplicity):
	\begin{widetext}
	\begin{eqnarray}
		\label{x123}
		&&B_{g, s}^{(\ell, 0)}(k_{1}, k_{2}, \mu) =\\
		&& \frac{(2 \ell + 1)}{V_{123}} \prod_{i =1}^{3} \int_{k_{i}} d^{3}q_{i} \delta_{D}(\vec{q}_{123})\int_{V_{s}} \frac{d^{3}\vec{x}_{1}} {\left( 2\pi \right) ^{3}} \frac{d^{3}\vec{x}_{2}} {\left( 2\pi \right) ^{3}} \frac{d^{3}\vec{x}_{3}} {\left( 2\pi \right) ^{3}}\,e^{-i\vec{q}_{1}\cdot\vec{x}_{1}}e^{-i\vec{q}_{2}\cdot\vec{x}_{2}} e^{i\vec{q}_{3}\cdot\vec{x}_{3}} \mathcal{L}_{\ell} (\hat{q}_{1}\cdot \hat{x}_{c}) \left\langle\delta_{g,s} \left( \vec{x}_{1}\right)\delta_{g,s} \left( \vec{x}_{2}\right) \delta_{g,s} \left( \vec{x}_{3}\right) \right\rangle \nonumber
	\end{eqnarray}
    \end{widetext}
 where the integration is  over positions of tracers inside the survey volume $V_s$. One way to evaluate this integral using Fast Fourier Transforms is to replace $x_{c}$ with $\hat{x}_{1}$ \cite{2015PhRvD..92h3532S}. However, for odd-parity signals the leading suppression factor is $(k_{n}x_{c})^{-1}$, and to this order one can 
 factorize the integral without such replacement, therefore being more accurate. Indeed, from the expansion in Eq.~\eqref{p23exp} we can see that:
\begin{equation}
	\hat{x}_{1} + \hat{x}_{2} + \hat{x}_{3} = 3 \,\hat{x}_{c} + \mathcal{O}(\epsilon_{1,2}^{2}), 
\end{equation}
which means the leading order correction to this sum is subleading for odd-party multipoles whose corrections were $\mathcal{O}(\epsilon_{1,2})$. Therefore, we can simply replace $\hat{x}_{c}$ with 1/3 ($\hat{x}_{1} + \hat{x}_{2} + \hat{x}_{3}$) and factorize the integrals. As an example, for dipole we can write: $\mathcal{L}_{1} (\hat{k}_{1}\cdot \hat{x}_{c}) = 1/3\, (\mathcal{L}_{1} (\hat{k}_{1}\cdot \hat{x}_{1}) + \mathcal{L}_{1} (\hat{k}_{1}\cdot\hat{x}_{2}) + \mathcal{L}_{1} (\hat{k}_{1}\cdot\hat{x}_{3}) )$ which simplifies Eq.~\eqref{x123} for numerical integrations with subleading errors compared to relativistic and wide-angle effects. This can be used for calculation of imaginary multipoles from data.

\subsection{Beyond Plane-Parallel Corrections to Even-Parity (Real) Multipoles}
	\label{b-radial}
	We now compute corrections beyond the plane-parallel approximation in Eq.~\eqref{exp22}, keeping terms up to second order in $\vec{u}$ and $\delta$ for the tree-level bispectrum. We also keep $1/x^2$ corrections because, as we will see, the leading order suppression factor for real multipoles is $1/(kx)^2$. We have:
	
	\begin{widetext}
		\begin{eqnarray}
			\label{real-gd}
			\delta_{g,s}(\vec{x})& =& \delta_{g}(\vec{x}) +f\,\hat{x}_{j}\, \partial_{x_{j}}\mathlarger{[}  (\vec{u}_{g}(\vec{x})\cdot\hat{x}) (1 + \delta_{g}(\vec{x}))\mathlarger{]}  +f\,\frac{2}{x} (\vec{u}_{g}(\vec{x}) \cdot\hat{x}) \mathlarger{[}1 + \delta_{g}(\vec{x})\mathlarger{]} +\\
			&& \frac{f^{2}}{2}\,\hat{x}_{k}\,\hat{x}_{j} \,\partial_{x_{j}} \partial_{x_{k}}\, \mathlarger{[} (\vec{u}_{g}(\vec{x})\cdot\hat{x})^2 \mathlarger{]} + f^2 \frac{2}{x}\,\hat{x}_{j}\, \partial_{x_{j}}\mathlarger{[}  (\vec{u}_{g}(\vec{x})\cdot\hat{x})^2 \mathlarger{]} + \frac{ f^2}{x^2}\,\mathlarger{[}  (\vec{u}_{g}(\vec{x})\cdot\hat{x})^2 \mathlarger{]} \,. \nonumber 
		\end{eqnarray}
	\end{widetext}
	Here we do not include relativistic corrections, which are suppressed by powers of $\mathcal{H}/k$  in Fourier space, but we keep inverse-distance terms. This is adequate for our purpose of comparing wide-angle corrections with other contributions because, again, inverse-distance and relativistic corrections are of the same order. In addition to that, wide-angle and inverse-distance terms combined comprise the leading corrections to multipoles of the plane-parallel bispectrum. 
	
	Similar to the odd-parity case, we use Eqs.~(\ref{gal-b}-\ref{kernel2}) to re-write Eq.~\eqref{real-gd} in the following form:
	\begin{widetext}
		\begin{eqnarray}
			\label{deltab2}
			\delta_{g,s}(\vec{x}) = \underset{\vec{q}_{1}}{\int}  e^{i \vec{q}_{1}\cdot\vec{x}}\ R_{1}^{\text{total}}(\vec{q}_{1}, \vec{x} )\, \delta_{L}(\vec{q}_{1})\,+ \underset{\vec{q}_{1}, \vec{q}_{2}}{\int} e^{i \vec{q}_{12}\cdot\vec{x}}\   R_{2'}^{\text{total}}(\vec{q}_{1},\vec{q}_{2}, \vec{x})\,\delta_{L}(\vec{q}_{1})\, \delta_{L}(\vec{q}_{2})
		\end{eqnarray}
	\end{widetext}
	where $R_{1}^{\text{total}} \equiv R_{1}^{\text{RSD}}+ R_{1}^{\text{ID}} $  and $R_{2'}^{\text{total}}\equiv R_{2}^{\text{RSD}}+R_{2'}^{\text{ID}}$. $R_{1, 2}^{\text{RSD}}$ and $R_{1}^{\text{ID}}$ are the same as Eqs.~\eqref{RSD1}, \eqref{RSD2} and \eqref{ID1}.  For $R_{2'}^{\text{ID}}$, the following term should be added to Eq.~\eqref{ID2} that was subleading in previous case, 
		\begin{eqnarray}
			\label{ID12}
			R_{2'}^{\text{ID}}(\vec{q}_{1},\vec{q}_{2}, \vec{x} )& =&R_{2}^{\text{ID}}(\vec{q}_{1},\vec{q}_{2}, \vec{x} ) - \frac{f^{2}}{x^2} \frac{(\vec{q}_{1}\cdot\hat{x})(\vec{q}_{2}\cdot\hat{x})}{q_{1}^{2}q_{2}^{2}} \nonumber \\ & & 
		\end{eqnarray}
	The relation for the local bispectrum is Eq.~\eqref{localbis} with $R_{2}^{\text{total}}$ replaced with $R_{2'}^{\text{total}}$. Also, we remind the reader that the plane-parallel parallel multipoles are obtained from Eq.~\eqref{ppb}.
	
	\subsection*{Inverse-distance Corrections}
	Similar to the previous cases, since we are separating inverse-distance corrections from wide-angle ones, we can set $\hat{x}_{1} = \hat{x}_{2} = \hat{x}_{c}$ for these contributions. Contrary to the imaginary case, there is mixing between wide-angle and inverse-distance corrections here. We will include these mixing terms later when we calculate wide-angle corrections so here we focus on inverse-distance terms alone. Therefore, since there is no $\vec{x}_{1}$ and $\vec{x}_{2}$ dependence in this case, the integrations over $\vec{x}_{13}$, $\vec{x}_{23}$ in Eq.~\eqref{localbis} turns into delta functions replacing $\vec{q}$'s with $\vec{k}$'s. This leads to the following expression for leading order real inverse-distance corrections:
		\begin{widetext}
		\begin{eqnarray}
			\label{IDR}
			B_{g, \mathcal{R}}^{\text{local,ID}}(\vec{k}_{1},\vec{k}_{2},\vec{x}_{c}) &= & \biggl \{R_{1}^{\text{RSD}}(\vec{k}_{1}, \vec{x}_{c})\,R_{1}^{\text{RSD}}(\vec{k}_{2}, \vec{x}_{c})\biggl[ - \frac{2f^{2}}{x_{c}^2} \frac{(\vec{k}_{1}\cdot \hat{x}_{c})(\vec{k}_{2}\cdot \hat{x}_{c})}{k_{1}^{2}k_{2}^{2}} \biggl ]  +  \nonumber \\
			&&+ R_{1}^{\text{ID}}(\vec{k}_{1}, \vec{x}_{c})\,R_{1}^{\text{ID}}(\vec{k}_{2}, \vec{x}_{c}) \biggl[ R_{2}^{\text{RSD}}(-\vec{k}_{1},-\vec{k}_{2}, \vec{x}_{c} )  + \,\, R_{2}^{\text{RSD}}(-\vec{k}_{2},-\vec{k}_{1}, \vec{x}_{c}) \biggl ]+  \\
			&&+ \left( R_{1}^{\text{RSD}}(\vec{k}_{1}, \vec{x}_{c}) R_{1}^{\text{ID}}(\vec{k}_{2}, \vec{x}_{c}) + R_{1}^{\text{ID}}(\vec{k}_{1}, \vec{x}_{c}) R_{1}^{\text{RSD}}(\vec{k}_{2}, \vec{x}_{c})\right) \times \nonumber \\
			&&\times \biggl[ R_{2}^{\text{ID}}(-\vec{k}_{1},-\vec{k}_{2}, \vec{x}_{c}) +  R_{2}^{\text{ID}}(-\vec{k}_{2},-\vec{k}_{1}, \vec{x}_{c})  \biggl] \, \biggl \} P_{L}(k_{1}) P_{L}(k_{2})\,\, \,\,\,+\,\, \nonumber \\ && +\,\, \begin{pmatrix} 
				\vec{k}_{1} \\
				\vec{k}_{2} \to \vec{k}_{3} 
			\end{pmatrix}\,\,\,\, + \,\,\,\, \begin{pmatrix}
				\vec{k}_{1} \to \vec{k}_{2} \\
				\vec{k}_{3}  
			\end{pmatrix} \,,  \nonumber 
		\end{eqnarray}
	\end{widetext}
where $R_{2}^{\text{RSD}}$'s in this equation are given in Eq.~\eqref{RSD2}. From this expression, it is straightforward to calculate the real local multipoles from Eq.~\eqref{estB1}:
	\begin{eqnarray}
	\label{estBR}
	&&\mathcal{B}_{\mathcal{R}, \text{ID}}^{(\ell,m)}({k}_{1},{k}_{2},\mu, x_{c}) =  
	\nonumber \\
	&&\frac{(2\ell + 1)}{4\pi} \int d\Omega_{k}\, B_{g, \mathcal{R}}^{\text{local,ID}}(\vec{k}_{1},\vec{k}_{2},\vec{x}_{c}) \, Y_{\ell m}^{*}(\theta,\phi) \,.
\end{eqnarray}

\subsection*{Wide-angle Corrections}
The last step is to find the real (even-parity) wide-angle corrections to the plane-parallel multipoles. Similar to the inverse-distance terms these are  suppressed by $1/(kx_{c})^2$. Therefore, from Eq.~\eqref{localbis}, we start from the following local bispectrum which includes RSD terms plus $R_{1, 2}^{\text{ID}}$ contributions that come with an imaginary $i/x$ factor. The latter, when mixed with wide-angle corrections,  become real and get an extra $1/x$ factor like other leading order real corrections. These are the mixing terms. We have:
	\begin{widetext}
	\begin{eqnarray}
		\label{wab2}
		&&B_{g, \mathcal{R}}^{\text{local, (pp+wa)}}(\vec{k}_{1},\vec{k}_{2},\vec{x}_{c}) = \Re \biggl \{ \underset{\vec{q}_{1}, \vec{q}_{2}, \vec{x}_{13},\vec{x}_{23}}{\int} e^{i(\vec{q}_{1} - \vec{k}_{1})\cdot\vec{x}_{13}} e^{i(\vec{q}_{2} - \vec{k}_{2})\cdot\vec{x}_{23}} \biggl (R_{1}^{\text{RSD}}(\vec{q}_{1}, \vec{x}_{1})\,R_{1}^{\text{RSD}}(\vec{q}_{2}, \vec{x}_{2}) \,\, +  \\
		&&+\,  R_{1}^{\text{RSD}}(\vec{q}_{1}, \vec{x}_{1})\,R_{1}^{\text{ID}}(\vec{q}_{2}, \vec{x}_{2}) + R_{1}^{\text{ID}}(\vec{q}_{1}, \vec{x}_{1})\,R_{1}^{\text{RSD}}(\vec{q}_{2}, \vec{x}_{2}) \biggl )\biggl[ R_{2}^{\text{RSD}}(-\vec{q}_{1},-\vec{q}_{2}, \vec{x}_{3} )+\,\,R_{2}^{\text{RSD}}(-\vec{q}_{2},-\vec{q}_{1}, \vec{x}_{3}) \biggl ]  \,\,+  \nonumber \\
		 && +R_{1}^{\text{RSD}}(\vec{q}_{1}, \vec{x}_{1})\,R_{1}^{\text{RSD}}(\vec{q}_{2}, \vec{x}_{2}) \bigg [  R_{2}^{\text{ID}}(-\vec{q}_{1},-\vec{q}_{2}, \vec{x}_{3} )+\,\,R_{2}^{\text{ID}}(-\vec{q}_{2},-\vec{q}_{1}, \vec{x}_{3})  \bigg] \biggl ) P_{L}(q_{1}) P_{L}(q_{2}) \biggl \}  \,\, + \nonumber \\ 
		 &&\,\,\, +\,\,\,\, \begin{pmatrix}
			\vec{k}_{1} \\
			\vec{k}_{2} \to \vec{k}_{3} 
		\end{pmatrix}\,\,\,\, + \,\,\,\, \begin{pmatrix}
			\vec{k}_{1} \to \vec{k}_{2} \\
			\vec{k}_{3}  
		\end{pmatrix} \,. \nonumber 
	\end{eqnarray}
\end{widetext}	
\begin{figure*}
	\centering
	\includegraphics[width=1\textwidth]{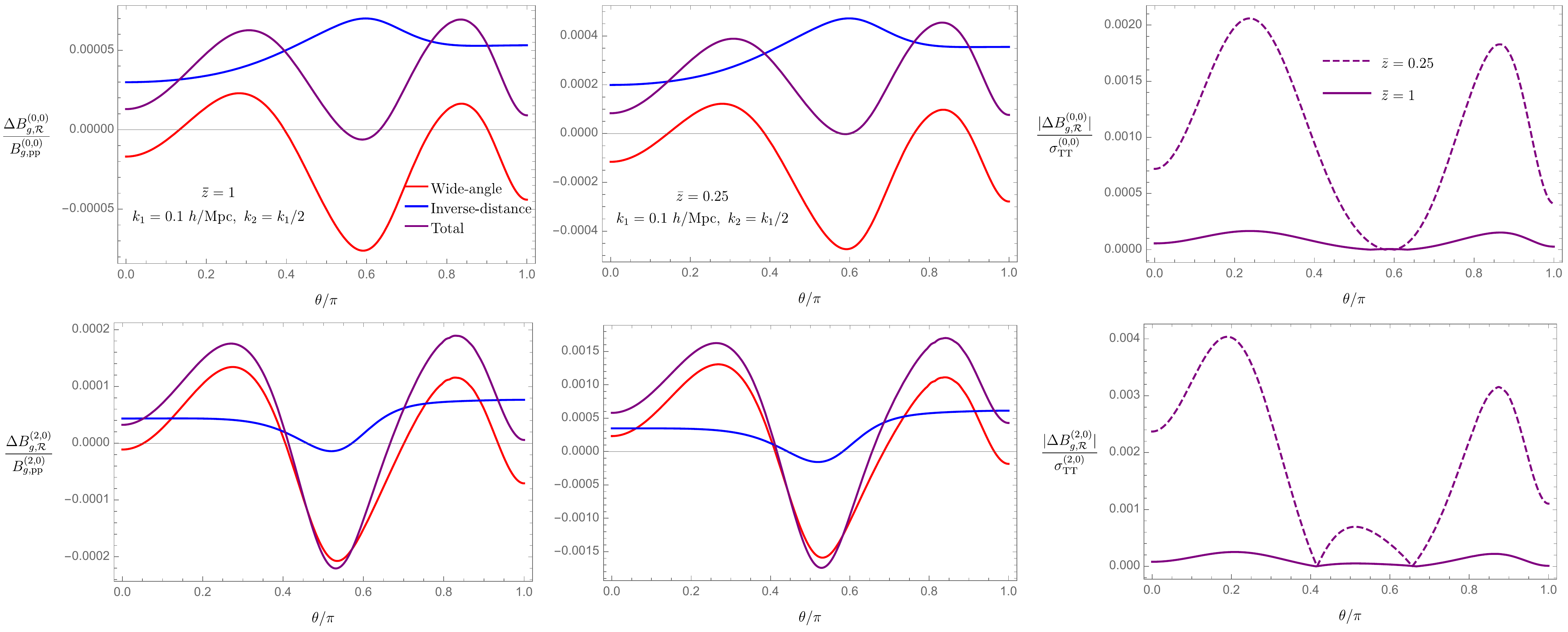}
	\caption{Corrections to plane-parallel multipoles of the bispectrum for different triangle shapes for monopole (top) and quadrupole (bottom) with $m = 0$ in all cases. Again, $\theta$ is the angle between $\vec{k}_{1}$ and $\vec{k}_{2}$. We assumed the same redshift bins as in the dipole case. The rightmost plots are ratios of total corrections to the corresponding sample variances calculated from Eq.~\eqref{varbi2}. We set $\Delta k = k_{f}$ for each bin.}
	\label{fig:figure5}
\end{figure*}
where ``pp" refers to plane-parallel contribution which is the special limit of the expression above. Similar to the imaginary case, to extract wide-angle corrections we consider the geometry of Fig.~\ref{fig:midpb} and make use of expansions as in Eq.~\eqref{p23exp} and Eq.~\eqref{p23expp} for unit vectors and $1/x_{n}$ terms in these expressions. In this case we keep terms up to second order in $\epsilon_{1}$ and $\epsilon_{2}$ and their mixing. Then, Eq.~\eqref{wab2} can be written as sum of terms that take the following general form: \vspace{-0.1cm}
\begin{eqnarray}
	\label{o2bn}
	&&\mathcal{I}_{b_{1},...,b_{6}}^{\mathcal{R}}(\vec{k}_{1}, \vec{k}_{2}, \vec{x}_{c}) =  \\
	&& \underset{\{\vec{q}_{1}, \vec{q}_{2}, \vec{x}_{13},\vec{x}_{23}\}}{\int} e^{i(\vec{q}_{1} - \vec{k}_{1})\cdot\vec{x}_{13}} e^{i(\vec{q}_{2} - \vec{k}_{2})\cdot\vec{x}_{23}}  \mathcal{C}_{b_{1},...,b_{6}}(\vec{q}_{1}, \vec{q}_{2}, \vec{x}_{c}) \times \nonumber \\
	&&\times (\boldsymbol \epsilon_{1x})^{b_{1}} (\boldsymbol \epsilon_{1y})^{b_{2}} (\boldsymbol \epsilon_{1z})^{b_{3}}(\boldsymbol \epsilon_{2x})^{b_{4}} (\boldsymbol \epsilon_{2y})^{b_{5}} (\boldsymbol \epsilon_{2z})^{b_{6}}  + \text{c.p} \nonumber \,.
\end{eqnarray}
where $b_{1}\, ... \,b_{6}$ can be $0$, $1$ or $2$ with the condition: $\sum_{j = 1}^{6} b_{j} \leq 2$ for leading order contributions. Again, we can pull $\epsilon_{1j,2 k}$ factors out of the integral to obtain:
\begin{eqnarray}
	\label{o3bn}
	&&\mathcal{I}_{b_{1}, ... , b_{6}}^{\mathcal{R}}(\vec{k}_{1}, \vec{k}_{2}, \vec{x}_{c}) = \nonumber \\
	&&\left( \frac{i^{(b_{1} + ... + b_{6})}}{x_{c}^{b_{1}} ... x_{c}^{b_{6}}} \right) \left(\frac{\partial^{b_{1}}}{\partial{k_{1x}^{b_{1}}}}\,...\, \frac{\partial^{b_{6}}}{\partial{k_{2z}^{b_{6}}}}\right) \,\,\mathcal{C}_{b_{1}, ... , b_{6}}(\vec{k}_{1},\vec{k}_{2}, \vec{x}_{c}) \nonumber \\ 
	&&\,\,\, + \,\,\, \text{c.p} \nonumber \,.
\end{eqnarray}	
The plane-parallel contributions, Eq.~\eqref{ppb}, are reproduced in the limit $\vec{\epsilon}_{1, 2} \to 0$ or $\hat{x}_{1} \simeq \hat{x}_{2} \simeq \hat{x}_{3} \simeq \hat{x}_{c}$ and $1/x_{c} \to \infty$. For such terms $b_{m} = 0$. These are followed by leading order wide-angle corrections which build up the corresponding local bispectrum:
	\begin{eqnarray}
	\label{o10b}
	B_{g,\mathcal{R}}^{\text{loc,wa}}(\vec{k}_{1},\vec{k}_{2},\vec{x}_{c}) = \sum_{\{b_{1}, b_{2}, ..., b_{6}\}} \mathcal{I}_{b_{1},  ... , b_{6}}^{\mathcal{R}}(\vec{k}_{1}, \vec{k}_{2}, \vec{x}_{c}) \hspace{10mm}
\end{eqnarray}
with $\sum_{j = 1}^{6} b_{j}  = $1 or 2 (in the case of 1 in Eq.~\eqref{o2bn} there is an extra $1/kx_{c}$ factor which comes from $\text{ID}$ terms). Finally, it is straightforward to calculate wide-angle multipoles using Eq.~\eqref{estB1} and Eq.~\eqref{o3bn}:
	\begin{eqnarray}
	\label{estBW}
	&&\mathcal{B}_{\mathcal{R}, \text{wa}}^{(\ell,m)}({k}_{1},{k}_{2},\mu, x_{c})  = 
	\nonumber \\
	&&\frac{(2\ell + 1)}{4\pi} \int d\Omega_{k}\, B_{g,\mathcal{R}}^{\text{loc,wa}}(\vec{k}_{1},\vec{k}_{2},\vec{x}_{c}) \, Y_{\ell m}^{*}(\theta,\phi) \,. \nonumber \\ 
\end{eqnarray}
Again, we do not write down the expressions explicitly because of the large number of terms involved in the expansion. 
\subsection*{Full Results}	
Using the  local multipoles from previous results into Eq.~\eqref{estB2}, we get the following contributions for the final multipoles:
	\begin{eqnarray}
		\label{finalbispectrum}
		B_{g, \mathcal{R}}^{(\ell, m)}({k}_{1}, {k}_{2}, \mu)  &=& B_{g, \text{pp}}^{(\ell, m)}({k}_{1}, {k}_{2}, \mu) + B_{\mathcal{R}, \text{wa}}^{(\ell,m)}({k}_{1},{k}_{2},\mu) \nonumber \\
		&&+ B_{\mathcal{R}, \text{ID}}^{(\ell,m)}({k}_{1},{k}_{2},\mu) \,.
	\end{eqnarray} 
	In this section we plot the relative magnitude of corrections to plane-parallel multipoles for $\ell = $ 0, 2 ($m = 0$) and compare them to cosmic variance. Figure~\ref{fig:figure5} shows examples of such quantities where $\Delta B_{g, \mathcal{R}}^{(\ell, 0)} =  B_{\mathcal{R}, \text{wa/ID}}^{(\ell,m)}({k}_{1},{k}_{2},\mu)$ and their sum. Plots are generated for the same two redshift bins used in odd-parity cases. Variances are calculated by following the same procedure that resulted in Eq.~\eqref{varbi}. For even values of $\ell$ (m = 0) one finds:
		\begin{widetext}
		\begin{eqnarray}
			\label{varbi2}
			(\sigma^{(\ell, 0)}_{B})^{2} =    s_{123} \frac{ V_{f}}{V_{123}} \frac{(2 \ell + 1)}{4 \pi} \int_{-1}^{1} d \eta_{1}\int_{0}^{2 \pi} d \phi_{12}\, \mathcal{L}_{\ell}(\eta_{1})^{2} \prod_{a = 1}^{3} \left [ P_{g, pp} (k_{a}, \eta_{a}) + \frac{1}{(2 \pi)^{3} \bar{n}_{g}} \right ]
		\end{eqnarray}
	\end{widetext}
  \begin{figure*}
  	\centering
  	\includegraphics[width=0.6\textwidth]{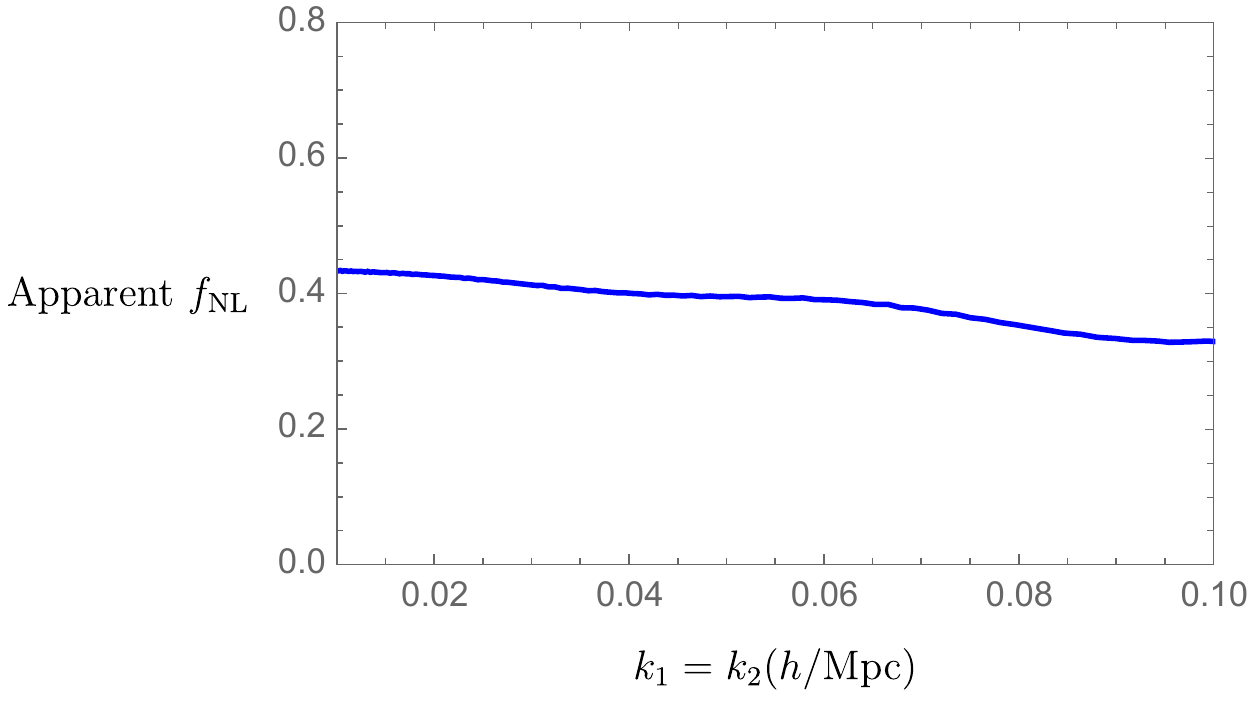}
  	\caption{An estimate of the apparent primordial local non-Gaussianity parameter $f_{\text{NL}}$ when ignoring wide-angle and inverse-distance effects. For simplicity, we have ignored stochasticity, $b_{1} = 1.5$ and $z = 1$. }
  	\label{fig:figurbfl}
  \end{figure*}
The relative sizes of these corrections are sub-percent and smaller compared to the imaginary case.  This is expected because of the additional suppression factor in this case. The ratio to cosmic variance also suggests that although these corrections become more important at lower redshifts, they can be safely neglected given the current and near future precision.  \vspace{-0.3cm}
\subsection*{Impact of wide-angle effects on primordial non-Gaussianity parameter measurements}
Finally, we comment on how big a contamination wide-angle effects are to local $f_{\text{NL}}$ constraints from galaxy bispectrum measurements. At leading order primordial non-gaussianity leaves its footprints in galaxy bispectrum with corrections linear in $f_{\text{NL}}$. Similar to what was done in power spectrum case, to have an idea about how important these contaminations are, we can find the ``apparent" $f_{\text{NL}}$ by comparing the linear (in $f_{\text{NL}}$) non-gaussian contribution to the bispectrum for isosceles triangles  in the  squeezed limit  with the local monopole of the wide-angle and inverse-distance corrections. From the general relation for the primordial non-gaussianity term in bispectrum \cite{DesJeoSch2016}, we take the special case $k_{1} = k_{2} \gg k_{3}$ which results in:

\begin{widetext}
\begin{eqnarray}
	\label{bifnl}
	B^{f}_{g}(k_1,k_2, k_3, z) &\approx&  \frac{3 f_{\text{NL}} \delta_{cr}(b_{1} -1) \Omega_{M0} H_0^2 }{D(z) T(k_{3})} \frac{P_{m}(k_{1}) P_{m}(k_{3})}{k_{3}^2}  \left [\frac{2 b_{1}^3}{\delta_{cr} (b_{1} - 1)} +2\, b_{1}^2 \left(\frac{(2 \delta_{cr} - 1)(b_{1} - 1) + \delta_{cr} b_{2}^{L}}{2\delta_{cr} (b_{1} - 1)} \right) \,+ \right . \nonumber \\
	&&\left. 2 b_{1} \left(\frac{10}{7} b_{1} + b_{2} - \frac{2}{3} b_{K^2} \right)  \right]
 \end{eqnarray}
\end{widetext}
where we have ignored stochasticity,  $b_{\text{K}^2} = \gamma_{2}$, $b_{2}^{L} = b_{2} - 8/21 (b_{1} - 1)$ is the Lagrangian quadratic bias. Figure~\ref{fig:figurbfl} shows plot of the apparant PNG parameter such a plot as a function of the isosceles triangle scale with $\theta = 0.95 \pi$. Other parameters are similar to previous plots and we have ignored stochasticity. This result implies that wide-angle and inverse-distance corrections should not be a significant problem for $f_{\text{NL}}$ measurements from the galaxy bispectrum when using the plane-parallel assumption until precision of $\Delta f_{\text{NL}} \sim 1$ is reached. 

\section{Summary and Discussion}
	\label{conclusions}

In this paper, we presented a simple analytic perturbative calculation of wide-angle and inverse-distance corrections for galaxy clustering statistics. Also, a simple procedure is provided for calculation of the evolution of  fluctuations in a relativistic context. At linear level, we computed  wide-angle corrections to multipoles of the galaxy power spectrum and compared these with relativistic contributions in linear case.  Figs~\ref{fig:Lensing} and~\ref{fig:SBE} show examples of such comparisons. These show that wide-angle effects are as important as relativistic effects and their relative dominance depends on the details of the galaxy sample. Also there is more sensitivity to magnification bias compared with evolution bias.

In addition, we calculated wide-angle and inverse-distance corrections (which are of the same order as local (non-integrated) relativistic corrections) to multipoles of galaxy power spectrum at nonlinear regime (1-loop in SPT) and showed that nonlinear effects change the relative magnitude of corrections after $k \gtrsim 0.1 $h/Mpc where loop effects start becoming important (see right panels in Fig.~\ref{fig:figure3}). This implies that the mixing between nonlinear and wide-angle/relativistic effects can be ignored. Similarly, the mixing between wide-angle/relativistic corrections and FoG effects at large scales are small as it is shown in Fig.~\ref{fig:figurefog}.
 
In general, the relative sizes of all these corrections are small  at $k \gtrsim 0.01 h$/Mpc. Therefore, we can conclude that the plane-parallel approximation remains a valid procedure for the power spectrum at these scales as long as we are not able to resolve well-below sub-percent uncertainties. However, at large scales, wide-angle and relativistic effects can become a contamination for local $f_{\text{NL}}$ measurements due to similar scaling. Figure~\ref{fig:figurefnl} shows an estimate of how large these contamination can be as a function of magnification bias. 

For the power spectrum case, because of our symmetric choice of the line of sight and the fact that we studied autocorrelations, i.e $\langle \delta_{g} \delta_{g} \rangle$, we didn't have any imaginary contribution to multipoles due to parity invariance. However, it is not  true in the case of cross correlations \cite{McDonald2009, Gaztanaga:2015jrs}. In those situations, the perturbative method developed here can be used for an efficient  analytic evaluation of wide-angle corrections to the power spectrum multipoles.

We then extended the method to compare wide-angle effects on multipoles of the local galaxy bispectrum. For the imaginary (odd-parity) case (see Fig.~\ref{fig:figure4} for two examples), we estimated the signal-to-noise ratio based on forecast parameters of Euclid and DESI surveys (Fig.~\ref{fig:sn}). We conclude that wide-angle effects are as important as relativistic contributions (more important in many cases) and should not be neglected if these signals are going to be used to constrain large-scale gravity. For even multipoles, Fig.~\ref{fig:figure5}, the same conclusion is true but signals are much less important as their ratio to cosmic variance is much smaller compared to similar ratios for plane parallel multipoles.  Also we can conclude that assuming  plane-parallel approximation should not be a significant problem for local $f_{\text{NL}}$ measurements from the galaxy bispectrum until $\Delta f_{\text{NL}} \sim 1$ is reached (figure above).

Some possible extensions of this work are worth mentioning. In this work we neglected window effects for simplicity. In general, as discussed in section \ref{window-geometry}, the results obtained here can be windowed by convolution in Eq.\eqref{window2}.  The actual mask usually have a complicated behavior depending on the geometry of the survey volume at a given redshift bin. In addition, the range of $k$-modes that we are interested in makes difference in how important these effects are. For example, multipoles of the window function itself can mix up with those of power spectrum or bispectrum and cause contamination. However, for $k \ge0.01$ h/Mpc, these window multipoles should not be that important (see for example~\cite{2015MNRAS.447.1789Y} for a discussion on this). Nonetheless, a more realistic comparison of observationally measured and calculated multipoles requires a detailed understanding of window function for a given survey (see e.g.~\cite{2019JCAP...03..040B} for a detailed calculation of window effects and their mixings with wide-angle effects in linear theory). In this work, we also assumed a uniform selection function which can be the most important source of systematic error as it limits the amount of accessible information on galaxy number density contrasts. This is, however, highly dependent on the specific survey under consideration.

Despite our simplifications, the generic conclusions of this paper are expected to apply to  more realistic galaxy survey settings: 1) use of the plane-parallel galaxy power spectrum (including nonlinear corrections) is justified provided we add deviations from it in linear theory (as in Eq. \eqref{nlnarw}), 2) deviations from the plane-parallel approximation for the galaxy bispectrum induce imaginary contributions that are important at large-scales when compared to relativistic effects.

\appendix
	
\section{RSD Kernels}
\label{Kernels}

Here we derive the RSD PT kernels including wide-angle and inverse-distance corrections. The starting point is  Eq.~\eqref{density1}, which can be rewritten by expanding the exponential as,
\begin{widetext}
\beq	
\delta_s(\k) = \sum_{m=1}^\infty {(f\mu_x(\k)k)^{m-1}\over (m-1)!} \underset{\x,\p_1\ldots \p_m}{\int} {\rm e}^{i(\p_1+\ldots+\p_m-\k)\cdot\x}	\Big[\delta(\p_1) +{{f\mu_x(\k)k}\over m} {\mu_x(\p_1)\over p_1}\, \Theta(\p_1) \Big] {\mu_x(\p_2)\over p_2}\Theta(\p_2)\ldots {\mu_x(\p_m)\over p_m} \Theta(\p_m)
\label{deltas}
\eeq	
where $\mu_x(\q)\equiv (\hat{q}\cdot\hx)$ and we have assumed no velocity bias and vorticity, $\u_g(\k)=\u(\k)=-i\k \Theta(\k)/k^2$. To find the RSD kernels, we must expand $\delta_s$ in terms of linear fluctuations $\delta_{\rm lin}(\q)$, as usual this is done through the PT expansion for the $\delta$ and $\Theta$ fields, namely

\beqa
\label{Fn}
\delta(\k) &=& \sum_{n=1}^\infty \underset{\q_1\ldots\q_n}{\int} \dD(\k-\q_1-\ldots-\q_n)\ F_n(\q_1,\ldots,\q_n)\, \delta_{\rm lin}(\q_1) \ldots \delta_{\rm lin}(\q_n) \\ 
\label{Gn}
\Theta(\k) &=& \sum_{n=1}^\infty \underset{\q_1\ldots\q_n}{\int} \dD(\k-\q_1-\ldots-\q_n)\ G_n(\q_1,\ldots,\q_n)\, \delta_{\rm lin}(\q_1) \ldots \delta_{\rm lin}(\q_n)  
\eeqa
Using these in Eq.~\eqref{deltas} and collecting terms with a given power of $\delta_{\rm lin}$ gives the RSD kernels $\widetilde{Z}_n$
\beq
\delta_s(\k) = \sum_{n=1}^\infty \underset{\q_1\ldots\q_n}{\int} \widetilde{Z}_n(\k;\q_1,\ldots,\q_n)\, \delta_{\rm lin}(\q_1) \ldots \delta_{\rm lin}(\q_n) 
\label{dsZntilde}
\eeq
where 
\beqa
\widetilde{Z}_n(\k;\q_1,\ldots,\q_n) &\equiv& \underset{\x}{\int} {\rm e}^{-i(\k-\q_1-\ldots-\q_n)\cdot\x} \sum_{m=1}^n {(f\mu_x(\k)k)^{m-1}\over (m-1)!} \underset{n_1+\ldots+n_m=n}{\sum_{n_1=1}^n \ldots \sum_{n_m=1}^n} \Big[ F_{n_1} + {f\mu_x(\k)k\over m} {\mu_x(\q^{(n_1)})\over q^{(n_1)}}\, G_{n_1} \Big] \nonumber \\
&&  \times {\mu_x(\q^{(n_2)})\over q^{(n_2)}}\, G_{n_2} \ldots  {\mu_x(\q^{(n_m)})\over q^{(n_m)}}\, G_{n_m}
\label{Zntilde}
\eeqa	
where e.g. $F_{n_1}$ depends on the first $n_1$ momenta $\q_1,\ldots,\q_{n_1}$, $G_{n_2}$	depends on the next $n_2$ momenta, etc, and $\q^{(n_1)}$ denotes the partial sum over the first $n_1$ momenta $\q^{(n_1)}\equiv \q_1+\ldots+\q_{n_1}$, $\q^{(n_2)}$ denotes the partial sum over the next $n_2$ momenta, and so on. Note that if we put the observer at infinity, we recover the plane-parallel approximation recursion relations for the kernels. Indeed, then the LOS is fixed (e.g. $\mu_x(\k) \to \hat{\k}\cdot \hat{z}$) and the integral over $\x$ gives the usual momentum conserving delta function (as in Eqs.~\ref{Fn}-\ref{Gn}) expected from the translation invariance of the plane-parallel case, thus $\widetilde{Z}_n=\dD(\k-\q_1-\ldots-\q_n)\, Z_n$  with $Z_n$ the usual RSD kernels in the plane-parallel approximation~\cite{1999ApJ...517..531S}. 

It is worth considering how this situation looks like in configuration space. 	Replacing the dummy integration variable  $\x$ by $\x'$ in Eq.~\eqref{Zntilde}, we can write the Fourier transform of Eq.~\eqref{dsZntilde} as,
\beqa
\delta_s(\x) &=& \sum_{n=1}^\infty 
	\underset{\x',\q_1\ldots\q_n}{\int} {\rm e}^{i(\q_1+\ldots+\q_n)\cdot\x'} \Bigg\{ \sum_{m=1}^n {(f\mu_{x'}(\k)k)^{m-1}\over (m-1)!} \underset{n_1+\ldots+n_m=n}{\sum_{n_1=1}^n \ldots \sum_{n_m=1}^n} \Big[ F_{n_1} + {f\mu_{x'}(\k)k\over m} {\mu_{x'}(\q^{(n_1)})\over q^{(n_1)}}\, G_{n_1} \Big] \nonumber \\
&&  \times {\mu_{x'}(\q^{(n_2)})\over q^{(n_2)}}\, G_{n_2} \ldots  {\mu_{x'}(\q^{(n_m)})\over q^{(n_m)}}\, G_{n_m}\Bigg\}_{\k=-i\nabla} \ (2\pi)^3\dD(\x-\x')  
\ \delta_{\rm lin}(\q_1) \ldots \delta_{\rm lin}(\q_n)
\label{dsx1ststep}
\eeqa
Now, the result of integrating over $\x'$ is to take derivatives of the plane wave and the object $\{\}$ with respect to $\x'$ and evaluate in $\x'=\x$. When the gradient operator acts only on the plane wave, it just corresponds to replacing $\k$ by $\q_1+\ldots+\q_n$ inside $\{\}$, i.e. it corresponds to the plane-parallel kernels where the LOS is set to be $\hx$. If these are expanded about a fixed direction, it generates the wide-angle corrections. When at least one gradient operator acts on $\{\}$, derivatives of the unit vectors inside $\mu_{x'}$ generate inverse-distance terms (see Eq.~\ref{invDisGen}), leading to the other set of  corrections to the plane-parallel results. This means we can write,

\beq
\delta_s(\x) = \sum_{n=1}^\infty \underset{\q_1\ldots\q_n}{\int} {\rm e}^{i(\q_1+\ldots+\q_n)\cdot\x}\ Z_n^{\hx}(\q_1,\ldots,\q_n) \, \delta_{\rm lin}(\q_1) \ldots \delta_{\rm lin}(\q_n) + {\rm inverse-distance~corrections}
\label{Znx}
\eeq
where $Z_n^{\hx}$ denotes the plane-parallel RSD PT kernels where the fixed LOS is replaced by $\hx$. This result was used in~\cite{WadSco2012} to calculate the impact of the LOS variation in long-wavelength modes on the power spectrum covariance matrix.

\end{widetext}	
	
	\bibliographystyle{apsrev4-1}
	\bibliography{References-new}

\begin{thebibliography}{110}%
\makeatletter
\providecommand \@ifxundefined [1]{%
 \@ifx{#1\undefined}
}%
\providecommand \@ifnum [1]{%
 \ifnum #1\expandafter \@firstoftwo
 \else \expandafter \@secondoftwo
 \fi
}%
\providecommand \@ifx [1]{%
 \ifx #1\expandafter \@firstoftwo
 \else \expandafter \@secondoftwo
 \fi
}%
\providecommand \natexlab [1]{#1}%
\providecommand \enquote  [1]{``#1''}%
\providecommand \bibnamefont  [1]{#1}%
\providecommand \bibfnamefont [1]{#1}%
\providecommand \citenamefont [1]{#1}%
\providecommand \href@noop [0]{\@secondoftwo}%
\providecommand \href [0]{\begingroup \@sanitize@url \@href}%
\providecommand \@href[1]{\@@startlink{#1}\@@href}%
\providecommand \@@href[1]{\endgroup#1\@@endlink}%
\providecommand \@sanitize@url [0]{\catcode `\\12\catcode `\$12\catcode
  `\&12\catcode `\#12\catcode `\^12\catcode `\_12\catcode `\%12\relax}%
\providecommand \@@startlink[1]{}%
\providecommand \@@endlink[0]{}%
\providecommand \url  [0]{\begingroup\@sanitize@url \@url }%
\providecommand \@url [1]{\endgroup\@href {#1}{\urlprefix }}%
\providecommand \urlprefix  [0]{URL }%
\providecommand \Eprint [0]{\href }%
\providecommand \doibase [0]{http://dx.doi.org/}%
\providecommand \selectlanguage [0]{\@gobble}%
\providecommand \bibinfo  [0]{\@secondoftwo}%
\providecommand \bibfield  [0]{\@secondoftwo}%
\providecommand \translation [1]{[#1]}%
\providecommand \BibitemOpen [0]{}%
\providecommand \bibitemStop [0]{}%
\providecommand \bibitemNoStop [0]{.\EOS\space}%
\providecommand \EOS [0]{\spacefactor3000\relax}%
\providecommand \BibitemShut  [1]{\csname bibitem#1\endcsname}%
\let\auto@bib@innerbib\@empty
\bibitem [{\citenamefont {{Peebles}}(1980)}]{Pee80}%
  \BibitemOpen
  \bibfield  {author} {\bibinfo {author} {\bibfnamefont {P.}~\bibnamefont
  {{Peebles}}},\ }\href@noop {} {\emph {\bibinfo {title} {{The large-scale
  structure of the universe}}}}\ (\bibinfo  {publisher} {Princeton University
  Press},\ \bibinfo {year} {1980})\BibitemShut {NoStop}%
\bibitem [{\citenamefont {{Bernardeau}}\ \emph {et~al.}(2002)\citenamefont
  {{Bernardeau}}, \citenamefont {{Colombi}}, \citenamefont {{Gazta{\~n}aga}},\
  and\ \citenamefont {{Scoccimarro}}}]{2002PhR...367....1B}%
  \BibitemOpen
  \bibfield  {author} {\bibinfo {author} {\bibfnamefont {F.}~\bibnamefont
  {{Bernardeau}}}, \bibinfo {author} {\bibfnamefont {S.}~\bibnamefont
  {{Colombi}}}, \bibinfo {author} {\bibfnamefont {E.}~\bibnamefont
  {{Gazta{\~n}aga}}}, \ and\ \bibinfo {author} {\bibfnamefont {R.}~\bibnamefont
  {{Scoccimarro}}},\ }\href {\doibase 10.1016/S0370-1573(02)00135-7} {\bibfield
   {journal} {\bibinfo  {journal} {Physics Reports}\ }\textbf {\bibinfo
  {volume} {367}},\ \bibinfo {pages} {1} (\bibinfo {year} {2002})},\ \Eprint
  {http://arxiv.org/abs/astro-ph/0112551} {arXiv:astro-ph/0112551 [astro-ph]}
  \BibitemShut {NoStop}%
\bibitem [{\citenamefont {{Baldauf}}(2020)}]{Bal20}%
  \BibitemOpen
  \bibfield  {author} {\bibinfo {author} {\bibfnamefont {T.}~\bibnamefont
  {{Baldauf}}},\ }\href@noop {} {\bibfield  {journal} {\bibinfo  {journal} {in:
  Effective Field Theory in Particle Physics and Cosmology, Edited by: Sacha
  Davidson, Paolo Gambino, Mikko Laine, Matthias Neubert and Christophe
  Salomon, Oxford University Press}\ } (\bibinfo {year} {{2020}})}\BibitemShut
  {NoStop}%
\bibitem [{\citenamefont {Kaiser}(1987)}]{10.1093/mnras/227.1.1}%
  \BibitemOpen
  \bibfield  {author} {\bibinfo {author} {\bibfnamefont {N.}~\bibnamefont
  {Kaiser}},\ }\href {\doibase 10.1093/mnras/227.1.1} {\bibfield  {journal}
  {\bibinfo  {journal} {MNRAS}\ }\textbf {\bibinfo {volume} {227}},\ \bibinfo
  {pages} {1} (\bibinfo {year} {1987})}\BibitemShut {NoStop}%
\bibitem [{\citenamefont {{Hamilton}}(1992)}]{1992ApJ...385L...5H}%
  \BibitemOpen
  \bibfield  {author} {\bibinfo {author} {\bibfnamefont {A.~J.~S.}\
  \bibnamefont {{Hamilton}}},\ }\href {\doibase 10.1086/186264} {\bibfield
  {journal} {\bibinfo  {journal} {ApJL}\ }\textbf {\bibinfo {volume} {385}},\
  \bibinfo {pages} {L5} (\bibinfo {year} {1992})}\BibitemShut {NoStop}%
\bibitem [{\citenamefont {{Desjacques}}\ \emph
  {et~al.}(2018{\natexlab{a}})\citenamefont {{Desjacques}}, \citenamefont
  {{Jeong}},\ and\ \citenamefont {{Schmidt}}}]{2018PhR...733....1D}%
  \BibitemOpen
  \bibfield  {author} {\bibinfo {author} {\bibfnamefont {V.}~\bibnamefont
  {{Desjacques}}}, \bibinfo {author} {\bibfnamefont {D.}~\bibnamefont
  {{Jeong}}}, \ and\ \bibinfo {author} {\bibfnamefont {F.}~\bibnamefont
  {{Schmidt}}},\ }\href {\doibase 10.1016/j.physrep.2017.12.002} {\bibfield
  {journal} {\bibinfo  {journal} {Physics Reports}\ }\textbf {\bibinfo {volume}
  {733}},\ \bibinfo {pages} {1} (\bibinfo {year} {2018}{\natexlab{a}})},\
  \Eprint {http://arxiv.org/abs/1611.09787} {arXiv:1611.09787 [astro-ph.CO]}
  \BibitemShut {NoStop}%
\bibitem [{\citenamefont {{Blake}}\ \emph
  {et~al.}(2011{\natexlab{a}})\citenamefont {{Blake}} \emph
  {et~al.}}]{2011MNRAS.418.1725B}%
  \BibitemOpen
  \bibfield  {author} {\bibinfo {author} {\bibfnamefont {C.}~\bibnamefont
  {{Blake}}} \emph {et~al.},\ }\href {\doibase
  10.1111/j.1365-2966.2011.19606.x} {\bibfield  {journal} {\bibinfo  {journal}
  {MNRAS}\ }\textbf {\bibinfo {volume} {418}},\ \bibinfo {pages} {1725}
  (\bibinfo {year} {2011}{\natexlab{a}})},\ \Eprint
  {http://arxiv.org/abs/1108.2637} {arXiv:1108.2637 [astro-ph.CO]} \BibitemShut
  {NoStop}%
\bibitem [{\citenamefont {{de la Torre}}\ \emph {et~al.}(2013)\citenamefont
  {{de la Torre}} \emph {et~al.}}]{2013A&A...557A..54D}%
  \BibitemOpen
  \bibfield  {author} {\bibinfo {author} {\bibfnamefont {S.}~\bibnamefont {{de
  la Torre}}} \emph {et~al.},\ }\href {\doibase 10.1051/0004-6361/201321463}
  {\bibfield  {journal} {\bibinfo  {journal} {AAP}\ }\textbf {\bibinfo {volume}
  {557}},\ \bibinfo {eid} {A54} (\bibinfo {year} {2013})},\ \Eprint
  {http://arxiv.org/abs/1303.2622} {arXiv:1303.2622 [astro-ph.CO]} \BibitemShut
  {NoStop}%
\bibitem [{\citenamefont {{Raccanelli}}\ \emph {et~al.}(2013)\citenamefont
  {{Raccanelli}} \emph {et~al.}}]{2013MNRAS.436...89R}%
  \BibitemOpen
  \bibfield  {author} {\bibinfo {author} {\bibfnamefont {A.}~\bibnamefont
  {{Raccanelli}}} \emph {et~al.},\ }\href {\doibase 10.1093/mnras/stt1517}
  {\bibfield  {journal} {\bibinfo  {journal} {MNRAS}\ }\textbf {\bibinfo
  {volume} {436}},\ \bibinfo {pages} {89} (\bibinfo {year} {2013})},\ \Eprint
  {http://arxiv.org/abs/1207.0500} {arXiv:1207.0500 [astro-ph.CO]} \BibitemShut
  {NoStop}%
\bibitem [{\citenamefont {{Reid}}\ \emph {et~al.}(2012)\citenamefont {{Reid}}
  \emph {et~al.}}]{2012MNRAS.426.2719R}%
  \BibitemOpen
  \bibfield  {author} {\bibinfo {author} {\bibfnamefont {B.~A.}\ \bibnamefont
  {{Reid}}} \emph {et~al.},\ }\href {\doibase 10.1111/j.1365-2966.2012.21779.x}
  {\bibfield  {journal} {\bibinfo  {journal} {MNRAS}\ }\textbf {\bibinfo
  {volume} {426}},\ \bibinfo {pages} {2719} (\bibinfo {year} {2012})},\ \Eprint
  {http://arxiv.org/abs/1203.6641} {arXiv:1203.6641 [astro-ph.CO]} \BibitemShut
  {NoStop}%
\bibitem [{\citenamefont {{Beutler}}\ \emph {et~al.}(2014)\citenamefont
  {{Beutler}} \emph {et~al.}}]{2014MNRAS.443.1065B}%
  \BibitemOpen
  \bibfield  {author} {\bibinfo {author} {\bibfnamefont {F.}~\bibnamefont
  {{Beutler}}} \emph {et~al.},\ }\href {\doibase 10.1093/mnras/stu1051}
  {\bibfield  {journal} {\bibinfo  {journal} {MNRAS}\ }\textbf {\bibinfo
  {volume} {443}},\ \bibinfo {pages} {1065} (\bibinfo {year} {2014})},\ \Eprint
  {http://arxiv.org/abs/1312.4611} {arXiv:1312.4611 [astro-ph.CO]} \BibitemShut
  {NoStop}%
\bibitem [{\citenamefont {{Chuang}}\ \emph {et~al.}(2016)\citenamefont
  {{Chuang}} \emph {et~al.}}]{2016MNRAS.461.3781C}%
  \BibitemOpen
  \bibfield  {author} {\bibinfo {author} {\bibfnamefont {C.-H.}\ \bibnamefont
  {{Chuang}}} \emph {et~al.},\ }\href {\doibase 10.1093/mnras/stw1535}
  {\bibfield  {journal} {\bibinfo  {journal} {MNRAS}\ }\textbf {\bibinfo
  {volume} {461}},\ \bibinfo {pages} {3781} (\bibinfo {year} {2016})},\ \Eprint
  {http://arxiv.org/abs/1312.4889} {arXiv:1312.4889 [astro-ph.CO]} \BibitemShut
  {NoStop}%
\bibitem [{\citenamefont {{Samushia}}\ \emph {et~al.}(2014)\citenamefont
  {{Samushia}} \emph {et~al.}}]{2014MNRAS.439.3504S}%
  \BibitemOpen
  \bibfield  {author} {\bibinfo {author} {\bibfnamefont {L.}~\bibnamefont
  {{Samushia}}} \emph {et~al.},\ }\href {\doibase 10.1093/mnras/stu197}
  {\bibfield  {journal} {\bibinfo  {journal} {MNRAS}\ }\textbf {\bibinfo
  {volume} {439}},\ \bibinfo {pages} {3504} (\bibinfo {year} {2014})},\ \Eprint
  {http://arxiv.org/abs/1312.4899} {arXiv:1312.4899 [astro-ph.CO]} \BibitemShut
  {NoStop}%
\bibitem [{\citenamefont {{S{\'a}nchez}}\ \emph {et~al.}(2014)\citenamefont
  {{S{\'a}nchez}} \emph {et~al.}}]{2014MNRAS.440.2692S}%
  \BibitemOpen
  \bibfield  {author} {\bibinfo {author} {\bibfnamefont {A.~G.}\ \bibnamefont
  {{S{\'a}nchez}}} \emph {et~al.},\ }\href {\doibase 10.1093/mnras/stu342}
  {\bibfield  {journal} {\bibinfo  {journal} {MNRAS}\ }\textbf {\bibinfo
  {volume} {440}},\ \bibinfo {pages} {2692} (\bibinfo {year} {2014})},\ \Eprint
  {http://arxiv.org/abs/1312.4854} {arXiv:1312.4854 [astro-ph.CO]} \BibitemShut
  {NoStop}%
\bibitem [{\citenamefont {{Benisty}}(2020)}]{2020arXiv200503751B}%
  \BibitemOpen
  \bibfield  {author} {\bibinfo {author} {\bibfnamefont {D.}~\bibnamefont
  {{Benisty}}},\ }\href@noop {} {\bibfield  {journal} {\bibinfo  {journal}
  {arXiv e-prints}\ ,\ \bibinfo {eid} {arXiv:2005.03751}} (\bibinfo {year}
  {2020})},\ \Eprint {http://arxiv.org/abs/2005.03751} {arXiv:2005.03751
  [astro-ph.CO]} \BibitemShut {NoStop}%
\bibitem [{\citenamefont {{Zhang}}\ \emph {et~al.}(2020)\citenamefont {{Zhang}}
  \emph {et~al.}}]{2020arXiv200712607Z}%
  \BibitemOpen
  \bibfield  {author} {\bibinfo {author} {\bibfnamefont {Y.}~\bibnamefont
  {{Zhang}}} \emph {et~al.},\ }\href@noop {} {\bibfield  {journal} {\bibinfo
  {journal} {arXiv e-prints}\ ,\ \bibinfo {eid} {arXiv:2007.12607}} (\bibinfo
  {year} {2020})},\ \Eprint {http://arxiv.org/abs/2007.12607} {arXiv:2007.12607
  [astro-ph.CO]} \BibitemShut {NoStop}%
\bibitem [{\citenamefont {{Gil-Mar{\'\i}n}}\ \emph {et~al.}(2020)\citenamefont
  {{Gil-Mar{\'\i}n}} \emph {et~al.}}]{2020arXiv200708994G}%
  \BibitemOpen
  \bibfield  {author} {\bibinfo {author} {\bibfnamefont {H.}~\bibnamefont
  {{Gil-Mar{\'\i}n}}} \emph {et~al.},\ }\href@noop {} {\bibfield  {journal}
  {\bibinfo  {journal} {arXiv e-prints}\ ,\ \bibinfo {eid} {arXiv:2007.08994}}
  (\bibinfo {year} {2020})},\ \Eprint {http://arxiv.org/abs/2007.08994}
  {arXiv:2007.08994 [astro-ph.CO]} \BibitemShut {NoStop}%
\bibitem [{\citenamefont {{Yoo}}(2009)}]{J.Yoo_2009}%
  \BibitemOpen
  \bibfield  {author} {\bibinfo {author} {\bibfnamefont {J.}~\bibnamefont
  {{Yoo}}},\ }\href {\doibase 10.1103/PhysRevD.79.023517} {\bibfield  {journal}
  {\bibinfo  {journal} {\prd}\ }\textbf {\bibinfo {volume} {79}},\ \bibinfo
  {eid} {023517} (\bibinfo {year} {2009})},\ \Eprint
  {http://arxiv.org/abs/0808.3138} {arXiv:0808.3138 [astro-ph]} \BibitemShut
  {NoStop}%
\bibitem [{\citenamefont {{Yoo}}\ \emph {et~al.}(2009)\citenamefont {{Yoo}},
  \citenamefont {{Fitzpatrick}},\ and\ \citenamefont
  {{Zaldarriaga}}}]{2009PhRvD..80h3514Y}%
  \BibitemOpen
  \bibfield  {author} {\bibinfo {author} {\bibfnamefont {J.}~\bibnamefont
  {{Yoo}}}, \bibinfo {author} {\bibfnamefont {A.~L.}\ \bibnamefont
  {{Fitzpatrick}}}, \ and\ \bibinfo {author} {\bibfnamefont {M.}~\bibnamefont
  {{Zaldarriaga}}},\ }\href {\doibase 10.1103/PhysRevD.80.083514} {\bibfield
  {journal} {\bibinfo  {journal} {Physical Review D}\ }\textbf {\bibinfo
  {volume} {80}},\ \bibinfo {eid} {083514} (\bibinfo {year} {2009})},\ \Eprint
  {http://arxiv.org/abs/0907.0707} {arXiv:0907.0707 [astro-ph.CO]} \BibitemShut
  {NoStop}%
\bibitem [{\citenamefont {{Yoo}}(2010)}]{2010PhRvD..82h3508Y}%
  \BibitemOpen
  \bibfield  {author} {\bibinfo {author} {\bibfnamefont {J.}~\bibnamefont
  {{Yoo}}},\ }\href {\doibase 10.1103/PhysRevD.82.083508} {\bibfield  {journal}
  {\bibinfo  {journal} {Physical Review D}\ }\textbf {\bibinfo {volume} {82}},\
  \bibinfo {eid} {083508} (\bibinfo {year} {2010})},\ \Eprint
  {http://arxiv.org/abs/1009.3021} {arXiv:1009.3021 [astro-ph.CO]} \BibitemShut
  {NoStop}%
\bibitem [{\citenamefont {{Challinor}}\ and\ \citenamefont
  {{Lewis}}(2011)}]{2011PhRvD..84d3516C}%
  \BibitemOpen
  \bibfield  {author} {\bibinfo {author} {\bibfnamefont {A.}~\bibnamefont
  {{Challinor}}}\ and\ \bibinfo {author} {\bibfnamefont {A.}~\bibnamefont
  {{Lewis}}},\ }\href {\doibase 10.1103/PhysRevD.84.043516} {\bibfield
  {journal} {\bibinfo  {journal} {Physical Review D}\ }\textbf {\bibinfo
  {volume} {84}},\ \bibinfo {eid} {043516} (\bibinfo {year} {2011})},\ \Eprint
  {http://arxiv.org/abs/1105.5292} {arXiv:1105.5292 [astro-ph.CO]} \BibitemShut
  {NoStop}%
\bibitem [{\citenamefont {{Bonvin}}\ and\ \citenamefont
  {{Durrer}}(2011)}]{2011PhRvD..84f3505B}%
  \BibitemOpen
  \bibfield  {author} {\bibinfo {author} {\bibfnamefont {C.}~\bibnamefont
  {{Bonvin}}}\ and\ \bibinfo {author} {\bibfnamefont {R.}~\bibnamefont
  {{Durrer}}},\ }\href {\doibase 10.1103/PhysRevD.84.063505} {\bibfield
  {journal} {\bibinfo  {journal} {Physical Review D}\ }\textbf {\bibinfo
  {volume} {84}},\ \bibinfo {eid} {063505} (\bibinfo {year} {2011})},\ \Eprint
  {http://arxiv.org/abs/1105.5280} {arXiv:1105.5280 [astro-ph.CO]} \BibitemShut
  {NoStop}%
\bibitem [{\citenamefont {{Jeong}}\ \emph {et~al.}(2012)\citenamefont
  {{Jeong}}, \citenamefont {{Schmidt}},\ and\ \citenamefont
  {{Hirata}}}]{2012PhRvD..85b3504J}%
  \BibitemOpen
  \bibfield  {author} {\bibinfo {author} {\bibfnamefont {D.}~\bibnamefont
  {{Jeong}}}, \bibinfo {author} {\bibfnamefont {F.}~\bibnamefont {{Schmidt}}},
  \ and\ \bibinfo {author} {\bibfnamefont {C.~M.}\ \bibnamefont {{Hirata}}},\
  }\href {\doibase 10.1103/PhysRevD.85.023504} {\bibfield  {journal} {\bibinfo
  {journal} {Physical Review D}\ }\textbf {\bibinfo {volume} {85}},\ \bibinfo
  {eid} {023504} (\bibinfo {year} {2012})},\ \Eprint
  {http://arxiv.org/abs/1107.5427} {arXiv:1107.5427 [astro-ph.CO]} \BibitemShut
  {NoStop}%
\bibitem [{\citenamefont {{Yoo}}\ and\ \citenamefont
  {{Zaldarriaga}}(2014)}]{2014PhRvD..90b3513Y}%
  \BibitemOpen
  \bibfield  {author} {\bibinfo {author} {\bibfnamefont {J.}~\bibnamefont
  {{Yoo}}}\ and\ \bibinfo {author} {\bibfnamefont {M.}~\bibnamefont
  {{Zaldarriaga}}},\ }\href {\doibase 10.1103/PhysRevD.90.023513} {\bibfield
  {journal} {\bibinfo  {journal} {Physical Review D}\ }\textbf {\bibinfo
  {volume} {90}},\ \bibinfo {eid} {023513} (\bibinfo {year} {2014})},\ \Eprint
  {http://arxiv.org/abs/1406.4140} {arXiv:1406.4140 [astro-ph.CO]} \BibitemShut
  {NoStop}%
\bibitem [{\citenamefont {{Bertacca}}\ \emph {et~al.}(2014)\citenamefont
  {{Bertacca}}, \citenamefont {{Maartens}},\ and\ \citenamefont
  {{Clarkson}}}]{2014JCAP...09..037B}%
  \BibitemOpen
  \bibfield  {author} {\bibinfo {author} {\bibfnamefont {D.}~\bibnamefont
  {{Bertacca}}}, \bibinfo {author} {\bibfnamefont {R.}~\bibnamefont
  {{Maartens}}}, \ and\ \bibinfo {author} {\bibfnamefont {C.}~\bibnamefont
  {{Clarkson}}},\ }\href {\doibase 10.1088/1475-7516/2014/09/037} {\bibfield
  {journal} {\bibinfo  {journal} {JCAP}\ }\textbf {\bibinfo {volume} {2014}},\
  \bibinfo {eid} {037} (\bibinfo {year} {2014})},\ \Eprint
  {http://arxiv.org/abs/1405.4403} {arXiv:1405.4403 [astro-ph.CO]} \BibitemShut
  {NoStop}%
\bibitem [{\citenamefont {{Di Dio}}\ and\ \citenamefont
  {{Seljak}}(2019)}]{2019JCAP...04..050D}%
  \BibitemOpen
  \bibfield  {author} {\bibinfo {author} {\bibfnamefont {E.}~\bibnamefont {{Di
  Dio}}}\ and\ \bibinfo {author} {\bibfnamefont {U.}~\bibnamefont {{Seljak}}},\
  }\href {\doibase 10.1088/1475-7516/2019/04/050} {\bibfield  {journal}
  {\bibinfo  {journal} {JCAP}\ }\textbf {\bibinfo {volume} {2019}},\ \bibinfo
  {eid} {050} (\bibinfo {year} {2019})},\ \Eprint
  {http://arxiv.org/abs/1811.03054} {arXiv:1811.03054 [astro-ph.CO]}
  \BibitemShut {NoStop}%
\bibitem [{\citenamefont {{Di Dio}}\ and\ \citenamefont
  {{Beutler}}(2020)}]{2020JCAP...09..058D}%
  \BibitemOpen
  \bibfield  {author} {\bibinfo {author} {\bibfnamefont {E.}~\bibnamefont {{Di
  Dio}}}\ and\ \bibinfo {author} {\bibfnamefont {F.}~\bibnamefont
  {{Beutler}}},\ }\href {\doibase 10.1088/1475-7516/2020/09/058} {\bibfield
  {journal} {\bibinfo  {journal} {Journal of Cosmology and Astroparticle
  Physics}\ }\textbf {\bibinfo {volume} {2020}},\ \bibinfo {eid} {058}
  (\bibinfo {year} {2020})},\ \Eprint {http://arxiv.org/abs/2004.07916}
  {arXiv:2004.07916 [astro-ph.CO]} \BibitemShut {NoStop}%
\bibitem [{\citenamefont {{Szalay}}\ \emph {et~al.}(1998)\citenamefont
  {{Szalay}}, \citenamefont {{Matsubara}},\ and\ \citenamefont
  {{Landy}}}]{1998ApJ...498L...1S}%
  \BibitemOpen
  \bibfield  {author} {\bibinfo {author} {\bibfnamefont {A.~S.}\ \bibnamefont
  {{Szalay}}}, \bibinfo {author} {\bibfnamefont {T.}~\bibnamefont
  {{Matsubara}}}, \ and\ \bibinfo {author} {\bibfnamefont {S.~D.}\ \bibnamefont
  {{Landy}}},\ }\href {\doibase 10.1086/311293} {\bibfield  {journal} {\bibinfo
   {journal} {ApJL}\ }\textbf {\bibinfo {volume} {498}},\ \bibinfo {pages} {L1}
  (\bibinfo {year} {1998})},\ \Eprint {http://arxiv.org/abs/astro-ph/9712007}
  {arXiv:astro-ph/9712007 [astro-ph]} \BibitemShut {NoStop}%
\bibitem [{\citenamefont {{Bharadwaj}}(1999)}]{1999ApJ...516..507B}%
  \BibitemOpen
  \bibfield  {author} {\bibinfo {author} {\bibfnamefont {S.}~\bibnamefont
  {{Bharadwaj}}},\ }\href {\doibase 10.1086/307118} {\bibfield  {journal}
  {\bibinfo  {journal} {ApJ}\ }\textbf {\bibinfo {volume} {516}},\ \bibinfo
  {pages} {507} (\bibinfo {year} {1999})},\ \Eprint
  {http://arxiv.org/abs/astro-ph/9812274} {arXiv:astro-ph/9812274 [astro-ph]}
  \BibitemShut {NoStop}%
\bibitem [{\citenamefont {{Matsubara}}(2000)}]{2000ApJ...535....1M}%
  \BibitemOpen
  \bibfield  {author} {\bibinfo {author} {\bibfnamefont {T.}~\bibnamefont
  {{Matsubara}}},\ }\href {\doibase 10.1086/308827} {\bibfield  {journal}
  {\bibinfo  {journal} {ApJ}\ }\textbf {\bibinfo {volume} {535}},\ \bibinfo
  {pages} {1} (\bibinfo {year} {2000})},\ \Eprint
  {http://arxiv.org/abs/astro-ph/9908056} {arXiv:astro-ph/9908056 [astro-ph]}
  \BibitemShut {NoStop}%
\bibitem [{\citenamefont {{Szapudi}}(2004)}]{2004ApJ...614...51S}%
  \BibitemOpen
  \bibfield  {author} {\bibinfo {author} {\bibfnamefont {I.}~\bibnamefont
  {{Szapudi}}},\ }\href {\doibase 10.1086/423168} {\bibfield  {journal}
  {\bibinfo  {journal} {ApJ}\ }\textbf {\bibinfo {volume} {614}},\ \bibinfo
  {pages} {51} (\bibinfo {year} {2004})},\ \Eprint
  {http://arxiv.org/abs/astro-ph/0404477} {arXiv:astro-ph/0404477 [astro-ph]}
  \BibitemShut {NoStop}%
\bibitem [{\citenamefont {{P{\'a}pai}}\ and\ \citenamefont
  {{Szapudi}}(2008)}]{2008MNRAS.389..292P}%
  \BibitemOpen
  \bibfield  {author} {\bibinfo {author} {\bibfnamefont {P.}~\bibnamefont
  {{P{\'a}pai}}}\ and\ \bibinfo {author} {\bibfnamefont {I.}~\bibnamefont
  {{Szapudi}}},\ }\href {\doibase 10.1111/j.1365-2966.2008.13572.x} {\bibfield
  {journal} {\bibinfo  {journal} {MNRAS}\ }\textbf {\bibinfo {volume} {389}},\
  \bibinfo {pages} {292} (\bibinfo {year} {2008})},\ \Eprint
  {http://arxiv.org/abs/0802.2940} {arXiv:0802.2940 [astro-ph]} \BibitemShut
  {NoStop}%
\bibitem [{\citenamefont {{Raccanelli}}\ \emph {et~al.}(2010)\citenamefont
  {{Raccanelli}}, \citenamefont {{Samushia}},\ and\ \citenamefont
  {{Percival}}}]{2010MNRAS.409.1525R}%
  \BibitemOpen
  \bibfield  {author} {\bibinfo {author} {\bibfnamefont {A.}~\bibnamefont
  {{Raccanelli}}}, \bibinfo {author} {\bibfnamefont {L.}~\bibnamefont
  {{Samushia}}}, \ and\ \bibinfo {author} {\bibfnamefont {W.~J.}\ \bibnamefont
  {{Percival}}},\ }\href {\doibase 10.1111/j.1365-2966.2010.17388.x} {\bibfield
   {journal} {\bibinfo  {journal} {MNRAS}\ }\textbf {\bibinfo {volume} {409}},\
  \bibinfo {pages} {1525} (\bibinfo {year} {2010})},\ \Eprint
  {http://arxiv.org/abs/1006.1652} {arXiv:1006.1652 [astro-ph.CO]} \BibitemShut
  {NoStop}%
\bibitem [{\citenamefont {{Yoo}}\ and\ \citenamefont
  {{Seljak}}(2015)}]{2015MNRAS.447.1789Y}%
  \BibitemOpen
  \bibfield  {author} {\bibinfo {author} {\bibfnamefont {J.}~\bibnamefont
  {{Yoo}}}\ and\ \bibinfo {author} {\bibfnamefont {U.}~\bibnamefont
  {{Seljak}}},\ }\href {\doibase 10.1093/mnras/stu2491} {\bibfield  {journal}
  {\bibinfo  {journal} {MNRAS}\ }\textbf {\bibinfo {volume} {447}},\ \bibinfo
  {pages} {1789} (\bibinfo {year} {2015})},\ \Eprint
  {http://arxiv.org/abs/1308.1093} {arXiv:1308.1093 [astro-ph.CO]} \BibitemShut
  {NoStop}%
\bibitem [{\citenamefont {{Reimberg}}\ \emph {et~al.}(2016)\citenamefont
  {{Reimberg}}, \citenamefont {{Bernardeau}},\ and\ \citenamefont
  {{Pitrou}}}]{2016JCAP...01..048R}%
  \BibitemOpen
  \bibfield  {author} {\bibinfo {author} {\bibfnamefont {P.}~\bibnamefont
  {{Reimberg}}}, \bibinfo {author} {\bibfnamefont {F.}~\bibnamefont
  {{Bernardeau}}}, \ and\ \bibinfo {author} {\bibfnamefont {C.}~\bibnamefont
  {{Pitrou}}},\ }\href {\doibase 10.1088/1475-7516/2016/01/048} {\bibfield
  {journal} {\bibinfo  {journal} {JCAP}\ }\textbf {\bibinfo {volume} {2016}},\
  \bibinfo {eid} {048} (\bibinfo {year} {2016})},\ \Eprint
  {http://arxiv.org/abs/1506.06596} {arXiv:1506.06596 [astro-ph.CO]}
  \BibitemShut {NoStop}%
\bibitem [{\citenamefont {{Castorina}}\ and\ \citenamefont
  {{White}}(2018{\natexlab{a}})}]{2018MNRAS.476.4403C}%
  \BibitemOpen
  \bibfield  {author} {\bibinfo {author} {\bibfnamefont {E.}~\bibnamefont
  {{Castorina}}}\ and\ \bibinfo {author} {\bibfnamefont {M.}~\bibnamefont
  {{White}}},\ }\href {\doibase 10.1093/mnras/sty410} {\bibfield  {journal}
  {\bibinfo  {journal} {MNRAS}\ }\textbf {\bibinfo {volume} {476}},\ \bibinfo
  {pages} {4403} (\bibinfo {year} {2018}{\natexlab{a}})},\ \Eprint
  {http://arxiv.org/abs/1709.09730} {arXiv:1709.09730 [astro-ph.CO]}
  \BibitemShut {NoStop}%
\bibitem [{\citenamefont {{Castorina}}\ and\ \citenamefont
  {{White}}(2020)}]{Castorina_2020}%
  \BibitemOpen
  \bibfield  {author} {\bibinfo {author} {\bibfnamefont {E.}~\bibnamefont
  {{Castorina}}}\ and\ \bibinfo {author} {\bibfnamefont {M.}~\bibnamefont
  {{White}}},\ }\href {\doibase 10.1093/mnras/staa2129} {\bibfield  {journal}
  {\bibinfo  {journal} {\mnras}\ }\textbf {\bibinfo {volume} {499}},\ \bibinfo
  {pages} {893} (\bibinfo {year} {2020})},\ \Eprint
  {http://arxiv.org/abs/1911.08353} {arXiv:1911.08353 [astro-ph.CO]}
  \BibitemShut {NoStop}%
\bibitem [{\citenamefont {{Shiraishi}}\ \emph
  {et~al.}(2021{\natexlab{a}})\citenamefont {{Shiraishi}}, \citenamefont
  {{Taruya}}, \citenamefont {{Okumura}},\ and\ \citenamefont
  {{Akitsu}}}]{2021MNRAS.503L...6S}%
  \BibitemOpen
  \bibfield  {author} {\bibinfo {author} {\bibfnamefont {M.}~\bibnamefont
  {{Shiraishi}}}, \bibinfo {author} {\bibfnamefont {A.}~\bibnamefont
  {{Taruya}}}, \bibinfo {author} {\bibfnamefont {T.}~\bibnamefont {{Okumura}}},
  \ and\ \bibinfo {author} {\bibfnamefont {K.}~\bibnamefont {{Akitsu}}},\
  }\href {\doibase 10.1093/mnrasl/slab009} {\bibfield  {journal} {\bibinfo
  {journal} {MNRAS}\ }\textbf {\bibinfo {volume} {503}},\ \bibinfo {pages} {L6}
  (\bibinfo {year} {2021}{\natexlab{a}})},\ \Eprint
  {http://arxiv.org/abs/2012.13290} {arXiv:2012.13290 [astro-ph.CO]}
  \BibitemShut {NoStop}%
\bibitem [{\citenamefont {{Bertacca}}\ \emph {et~al.}(2012)\citenamefont
  {{Bertacca}}, \citenamefont {{Maartens}}, \citenamefont {{Raccanelli}},\ and\
  \citenamefont {{Clarkson}}}]{Bertacca_2012}%
  \BibitemOpen
  \bibfield  {author} {\bibinfo {author} {\bibfnamefont {D.}~\bibnamefont
  {{Bertacca}}}, \bibinfo {author} {\bibfnamefont {R.}~\bibnamefont
  {{Maartens}}}, \bibinfo {author} {\bibfnamefont {A.}~\bibnamefont
  {{Raccanelli}}}, \ and\ \bibinfo {author} {\bibfnamefont {C.}~\bibnamefont
  {{Clarkson}}},\ }\href {\doibase 10.1088/1475-7516/2012/10/025} {\bibfield
  {journal} {\bibinfo  {journal} {\jcap}\ }\textbf {\bibinfo {volume} {2012}},\
  \bibinfo {eid} {025} (\bibinfo {year} {2012})},\ \Eprint
  {http://arxiv.org/abs/1205.5221} {arXiv:1205.5221 [astro-ph.CO]} \BibitemShut
  {NoStop}%
\bibitem [{\citenamefont {{Raccanelli}}\ \emph {et~al.}(2018)\citenamefont
  {{Raccanelli}}, \citenamefont {{Bertacca}}, \citenamefont {{Jeong}},
  \citenamefont {{Neyrinck}},\ and\ \citenamefont
  {{Szalay}}}]{raccanelli2016doppler}%
  \BibitemOpen
  \bibfield  {author} {\bibinfo {author} {\bibfnamefont {A.}~\bibnamefont
  {{Raccanelli}}}, \bibinfo {author} {\bibfnamefont {D.}~\bibnamefont
  {{Bertacca}}}, \bibinfo {author} {\bibfnamefont {D.}~\bibnamefont {{Jeong}}},
  \bibinfo {author} {\bibfnamefont {M.~C.}\ \bibnamefont {{Neyrinck}}}, \ and\
  \bibinfo {author} {\bibfnamefont {A.~S.}\ \bibnamefont {{Szalay}}},\ }\href
  {\doibase 10.1016/j.dark.2017.12.003} {\bibfield  {journal} {\bibinfo
  {journal} {Physics of the Dark Universe}\ }\textbf {\bibinfo {volume} {19}},\
  \bibinfo {pages} {109} (\bibinfo {year} {2018})},\ \Eprint
  {http://arxiv.org/abs/1602.03186} {arXiv:1602.03186 [astro-ph.CO]}
  \BibitemShut {NoStop}%
\bibitem [{\citenamefont {{Tansella}}\ \emph {et~al.}(2018)\citenamefont
  {{Tansella}}, \citenamefont {{Bonvin}}, \citenamefont {{Durrer}},
  \citenamefont {{Ghosh}},\ and\ \citenamefont {{Sellentin}}}]{Tansella_2018}%
  \BibitemOpen
  \bibfield  {author} {\bibinfo {author} {\bibfnamefont {V.}~\bibnamefont
  {{Tansella}}}, \bibinfo {author} {\bibfnamefont {C.}~\bibnamefont
  {{Bonvin}}}, \bibinfo {author} {\bibfnamefont {R.}~\bibnamefont {{Durrer}}},
  \bibinfo {author} {\bibfnamefont {B.}~\bibnamefont {{Ghosh}}}, \ and\
  \bibinfo {author} {\bibfnamefont {E.}~\bibnamefont {{Sellentin}}},\ }\href
  {\doibase 10.1088/1475-7516/2018/03/019} {\bibfield  {journal} {\bibinfo
  {journal} {\jcap}\ }\textbf {\bibinfo {volume} {2018}},\ \bibinfo {eid} {019}
  (\bibinfo {year} {2018})},\ \Eprint {http://arxiv.org/abs/1708.00492}
  {arXiv:1708.00492 [astro-ph.CO]} \BibitemShut {NoStop}%
\bibitem [{\citenamefont {{Beutler}}\ and\ \citenamefont {{Di
  Dio}}(2020)}]{Beutler_2020}%
  \BibitemOpen
  \bibfield  {author} {\bibinfo {author} {\bibfnamefont {F.}~\bibnamefont
  {{Beutler}}}\ and\ \bibinfo {author} {\bibfnamefont {E.}~\bibnamefont {{Di
  Dio}}},\ }\href {\doibase 10.1088/1475-7516/2020/07/048} {\bibfield
  {journal} {\bibinfo  {journal} {\jcap}\ }\textbf {\bibinfo {volume} {2020}},\
  \bibinfo {eid} {048} (\bibinfo {year} {2020})},\ \Eprint
  {http://arxiv.org/abs/2004.08014} {arXiv:2004.08014 [astro-ph.CO]}
  \BibitemShut {NoStop}%
\bibitem [{\citenamefont {{Grimm}}\ \emph {et~al.}(2020)\citenamefont
  {{Grimm}}, \citenamefont {{Scaccabarozzi}}, \citenamefont {{Yoo}},
  \citenamefont {{Biern}},\ and\ \citenamefont {{Gong}}}]{Grimm.et.al.2020}%
  \BibitemOpen
  \bibfield  {author} {\bibinfo {author} {\bibfnamefont {N.}~\bibnamefont
  {{Grimm}}}, \bibinfo {author} {\bibfnamefont {F.}~\bibnamefont
  {{Scaccabarozzi}}}, \bibinfo {author} {\bibfnamefont {J.}~\bibnamefont
  {{Yoo}}}, \bibinfo {author} {\bibfnamefont {S.~G.}\ \bibnamefont {{Biern}}},
  \ and\ \bibinfo {author} {\bibfnamefont {J.-O.}\ \bibnamefont {{Gong}}},\
  }\href {\doibase 10.1088/1475-7516/2020/11/064} {\bibfield  {journal}
  {\bibinfo  {journal} {\jcap}\ }\textbf {\bibinfo {volume} {2020}},\ \bibinfo
  {eid} {064} (\bibinfo {year} {2020})},\ \Eprint
  {http://arxiv.org/abs/2005.06484} {arXiv:2005.06484 [astro-ph.CO]}
  \BibitemShut {NoStop}%
\bibitem [{\citenamefont {{Castorina}}\ and\ \citenamefont {{Di
  Dio}}(2022)}]{Castorina-Dio-2021}%
  \BibitemOpen
  \bibfield  {author} {\bibinfo {author} {\bibfnamefont {E.}~\bibnamefont
  {{Castorina}}}\ and\ \bibinfo {author} {\bibfnamefont {E.}~\bibnamefont {{Di
  Dio}}},\ }\href {\doibase 10.1088/1475-7516/2022/01/061} {\bibfield
  {journal} {\bibinfo  {journal} {\jcap}\ }\textbf {\bibinfo {volume} {2022}},\
  \bibinfo {eid} {061} (\bibinfo {year} {2022})},\ \Eprint
  {http://arxiv.org/abs/2106.08857} {arXiv:2106.08857 [astro-ph.CO]}
  \BibitemShut {NoStop}%
\bibitem [{\citenamefont {{Elkhashab}}\ \emph {et~al.}(2022)\citenamefont
  {{Elkhashab}}, \citenamefont {{Porciani}},\ and\ \citenamefont
  {{Bertacca}}}]{Elkhashab.et.al2021}%
  \BibitemOpen
  \bibfield  {author} {\bibinfo {author} {\bibfnamefont {M.~Y.}\ \bibnamefont
  {{Elkhashab}}}, \bibinfo {author} {\bibfnamefont {C.}~\bibnamefont
  {{Porciani}}}, \ and\ \bibinfo {author} {\bibfnamefont {D.}~\bibnamefont
  {{Bertacca}}},\ }\href {\doibase 10.1093/mnras/stab3010} {\bibfield
  {journal} {\bibinfo  {journal} {\mnras}\ }\textbf {\bibinfo {volume} {509}},\
  \bibinfo {pages} {1626} (\bibinfo {year} {2022})},\ \Eprint
  {http://arxiv.org/abs/2108.13424} {arXiv:2108.13424 [astro-ph.CO]}
  \BibitemShut {NoStop}%
\bibitem [{\citenamefont {{Dalal}}\ \emph {et~al.}(2008)\citenamefont
  {{Dalal}}, \citenamefont {{Dor{\'e}}}, \citenamefont {{Huterer}},\ and\
  \citenamefont {{Shirokov}}}]{Dalal.et.al.2007}%
  \BibitemOpen
  \bibfield  {author} {\bibinfo {author} {\bibfnamefont {N.}~\bibnamefont
  {{Dalal}}}, \bibinfo {author} {\bibfnamefont {O.}~\bibnamefont {{Dor{\'e}}}},
  \bibinfo {author} {\bibfnamefont {D.}~\bibnamefont {{Huterer}}}, \ and\
  \bibinfo {author} {\bibfnamefont {A.}~\bibnamefont {{Shirokov}}},\ }\href
  {\doibase 10.1103/PhysRevD.77.123514} {\bibfield  {journal} {\bibinfo
  {journal} {\prd}\ }\textbf {\bibinfo {volume} {77}},\ \bibinfo {eid} {123514}
  (\bibinfo {year} {2008})},\ \Eprint {http://arxiv.org/abs/0710.4560}
  {arXiv:0710.4560 [astro-ph]} \BibitemShut {NoStop}%
\bibitem [{\citenamefont {{Slosar}}\ \emph {et~al.}(2008)\citenamefont
  {{Slosar}}, \citenamefont {{Hirata}}, \citenamefont {{Seljak}}, \citenamefont
  {{Ho}},\ and\ \citenamefont {{Padmanabhan}}}]{Slosar.et.al.2008}%
  \BibitemOpen
  \bibfield  {author} {\bibinfo {author} {\bibfnamefont {A.}~\bibnamefont
  {{Slosar}}}, \bibinfo {author} {\bibfnamefont {C.}~\bibnamefont {{Hirata}}},
  \bibinfo {author} {\bibfnamefont {U.}~\bibnamefont {{Seljak}}}, \bibinfo
  {author} {\bibfnamefont {S.}~\bibnamefont {{Ho}}}, \ and\ \bibinfo {author}
  {\bibfnamefont {N.}~\bibnamefont {{Padmanabhan}}},\ }\href {\doibase
  10.1088/1475-7516/2008/08/031} {\bibfield  {journal} {\bibinfo  {journal}
  {\jcap}\ }\textbf {\bibinfo {volume} {2008}},\ \bibinfo {eid} {031} (\bibinfo
  {year} {2008})},\ \Eprint {http://arxiv.org/abs/0805.3580} {arXiv:0805.3580
  [astro-ph]} \BibitemShut {NoStop}%
\bibitem [{\citenamefont {{Bruni}}\ \emph {et~al.}(2012)\citenamefont
  {{Bruni}}, \citenamefont {{Crittenden}}, \citenamefont {{Koyama}},
  \citenamefont {{Maartens}}, \citenamefont {{Pitrou}},\ and\ \citenamefont
  {{Wands}}}]{Bruni.et.al.2012}%
  \BibitemOpen
  \bibfield  {author} {\bibinfo {author} {\bibfnamefont {M.}~\bibnamefont
  {{Bruni}}}, \bibinfo {author} {\bibfnamefont {R.}~\bibnamefont
  {{Crittenden}}}, \bibinfo {author} {\bibfnamefont {K.}~\bibnamefont
  {{Koyama}}}, \bibinfo {author} {\bibfnamefont {R.}~\bibnamefont
  {{Maartens}}}, \bibinfo {author} {\bibfnamefont {C.}~\bibnamefont
  {{Pitrou}}}, \ and\ \bibinfo {author} {\bibfnamefont {D.}~\bibnamefont
  {{Wands}}},\ }\href {\doibase 10.1103/PhysRevD.85.041301} {\bibfield
  {journal} {\bibinfo  {journal} {\prd}\ }\textbf {\bibinfo {volume} {85}},\
  \bibinfo {eid} {041301} (\bibinfo {year} {2012})},\ \Eprint
  {http://arxiv.org/abs/1106.3999} {arXiv:1106.3999 [astro-ph.CO]} \BibitemShut
  {NoStop}%
\bibitem [{\citenamefont {{Camera}}\ \emph {et~al.}(2015)\citenamefont
  {{Camera}}, \citenamefont {{Maartens}},\ and\ \citenamefont
  {{Santos}}}]{Camera.et.al.2015}%
  \BibitemOpen
  \bibfield  {author} {\bibinfo {author} {\bibfnamefont {S.}~\bibnamefont
  {{Camera}}}, \bibinfo {author} {\bibfnamefont {R.}~\bibnamefont
  {{Maartens}}}, \ and\ \bibinfo {author} {\bibfnamefont {M.~G.}\ \bibnamefont
  {{Santos}}},\ }\href {\doibase 10.1093/mnrasl/slv069} {\bibfield  {journal}
  {\bibinfo  {journal} {\mnras}\ }\textbf {\bibinfo {volume} {451}},\ \bibinfo
  {pages} {L80} (\bibinfo {year} {2015})},\ \Eprint
  {http://arxiv.org/abs/1412.4781} {arXiv:1412.4781 [astro-ph.CO]} \BibitemShut
  {NoStop}%
\bibitem [{\citenamefont {{Kehagias}}\ \emph {et~al.}(2015)\citenamefont
  {{Kehagias}}, \citenamefont {{Dizgah}}, \citenamefont {{Nore{\~n}a}},
  \citenamefont {{Perrier}},\ and\ \citenamefont
  {{Riotto}}}]{Kehagias.et.al.2015}%
  \BibitemOpen
  \bibfield  {author} {\bibinfo {author} {\bibfnamefont {A.}~\bibnamefont
  {{Kehagias}}}, \bibinfo {author} {\bibfnamefont {A.~M.}\ \bibnamefont
  {{Dizgah}}}, \bibinfo {author} {\bibfnamefont {J.}~\bibnamefont
  {{Nore{\~n}a}}}, \bibinfo {author} {\bibfnamefont {H.}~\bibnamefont
  {{Perrier}}}, \ and\ \bibinfo {author} {\bibfnamefont {A.}~\bibnamefont
  {{Riotto}}},\ }\href {\doibase 10.1088/1475-7516/2015/08/018} {\bibfield
  {journal} {\bibinfo  {journal} {\jcap}\ }\textbf {\bibinfo {volume} {2015}},\
  \bibinfo {pages} {018} (\bibinfo {year} {2015})},\ \Eprint
  {http://arxiv.org/abs/1503.04467} {arXiv:1503.04467 [astro-ph.CO]}
  \BibitemShut {NoStop}%
\bibitem [{\citenamefont {{Lorenz}}\ \emph {et~al.}(2018)\citenamefont
  {{Lorenz}}, \citenamefont {{Alonso}},\ and\ \citenamefont
  {{Ferreira}}}]{Lorenz.et.al.2018}%
  \BibitemOpen
  \bibfield  {author} {\bibinfo {author} {\bibfnamefont {C.~S.}\ \bibnamefont
  {{Lorenz}}}, \bibinfo {author} {\bibfnamefont {D.}~\bibnamefont {{Alonso}}},
  \ and\ \bibinfo {author} {\bibfnamefont {P.~G.}\ \bibnamefont {{Ferreira}}},\
  }\href {\doibase 10.1103/PhysRevD.97.023537} {\bibfield  {journal} {\bibinfo
  {journal} {\prd}\ }\textbf {\bibinfo {volume} {97}},\ \bibinfo {eid} {023537}
  (\bibinfo {year} {2018})},\ \Eprint {http://arxiv.org/abs/1710.02477}
  {arXiv:1710.02477 [astro-ph.CO]} \BibitemShut {NoStop}%
\bibitem [{\citenamefont {{Contreras}}\ \emph {et~al.}(2019)\citenamefont
  {{Contreras}}, \citenamefont {{Johnson}},\ and\ \citenamefont
  {{Mertens}}}]{Contreras.et.al.2019}%
  \BibitemOpen
  \bibfield  {author} {\bibinfo {author} {\bibfnamefont {D.}~\bibnamefont
  {{Contreras}}}, \bibinfo {author} {\bibfnamefont {M.~C.}\ \bibnamefont
  {{Johnson}}}, \ and\ \bibinfo {author} {\bibfnamefont {J.~B.}\ \bibnamefont
  {{Mertens}}},\ }\href {\doibase 10.1088/1475-7516/2019/10/024} {\bibfield
  {journal} {\bibinfo  {journal} {\jcap}\ }\textbf {\bibinfo {volume} {2019}},\
  \bibinfo {eid} {024} (\bibinfo {year} {2019})},\ \Eprint
  {http://arxiv.org/abs/1904.10033} {arXiv:1904.10033 [astro-ph.CO]}
  \BibitemShut {NoStop}%
\bibitem [{\citenamefont {{Bernal}}\ \emph {et~al.}(2020)\citenamefont
  {{Bernal}}, \citenamefont {{Bellomo}}, \citenamefont {{Raccanelli}},\ and\
  \citenamefont {{Verde}}}]{Bernal.et.al.2020}%
  \BibitemOpen
  \bibfield  {author} {\bibinfo {author} {\bibfnamefont {J.~L.}\ \bibnamefont
  {{Bernal}}}, \bibinfo {author} {\bibfnamefont {N.}~\bibnamefont {{Bellomo}}},
  \bibinfo {author} {\bibfnamefont {A.}~\bibnamefont {{Raccanelli}}}, \ and\
  \bibinfo {author} {\bibfnamefont {L.}~\bibnamefont {{Verde}}},\ }\href
  {\doibase 10.1088/1475-7516/2020/10/017} {\bibfield  {journal} {\bibinfo
  {journal} {\jcap}\ }\textbf {\bibinfo {volume} {2020}},\ \bibinfo {eid} {017}
  (\bibinfo {year} {2020})},\ \Eprint {http://arxiv.org/abs/2005.09666}
  {arXiv:2005.09666 [astro-ph.CO]} \BibitemShut {NoStop}%
\bibitem [{\citenamefont {{Wang}}\ \emph {et~al.}(2020)\citenamefont {{Wang}},
  \citenamefont {{Beutler}},\ and\ \citenamefont {{Bacon}}}]{Wang.et.al.2020}%
  \BibitemOpen
  \bibfield  {author} {\bibinfo {author} {\bibfnamefont {M.~S.}\ \bibnamefont
  {{Wang}}}, \bibinfo {author} {\bibfnamefont {F.}~\bibnamefont {{Beutler}}}, \
  and\ \bibinfo {author} {\bibfnamefont {D.}~\bibnamefont {{Bacon}}},\ }\href
  {\doibase 10.1093/mnras/staa2998} {\bibfield  {journal} {\bibinfo  {journal}
  {\mnras}\ }\textbf {\bibinfo {volume} {499}},\ \bibinfo {pages} {2598}
  (\bibinfo {year} {2020})},\ \Eprint {http://arxiv.org/abs/2007.01802}
  {arXiv:2007.01802 [astro-ph.CO]} \BibitemShut {NoStop}%
\bibitem [{\citenamefont {{Maartens}}\ \emph {et~al.}(2021)\citenamefont
  {{Maartens}}, \citenamefont {{Jolicoeur}}, \citenamefont {{Umeh}},
  \citenamefont {{De Weerd}},\ and\ \citenamefont
  {{Clarkson}}}]{Maartens.et.al.2021}%
  \BibitemOpen
  \bibfield  {author} {\bibinfo {author} {\bibfnamefont {R.}~\bibnamefont
  {{Maartens}}}, \bibinfo {author} {\bibfnamefont {S.}~\bibnamefont
  {{Jolicoeur}}}, \bibinfo {author} {\bibfnamefont {O.}~\bibnamefont {{Umeh}}},
  \bibinfo {author} {\bibfnamefont {E.~M.}\ \bibnamefont {{De Weerd}}}, \ and\
  \bibinfo {author} {\bibfnamefont {C.}~\bibnamefont {{Clarkson}}},\ }\href
  {\doibase 10.1088/1475-7516/2021/04/013} {\bibfield  {journal} {\bibinfo
  {journal} {\jcap}\ }\textbf {\bibinfo {volume} {2021}},\ \bibinfo {eid} {013}
  (\bibinfo {year} {2021})},\ \Eprint {http://arxiv.org/abs/2011.13660}
  {arXiv:2011.13660 [astro-ph.CO]} \BibitemShut {NoStop}%
\bibitem [{\citenamefont {{Viljoen}}\ \emph {et~al.}(2021)\citenamefont
  {{Viljoen}}, \citenamefont {{Fonseca}},\ and\ \citenamefont
  {{Maartens}}}]{Viljoen.et.al.2021}%
  \BibitemOpen
  \bibfield  {author} {\bibinfo {author} {\bibfnamefont {J.-A.}\ \bibnamefont
  {{Viljoen}}}, \bibinfo {author} {\bibfnamefont {J.}~\bibnamefont
  {{Fonseca}}}, \ and\ \bibinfo {author} {\bibfnamefont {R.}~\bibnamefont
  {{Maartens}}},\ }\href {\doibase 10.1088/1475-7516/2021/12/004} {\bibfield
  {journal} {\bibinfo  {journal} {\jcap}\ }\textbf {\bibinfo {volume} {2021}},\
  \bibinfo {eid} {004} (\bibinfo {year} {2021})},\ \Eprint
  {http://arxiv.org/abs/2108.05746} {arXiv:2108.05746 [astro-ph.CO]}
  \BibitemShut {NoStop}%
\bibitem [{\citenamefont {{DESI Collaboration}}\ \emph
  {et~al.}(2016)\citenamefont {{DESI Collaboration}}, \citenamefont
  {{Aghamousa}}, \citenamefont {{Aguilar}} \emph
  {et~al.}}]{2016arXiv161100036D}%
  \BibitemOpen
  \bibfield  {author} {\bibinfo {author} {\bibnamefont {{DESI Collaboration}}},
  \bibinfo {author} {\bibfnamefont {A.}~\bibnamefont {{Aghamousa}}}, \bibinfo
  {author} {\bibfnamefont {J.}~\bibnamefont {{Aguilar}}},  \emph {et~al.},\
  }\href@noop {} {\bibfield  {journal} {\bibinfo  {journal} {arXiv e-prints}\
  ,\ \bibinfo {eid} {arXiv:1611.00036}} (\bibinfo {year} {2016})},\ \Eprint
  {http://arxiv.org/abs/1611.00036} {arXiv:1611.00036 [astro-ph.IM]}
  \BibitemShut {NoStop}%
\bibitem [{\citenamefont {{eBOSS Collaboration}}\ \emph
  {et~al.}(2020)\citenamefont {{eBOSS Collaboration}}, \citenamefont {Alam},
  \citenamefont {Aubert} \emph {et~al.}}]{2020arXiv200708991E}%
  \BibitemOpen
  \bibfield  {author} {\bibinfo {author} {\bibnamefont {{eBOSS
  Collaboration}}}, \bibinfo {author} {\bibfnamefont {S.}~\bibnamefont {Alam}},
  \bibinfo {author} {\bibfnamefont {M.}~\bibnamefont {Aubert}},  \emph
  {et~al.},\ }\href@noop {} {\bibfield  {journal} {\bibinfo  {journal} {arXiv
  e-prints}\ ,\ \bibinfo {eid} {arXiv:2007.08991}} (\bibinfo {year} {2020})},\
  \Eprint {http://arxiv.org/abs/2007.08991} {arXiv:2007.08991 [astro-ph.CO]}
  \BibitemShut {NoStop}%
\bibitem [{\citenamefont {{Amendola}}\ \emph {et~al.}(2018)\citenamefont
  {{Amendola}} \emph {et~al.}}]{2018LRR....21....2A}%
  \BibitemOpen
  \bibfield  {author} {\bibinfo {author} {\bibfnamefont {L.}~\bibnamefont
  {{Amendola}}} \emph {et~al.},\ }\href {\doibase 10.1007/s41114-017-0010-3}
  {\bibfield  {journal} {\bibinfo  {journal} {Living Reviews in Relativity}\
  }\textbf {\bibinfo {volume} {21}},\ \bibinfo {eid} {2} (\bibinfo {year}
  {2018})},\ \Eprint {http://arxiv.org/abs/1606.00180} {arXiv:1606.00180
  [astro-ph.CO]} \BibitemShut {NoStop}%
\bibitem [{\citenamefont {{Spergel}}\ \emph {et~al.}(2015)\citenamefont
  {{Spergel}}, \citenamefont {{Gehrels}}, \citenamefont {{Baltay}},
  \citenamefont {{Bennett}} \emph {et~al.}}]{2015arXiv150303757S}%
  \BibitemOpen
  \bibfield  {author} {\bibinfo {author} {\bibfnamefont {D.}~\bibnamefont
  {{Spergel}}}, \bibinfo {author} {\bibfnamefont {N.}~\bibnamefont
  {{Gehrels}}}, \bibinfo {author} {\bibfnamefont {C.}~\bibnamefont {{Baltay}}},
  \bibinfo {author} {\bibfnamefont {D.}~\bibnamefont {{Bennett}}},  \emph
  {et~al.},\ }\href@noop {} {\bibfield  {journal} {\bibinfo  {journal} {arXiv
  e-prints}\ ,\ \bibinfo {eid} {arXiv:1503.03757}} (\bibinfo {year} {2015})},\
  \Eprint {http://arxiv.org/abs/1503.03757} {arXiv:1503.03757 [astro-ph.IM]}
  \BibitemShut {NoStop}%
\bibitem [{\citenamefont {Abbott}\ \emph {et~al.}(2020)\citenamefont {Abbott},
  \citenamefont {Aguena}, \citenamefont {Alarcon}, \citenamefont {Allam},
  \citenamefont {Allen}, \citenamefont {Annis}, \citenamefont {Avila},
  \citenamefont {Bacon}, \citenamefont {Bechtol}, \citenamefont {Bermeo},\ and\
  \citenamefont {et~al.}}]{Abbott_2020}%
  \BibitemOpen
  \bibfield  {author} {\bibinfo {author} {\bibfnamefont {T.}~\bibnamefont
  {Abbott}}, \bibinfo {author} {\bibfnamefont {M.}~\bibnamefont {Aguena}},
  \bibinfo {author} {\bibfnamefont {A.}~\bibnamefont {Alarcon}}, \bibinfo
  {author} {\bibfnamefont {S.}~\bibnamefont {Allam}}, \bibinfo {author}
  {\bibfnamefont {S.}~\bibnamefont {Allen}}, \bibinfo {author} {\bibfnamefont
  {J.}~\bibnamefont {Annis}}, \bibinfo {author} {\bibfnamefont
  {S.}~\bibnamefont {Avila}}, \bibinfo {author} {\bibfnamefont
  {D.}~\bibnamefont {Bacon}}, \bibinfo {author} {\bibfnamefont
  {K.}~\bibnamefont {Bechtol}}, \bibinfo {author} {\bibfnamefont
  {A.}~\bibnamefont {Bermeo}}, \ and\ \bibinfo {author} {\bibnamefont
  {et~al.}},\ }\href {\doibase 10.1103/physrevd.102.023509} {\bibfield
  {journal} {\bibinfo  {journal} {Physical Review D}\ }\textbf {\bibinfo
  {volume} {102}} (\bibinfo {year} {2020}),\
  10.1103/physrevd.102.023509}\BibitemShut {NoStop}%
\bibitem [{\citenamefont {{Dor{\'e}}}\ \emph {et~al.}(2014)\citenamefont
  {{Dor{\'e}}} \emph {et~al.}}]{2014arXiv1412.4872D}%
  \BibitemOpen
  \bibfield  {author} {\bibinfo {author} {\bibfnamefont {O.}~\bibnamefont
  {{Dor{\'e}}}} \emph {et~al.},\ }\href@noop {} {\bibfield  {journal} {\bibinfo
   {journal} {arXiv e-prints}\ ,\ \bibinfo {eid} {arXiv:1412.4872}} (\bibinfo
  {year} {2014})},\ \Eprint {http://arxiv.org/abs/1412.4872} {arXiv:1412.4872
  [astro-ph.CO]} \BibitemShut {NoStop}%
\bibitem [{\citenamefont {{Castorina}}\ and\ \citenamefont
  {{White}}(2018{\natexlab{b}})}]{2018MNRAS.479..741C}%
  \BibitemOpen
  \bibfield  {author} {\bibinfo {author} {\bibfnamefont {E.}~\bibnamefont
  {{Castorina}}}\ and\ \bibinfo {author} {\bibfnamefont {M.}~\bibnamefont
  {{White}}},\ }\href {\doibase 10.1093/mnras/sty1437} {\bibfield  {journal}
  {\bibinfo  {journal} {MNRAS}\ }\textbf {\bibinfo {volume} {479}},\ \bibinfo
  {pages} {741} (\bibinfo {year} {2018}{\natexlab{b}})},\ \Eprint
  {http://arxiv.org/abs/1803.08185} {arXiv:1803.08185 [astro-ph.CO]}
  \BibitemShut {NoStop}%
\bibitem [{\citenamefont {{Taruya}}\ \emph {et~al.}(2020)\citenamefont
  {{Taruya}}, \citenamefont {{Saga}}, \citenamefont {{Breton}}, \citenamefont
  {{Rasera}},\ and\ \citenamefont {{Fujita}}}]{Taruya_2019}%
  \BibitemOpen
  \bibfield  {author} {\bibinfo {author} {\bibfnamefont {A.}~\bibnamefont
  {{Taruya}}}, \bibinfo {author} {\bibfnamefont {S.}~\bibnamefont {{Saga}}},
  \bibinfo {author} {\bibfnamefont {M.-A.}\ \bibnamefont {{Breton}}}, \bibinfo
  {author} {\bibfnamefont {Y.}~\bibnamefont {{Rasera}}}, \ and\ \bibinfo
  {author} {\bibfnamefont {T.}~\bibnamefont {{Fujita}}},\ }\href {\doibase
  10.1093/mnras/stz3272} {\bibfield  {journal} {\bibinfo  {journal} {\mnras}\
  }\textbf {\bibinfo {volume} {491}},\ \bibinfo {pages} {4162} (\bibinfo {year}
  {2020})},\ \Eprint {http://arxiv.org/abs/1908.03854} {arXiv:1908.03854
  [astro-ph.CO]} \BibitemShut {NoStop}%
\bibitem [{\citenamefont {{Grasshorn Gebhardt}}\ and\ \citenamefont
  {{Jeong}}(2020)}]{gebhardt2020nonlinear}%
  \BibitemOpen
  \bibfield  {author} {\bibinfo {author} {\bibfnamefont {H.~S.}\ \bibnamefont
  {{Grasshorn Gebhardt}}}\ and\ \bibinfo {author} {\bibfnamefont
  {D.}~\bibnamefont {{Jeong}}},\ }\href {\doibase 10.1103/PhysRevD.102.083521}
  {\bibfield  {journal} {\bibinfo  {journal} {\prd}\ }\textbf {\bibinfo
  {volume} {102}},\ \bibinfo {eid} {083521} (\bibinfo {year} {2020})},\ \Eprint
  {http://arxiv.org/abs/2008.08706} {arXiv:2008.08706 [astro-ph.CO]}
  \BibitemShut {NoStop}%
\bibitem [{\citenamefont {{Scoccimarro}}\ \emph {et~al.}(1999)\citenamefont
  {{Scoccimarro}}, \citenamefont {{Couchman}},\ and\ \citenamefont
  {{Frieman}}}]{1999ApJ...517..531S}%
  \BibitemOpen
  \bibfield  {author} {\bibinfo {author} {\bibfnamefont {R.}~\bibnamefont
  {{Scoccimarro}}}, \bibinfo {author} {\bibfnamefont {H.~M.~P.}\ \bibnamefont
  {{Couchman}}}, \ and\ \bibinfo {author} {\bibfnamefont {J.~A.}\ \bibnamefont
  {{Frieman}}},\ }\href {\doibase 10.1086/307220} {\bibfield  {journal}
  {\bibinfo  {journal} {ApJ}\ }\textbf {\bibinfo {volume} {517}},\ \bibinfo
  {pages} {531} (\bibinfo {year} {1999})},\ \Eprint
  {http://arxiv.org/abs/astro-ph/9808305} {arXiv:astro-ph/9808305 [astro-ph]}
  \BibitemShut {NoStop}%
\bibitem [{\citenamefont {{Tellarini}}\ \emph {et~al.}(2016)\citenamefont
  {{Tellarini}}, \citenamefont {{Ross}}, \citenamefont {{Tasinato}},\ and\
  \citenamefont {{Wands}}}]{2016JCAP...06..014T}%
  \BibitemOpen
  \bibfield  {author} {\bibinfo {author} {\bibfnamefont {M.}~\bibnamefont
  {{Tellarini}}}, \bibinfo {author} {\bibfnamefont {A.~J.}\ \bibnamefont
  {{Ross}}}, \bibinfo {author} {\bibfnamefont {G.}~\bibnamefont {{Tasinato}}},
  \ and\ \bibinfo {author} {\bibfnamefont {D.}~\bibnamefont {{Wands}}},\ }\href
  {\doibase 10.1088/1475-7516/2016/06/014} {\bibfield  {journal} {\bibinfo
  {journal} {JCAP}\ }\textbf {\bibinfo {volume} {2016}},\ \bibinfo {eid} {014}
  (\bibinfo {year} {2016})},\ \Eprint {http://arxiv.org/abs/1603.06814}
  {arXiv:1603.06814 [astro-ph.CO]} \BibitemShut {NoStop}%
\bibitem [{\citenamefont {{Slepian}}\ and\ \citenamefont
  {{Eisenstein}}(2017)}]{2017MNRAS.469.2059S}%
  \BibitemOpen
  \bibfield  {author} {\bibinfo {author} {\bibfnamefont {Z.}~\bibnamefont
  {{Slepian}}}\ and\ \bibinfo {author} {\bibfnamefont {D.~J.}\ \bibnamefont
  {{Eisenstein}}},\ }\href {\doibase 10.1093/mnras/stx490} {\bibfield
  {journal} {\bibinfo  {journal} {MNRAS}\ }\textbf {\bibinfo {volume} {469}},\
  \bibinfo {pages} {2059} (\bibinfo {year} {2017})},\ \Eprint
  {http://arxiv.org/abs/1607.03109} {arXiv:1607.03109 [astro-ph.CO]}
  \BibitemShut {NoStop}%
\bibitem [{\citenamefont {{Hashimoto}}\ \emph {et~al.}(2017)\citenamefont
  {{Hashimoto}}, \citenamefont {{Rasera}},\ and\ \citenamefont
  {{Taruya}}}]{2017PhRvD..96d3526H}%
  \BibitemOpen
  \bibfield  {author} {\bibinfo {author} {\bibfnamefont {I.}~\bibnamefont
  {{Hashimoto}}}, \bibinfo {author} {\bibfnamefont {Y.}~\bibnamefont
  {{Rasera}}}, \ and\ \bibinfo {author} {\bibfnamefont {A.}~\bibnamefont
  {{Taruya}}},\ }\href {\doibase 10.1103/PhysRevD.96.043526} {\bibfield
  {journal} {\bibinfo  {journal} {Physical Review D}\ }\textbf {\bibinfo
  {volume} {96}},\ \bibinfo {eid} {043526} (\bibinfo {year} {2017})},\ \Eprint
  {http://arxiv.org/abs/1705.02574} {arXiv:1705.02574 [astro-ph.CO]}
  \BibitemShut {NoStop}%
\bibitem [{\citenamefont {{Desjacques}}\ \emph
  {et~al.}(2018{\natexlab{b}})\citenamefont {{Desjacques}}, \citenamefont
  {{Jeong}},\ and\ \citenamefont {{Schmidt}}}]{2018JCAP...12..035D}%
  \BibitemOpen
  \bibfield  {author} {\bibinfo {author} {\bibfnamefont {V.}~\bibnamefont
  {{Desjacques}}}, \bibinfo {author} {\bibfnamefont {D.}~\bibnamefont
  {{Jeong}}}, \ and\ \bibinfo {author} {\bibfnamefont {F.}~\bibnamefont
  {{Schmidt}}},\ }\href {\doibase 10.1088/1475-7516/2018/12/035} {\bibfield
  {journal} {\bibinfo  {journal} {JCAP}\ }\textbf {\bibinfo {volume} {2018}},\
  \bibinfo {eid} {035} (\bibinfo {year} {2018}{\natexlab{b}})},\ \Eprint
  {http://arxiv.org/abs/1806.04015} {arXiv:1806.04015 [astro-ph.CO]}
  \BibitemShut {NoStop}%
\bibitem [{\citenamefont {{Yankelevich}}\ and\ \citenamefont
  {{Porciani}}(2019)}]{2019MNRAS.483.2078Y}%
  \BibitemOpen
  \bibfield  {author} {\bibinfo {author} {\bibfnamefont {V.}~\bibnamefont
  {{Yankelevich}}}\ and\ \bibinfo {author} {\bibfnamefont {C.}~\bibnamefont
  {{Porciani}}},\ }\href {\doibase 10.1093/mnras/sty3143} {\bibfield  {journal}
  {\bibinfo  {journal} {MNRAS}\ }\textbf {\bibinfo {volume} {483}},\ \bibinfo
  {pages} {2078} (\bibinfo {year} {2019})},\ \Eprint
  {http://arxiv.org/abs/1807.07076} {arXiv:1807.07076 [astro-ph.CO]}
  \BibitemShut {NoStop}%
\bibitem [{\citenamefont {{Bharadwaj}}\ \emph {et~al.}(2020)\citenamefont
  {{Bharadwaj}}, \citenamefont {{Mazumdar}},\ and\ \citenamefont
  {{Sarkar}}}]{2020MNRAS.493..594B}%
  \BibitemOpen
  \bibfield  {author} {\bibinfo {author} {\bibfnamefont {S.}~\bibnamefont
  {{Bharadwaj}}}, \bibinfo {author} {\bibfnamefont {A.}~\bibnamefont
  {{Mazumdar}}}, \ and\ \bibinfo {author} {\bibfnamefont {D.}~\bibnamefont
  {{Sarkar}}},\ }\href {\doibase 10.1093/mnras/staa279} {\bibfield  {journal}
  {\bibinfo  {journal} {MNRAS}\ }\textbf {\bibinfo {volume} {493}},\ \bibinfo
  {pages} {594} (\bibinfo {year} {2020})},\ \Eprint
  {http://arxiv.org/abs/2001.10243} {arXiv:2001.10243 [astro-ph.CO]}
  \BibitemShut {NoStop}%
\bibitem [{\citenamefont {{Mazumdar}}\ \emph {et~al.}(2020)\citenamefont
  {{Mazumdar}}, \citenamefont {{Bharadwaj}},\ and\ \citenamefont
  {{Sarkar}}}]{2020arXiv200507066M}%
  \BibitemOpen
  \bibfield  {author} {\bibinfo {author} {\bibfnamefont {A.}~\bibnamefont
  {{Mazumdar}}}, \bibinfo {author} {\bibfnamefont {S.}~\bibnamefont
  {{Bharadwaj}}}, \ and\ \bibinfo {author} {\bibfnamefont {D.}~\bibnamefont
  {{Sarkar}}},\ }\href {\doibase 10.1093/mnras/staa2548} {\bibfield  {journal}
  {\bibinfo  {journal} {\mnras}\ }\textbf {\bibinfo {volume} {498}},\ \bibinfo
  {pages} {3975} (\bibinfo {year} {2020})},\ \Eprint
  {http://arxiv.org/abs/2005.07066} {arXiv:2005.07066 [astro-ph.CO]}
  \BibitemShut {NoStop}%
\bibitem [{\citenamefont {{Bertacca}}\ \emph {et~al.}(2018)\citenamefont
  {{Bertacca}}, \citenamefont {{Raccanelli}}, \citenamefont {{Bartolo}},
  \citenamefont {{Liguori}}, \citenamefont {{Matarrese}},\ and\ \citenamefont
  {{Verde}}}]{2018PhRvD..97b3531B}%
  \BibitemOpen
  \bibfield  {author} {\bibinfo {author} {\bibfnamefont {D.}~\bibnamefont
  {{Bertacca}}}, \bibinfo {author} {\bibfnamefont {A.}~\bibnamefont
  {{Raccanelli}}}, \bibinfo {author} {\bibfnamefont {N.}~\bibnamefont
  {{Bartolo}}}, \bibinfo {author} {\bibfnamefont {M.}~\bibnamefont
  {{Liguori}}}, \bibinfo {author} {\bibfnamefont {S.}~\bibnamefont
  {{Matarrese}}}, \ and\ \bibinfo {author} {\bibfnamefont {L.}~\bibnamefont
  {{Verde}}},\ }\href {\doibase 10.1103/PhysRevD.97.023531} {\bibfield
  {journal} {\bibinfo  {journal} {Physical Review D}\ }\textbf {\bibinfo
  {volume} {97}},\ \bibinfo {eid} {023531} (\bibinfo {year} {2018})},\ \Eprint
  {http://arxiv.org/abs/1705.09306} {arXiv:1705.09306 [astro-ph.CO]}
  \BibitemShut {NoStop}%
\bibitem [{\citenamefont {{Di Dio}}\ \emph {et~al.}(2019)\citenamefont {{Di
  Dio}}, \citenamefont {{Durrer}}, \citenamefont {{Maartens}}, \citenamefont
  {{Montanari}},\ and\ \citenamefont {{Umeh}}}]{2019JCAP...04..053D}%
  \BibitemOpen
  \bibfield  {author} {\bibinfo {author} {\bibfnamefont {E.}~\bibnamefont {{Di
  Dio}}}, \bibinfo {author} {\bibfnamefont {R.}~\bibnamefont {{Durrer}}},
  \bibinfo {author} {\bibfnamefont {R.}~\bibnamefont {{Maartens}}}, \bibinfo
  {author} {\bibfnamefont {F.}~\bibnamefont {{Montanari}}}, \ and\ \bibinfo
  {author} {\bibfnamefont {O.}~\bibnamefont {{Umeh}}},\ }\href {\doibase
  10.1088/1475-7516/2019/04/053} {\bibfield  {journal} {\bibinfo  {journal}
  {JCAP}\ }\textbf {\bibinfo {volume} {2019}},\ \bibinfo {eid} {053} (\bibinfo
  {year} {2019})},\ \Eprint {http://arxiv.org/abs/1812.09297} {arXiv:1812.09297
  [astro-ph.CO]} \BibitemShut {NoStop}%
\bibitem [{\citenamefont {{Clarkson}}\ \emph {et~al.}(2019)\citenamefont
  {{Clarkson}}, \citenamefont {{de Weerd}}, \citenamefont {{Jolicoeur}},
  \citenamefont {{Maartens}},\ and\ \citenamefont
  {{Umeh}}}]{2019MNRAS.486L.101C}%
  \BibitemOpen
  \bibfield  {author} {\bibinfo {author} {\bibfnamefont {C.}~\bibnamefont
  {{Clarkson}}}, \bibinfo {author} {\bibfnamefont {E.~M.}\ \bibnamefont {{de
  Weerd}}}, \bibinfo {author} {\bibfnamefont {S.}~\bibnamefont {{Jolicoeur}}},
  \bibinfo {author} {\bibfnamefont {R.}~\bibnamefont {{Maartens}}}, \ and\
  \bibinfo {author} {\bibfnamefont {O.}~\bibnamefont {{Umeh}}},\ }\href
  {\doibase 10.1093/mnrasl/slz066} {\bibfield  {journal} {\bibinfo  {journal}
  {MNRAS}\ }\textbf {\bibinfo {volume} {486}},\ \bibinfo {pages} {L101}
  (\bibinfo {year} {2019})},\ \Eprint {http://arxiv.org/abs/1812.09512}
  {arXiv:1812.09512 [astro-ph.CO]} \BibitemShut {NoStop}%
\bibitem [{\citenamefont {{Jeong}}\ and\ \citenamefont
  {{Schmidt}}(2019)}]{2019arXiv190605198J}%
  \BibitemOpen
  \bibfield  {author} {\bibinfo {author} {\bibfnamefont {D.}~\bibnamefont
  {{Jeong}}}\ and\ \bibinfo {author} {\bibfnamefont {F.}~\bibnamefont
  {{Schmidt}}},\ }\href@noop {} {\bibfield  {journal} {\bibinfo  {journal}
  {arXiv e-prints}\ ,\ \bibinfo {eid} {arXiv:1906.05198}} (\bibinfo {year}
  {2019})},\ \Eprint {http://arxiv.org/abs/1906.05198} {arXiv:1906.05198
  [astro-ph.CO]} \BibitemShut {NoStop}%
\bibitem [{\citenamefont {{Maartens}}\ \emph {et~al.}(2020)\citenamefont
  {{Maartens}}, \citenamefont {{Jolicoeur}}, \citenamefont {{Umeh}},
  \citenamefont {{De Weerd}}, \citenamefont {{Clarkson}},\ and\ \citenamefont
  {{Camera}}}]{2020JCAP...03..065M}%
  \BibitemOpen
  \bibfield  {author} {\bibinfo {author} {\bibfnamefont {R.}~\bibnamefont
  {{Maartens}}}, \bibinfo {author} {\bibfnamefont {S.}~\bibnamefont
  {{Jolicoeur}}}, \bibinfo {author} {\bibfnamefont {O.}~\bibnamefont {{Umeh}}},
  \bibinfo {author} {\bibfnamefont {E.~M.}\ \bibnamefont {{De Weerd}}},
  \bibinfo {author} {\bibfnamefont {C.}~\bibnamefont {{Clarkson}}}, \ and\
  \bibinfo {author} {\bibfnamefont {S.}~\bibnamefont {{Camera}}},\ }\href
  {\doibase 10.1088/1475-7516/2020/03/065} {\bibfield  {journal} {\bibinfo
  {journal} {JCAP}\ }\textbf {\bibinfo {volume} {2020}},\ \bibinfo {eid} {065}
  (\bibinfo {year} {2020})},\ \Eprint {http://arxiv.org/abs/1911.02398}
  {arXiv:1911.02398 [astro-ph.CO]} \BibitemShut {NoStop}%
\bibitem [{\citenamefont {{Durrer}}\ \emph {et~al.}(2020)\citenamefont
  {{Durrer}}, \citenamefont {{Jalilvand}}, \citenamefont {{Kothari}},
  \citenamefont {{Maartens}},\ and\ \citenamefont
  {{Montanari}}}]{2020arXiv200802266D}%
  \BibitemOpen
  \bibfield  {author} {\bibinfo {author} {\bibfnamefont {R.}~\bibnamefont
  {{Durrer}}}, \bibinfo {author} {\bibfnamefont {M.}~\bibnamefont
  {{Jalilvand}}}, \bibinfo {author} {\bibfnamefont {R.}~\bibnamefont
  {{Kothari}}}, \bibinfo {author} {\bibfnamefont {R.}~\bibnamefont
  {{Maartens}}}, \ and\ \bibinfo {author} {\bibfnamefont {F.}~\bibnamefont
  {{Montanari}}},\ }\href {\doibase 10.1088/1475-7516/2020/12/003} {\bibfield
  {journal} {\bibinfo  {journal} {\jcap}\ }\textbf {\bibinfo {volume} {2020}},\
  \bibinfo {eid} {003} (\bibinfo {year} {2020})},\ \Eprint
  {http://arxiv.org/abs/2008.02266} {arXiv:2008.02266 [astro-ph.CO]}
  \BibitemShut {NoStop}%
\bibitem [{\citenamefont {{Jolicoeur}}\ \emph {et~al.}(2021)\citenamefont
  {{Jolicoeur}}, \citenamefont {{Maartens}}, \citenamefont {{De Weerd}},
  \citenamefont {{Umeh}}, \citenamefont {{Clarkson}},\ and\ \citenamefont
  {{Camera}}}]{2020arXiv200906197J}%
  \BibitemOpen
  \bibfield  {author} {\bibinfo {author} {\bibfnamefont {S.}~\bibnamefont
  {{Jolicoeur}}}, \bibinfo {author} {\bibfnamefont {R.}~\bibnamefont
  {{Maartens}}}, \bibinfo {author} {\bibfnamefont {E.~M.}\ \bibnamefont {{De
  Weerd}}}, \bibinfo {author} {\bibfnamefont {O.}~\bibnamefont {{Umeh}}},
  \bibinfo {author} {\bibfnamefont {C.}~\bibnamefont {{Clarkson}}}, \ and\
  \bibinfo {author} {\bibfnamefont {S.}~\bibnamefont {{Camera}}},\ }\href
  {\doibase 10.1088/1475-7516/2021/06/039} {\bibfield  {journal} {\bibinfo
  {journal} {\jcap}\ }\textbf {\bibinfo {volume} {2021}},\ \bibinfo {eid} {039}
  (\bibinfo {year} {2021})},\ \Eprint {http://arxiv.org/abs/2009.06197}
  {arXiv:2009.06197 [astro-ph.CO]} \BibitemShut {NoStop}%
\bibitem [{\citenamefont {{Garcia}}\ and\ \citenamefont
  {{Slepian}}(2020)}]{2020arXiv201103503G}%
  \BibitemOpen
  \bibfield  {author} {\bibinfo {author} {\bibfnamefont {K.}~\bibnamefont
  {{Garcia}}}\ and\ \bibinfo {author} {\bibfnamefont {Z.}~\bibnamefont
  {{Slepian}}},\ }\href@noop {} {\bibfield  {journal} {\bibinfo  {journal}
  {arXiv e-prints}\ ,\ \bibinfo {eid} {arXiv:2011.03503}} (\bibinfo {year}
  {2020})},\ \Eprint {http://arxiv.org/abs/2011.03503} {arXiv:2011.03503
  [astro-ph.CO]} \BibitemShut {NoStop}%
\bibitem [{\citenamefont {{Pardede}}\ \emph {et~al.}(2023)\citenamefont
  {{Pardede}}, \citenamefont {{Di Dio}},\ and\ \citenamefont
  {{Castorina}}}]{Pardede.et.al.2023}%
  \BibitemOpen
  \bibfield  {author} {\bibinfo {author} {\bibfnamefont {K.}~\bibnamefont
  {{Pardede}}}, \bibinfo {author} {\bibfnamefont {E.}~\bibnamefont {{Di Dio}}},
  \ and\ \bibinfo {author} {\bibfnamefont {E.}~\bibnamefont {{Castorina}}},\
  }\href {\doibase 10.48550/arXiv.2302.12789} {\bibfield  {journal} {\bibinfo
  {journal} {arXiv e-prints}\ ,\ \bibinfo {eid} {arXiv:2302.12789}} (\bibinfo
  {year} {2023})},\ \Eprint {http://arxiv.org/abs/2302.12789} {arXiv:2302.12789
  [astro-ph.CO]} \BibitemShut {NoStop}%
\bibitem [{\citenamefont {{Scoccimarro}}(2015)}]{2015PhRvD..92h3532S}%
  \BibitemOpen
  \bibfield  {author} {\bibinfo {author} {\bibfnamefont {R.}~\bibnamefont
  {{Scoccimarro}}},\ }\href {\doibase 10.1103/PhysRevD.92.083532} {\bibfield
  {journal} {\bibinfo  {journal} {Physical Review D}\ }\textbf {\bibinfo
  {volume} {92}},\ \bibinfo {eid} {083532} (\bibinfo {year} {2015})},\ \Eprint
  {http://arxiv.org/abs/1506.02729} {arXiv:1506.02729 [astro-ph.CO]}
  \BibitemShut {NoStop}%
\bibitem [{\citenamefont {{Yamamoto}}\ \emph {et~al.}(2006)\citenamefont
  {{Yamamoto}}, \citenamefont {{Nakamichi}}, \citenamefont {{Kamino}},
  \citenamefont {{Bassett}},\ and\ \citenamefont
  {{Nishioka}}}]{2006PASJ...58...93Y}%
  \BibitemOpen
  \bibfield  {author} {\bibinfo {author} {\bibfnamefont {K.}~\bibnamefont
  {{Yamamoto}}}, \bibinfo {author} {\bibfnamefont {M.}~\bibnamefont
  {{Nakamichi}}}, \bibinfo {author} {\bibfnamefont {A.}~\bibnamefont
  {{Kamino}}}, \bibinfo {author} {\bibfnamefont {B.~A.}\ \bibnamefont
  {{Bassett}}}, \ and\ \bibinfo {author} {\bibfnamefont {H.}~\bibnamefont
  {{Nishioka}}},\ }\href {\doibase 10.1093/pasj/58.1.93} {\bibfield  {journal}
  {\bibinfo  {journal} {Publications of the Astronomical Society of Japan}\
  }\textbf {\bibinfo {volume} {58}},\ \bibinfo {pages} {93} (\bibinfo {year}
  {2006})},\ \Eprint {http://arxiv.org/abs/astro-ph/0505115}
  {arXiv:astro-ph/0505115 [astro-ph]} \BibitemShut {NoStop}%
\bibitem [{\citenamefont {{Blake}}\ \emph
  {et~al.}(2011{\natexlab{b}})\citenamefont {{Blake}} \emph
  {et~al.}}]{2011MNRAS.415.2892B}%
  \BibitemOpen
  \bibfield  {author} {\bibinfo {author} {\bibfnamefont {C.}~\bibnamefont
  {{Blake}}} \emph {et~al.},\ }\href {\doibase
  10.1111/j.1365-2966.2011.19077.x} {\bibfield  {journal} {\bibinfo  {journal}
  {MNRAS}\ }\textbf {\bibinfo {volume} {415}},\ \bibinfo {pages} {2892}
  (\bibinfo {year} {2011}{\natexlab{b}})},\ \Eprint
  {http://arxiv.org/abs/1105.2862} {arXiv:1105.2862 [astro-ph.CO]} \BibitemShut
  {NoStop}%
\bibitem [{\citenamefont {{Samushia}}\ \emph {et~al.}(2015)\citenamefont
  {{Samushia}}, \citenamefont {{Branchini}},\ and\ \citenamefont
  {{Percival}}}]{2015MNRAS.452.3704S}%
  \BibitemOpen
  \bibfield  {author} {\bibinfo {author} {\bibfnamefont {L.}~\bibnamefont
  {{Samushia}}}, \bibinfo {author} {\bibfnamefont {E.}~\bibnamefont
  {{Branchini}}}, \ and\ \bibinfo {author} {\bibfnamefont {W.~J.}\ \bibnamefont
  {{Percival}}},\ }\href {\doibase 10.1093/mnras/stv1568} {\bibfield  {journal}
  {\bibinfo  {journal} {Monthly Notices of the Royal Astronomical Society}\
  }\textbf {\bibinfo {volume} {452}},\ \bibinfo {pages} {3704} (\bibinfo {year}
  {2015})},\ \Eprint {http://arxiv.org/abs/1504.02135} {arXiv:1504.02135
  [astro-ph.CO]} \BibitemShut {NoStop}%
\bibitem [{\citenamefont {{Bianchi}}\ \emph {et~al.}(2015)\citenamefont
  {{Bianchi}}, \citenamefont {{Gil-Mar{\'\i}n}}, \citenamefont {{Ruggeri}},\
  and\ \citenamefont {{Percival}}}]{BiaGilRug1510}%
  \BibitemOpen
  \bibfield  {author} {\bibinfo {author} {\bibfnamefont {D.}~\bibnamefont
  {{Bianchi}}}, \bibinfo {author} {\bibfnamefont {H.}~\bibnamefont
  {{Gil-Mar{\'\i}n}}}, \bibinfo {author} {\bibfnamefont {R.}~\bibnamefont
  {{Ruggeri}}}, \ and\ \bibinfo {author} {\bibfnamefont {W.~J.}\ \bibnamefont
  {{Percival}}},\ }\href {\doibase 10.1093/mnrasl/slv090} {\bibfield  {journal}
  {\bibinfo  {journal} {MNRAS}\ }\textbf {\bibinfo {volume} {453}},\ \bibinfo
  {pages} {L11} (\bibinfo {year} {2015})},\ \Eprint
  {http://arxiv.org/abs/1505.05341} {arXiv:1505.05341 [astro-ph.CO]}
  \BibitemShut {NoStop}%
\bibitem [{\citenamefont {{Scaccabarozzi}}\ \emph {et~al.}(2018)\citenamefont
  {{Scaccabarozzi}}, \citenamefont {{Yoo}},\ and\ \citenamefont
  {{Biern}}}]{Scaccabarozzi.et.al2018}%
  \BibitemOpen
  \bibfield  {author} {\bibinfo {author} {\bibfnamefont {F.}~\bibnamefont
  {{Scaccabarozzi}}}, \bibinfo {author} {\bibfnamefont {J.}~\bibnamefont
  {{Yoo}}}, \ and\ \bibinfo {author} {\bibfnamefont {S.~G.}\ \bibnamefont
  {{Biern}}},\ }\href {\doibase 10.1088/1475-7516/2018/10/024} {\bibfield
  {journal} {\bibinfo  {journal} {\jcap}\ }\textbf {\bibinfo {volume} {2018}},\
  \bibinfo {eid} {024} (\bibinfo {year} {2018})},\ \Eprint
  {http://arxiv.org/abs/1807.09796} {arXiv:1807.09796 [astro-ph.CO]}
  \BibitemShut {NoStop}%
\bibitem [{\citenamefont {{Yoo}}\ and\ \citenamefont
  {{Gong}}(2016)}]{Yoo.Gong.2016}%
  \BibitemOpen
  \bibfield  {author} {\bibinfo {author} {\bibfnamefont {J.}~\bibnamefont
  {{Yoo}}}\ and\ \bibinfo {author} {\bibfnamefont {J.-O.}\ \bibnamefont
  {{Gong}}},\ }\href {\doibase 10.1088/1475-7516/2016/07/017} {\bibfield
  {journal} {\bibinfo  {journal} {\jcap}\ }\textbf {\bibinfo {volume} {2016}},\
  \bibinfo {eid} {017} (\bibinfo {year} {2016})},\ \Eprint
  {http://arxiv.org/abs/1602.06300} {arXiv:1602.06300 [gr-qc]} \BibitemShut
  {NoStop}%
\bibitem [{\citenamefont {Ma}\ and\ \citenamefont
  {Bertschinger}(1995)}]{Ma_1995}%
  \BibitemOpen
  \bibfield  {author} {\bibinfo {author} {\bibfnamefont {C.-P.}\ \bibnamefont
  {Ma}}\ and\ \bibinfo {author} {\bibfnamefont {E.}~\bibnamefont
  {Bertschinger}},\ }\href {\doibase 10.1086/176550} {\bibfield  {journal}
  {\bibinfo  {journal} {The Astrophysical Journal}\ }\textbf {\bibinfo {volume}
  {455}},\ \bibinfo {pages} {7} (\bibinfo {year} {1995})}\BibitemShut {NoStop}%
\bibitem [{\citenamefont {{Mitsou}}\ \emph {et~al.}(2023)\citenamefont
  {{Mitsou}}, \citenamefont {{Yoo}},\ and\ \citenamefont
  {{Magi}}}]{Mitsou.et.al.2023}%
  \BibitemOpen
  \bibfield  {author} {\bibinfo {author} {\bibfnamefont {E.}~\bibnamefont
  {{Mitsou}}}, \bibinfo {author} {\bibfnamefont {J.}~\bibnamefont {{Yoo}}}, \
  and\ \bibinfo {author} {\bibfnamefont {M.}~\bibnamefont {{Magi}}},\ }\href
  {\doibase 10.48550/arXiv.2302.00427} {\bibfield  {journal} {\bibinfo
  {journal} {arXiv e-prints}\ ,\ \bibinfo {eid} {arXiv:2302.00427}} (\bibinfo
  {year} {2023})},\ \Eprint {http://arxiv.org/abs/2302.00427} {arXiv:2302.00427
  [gr-qc]} \BibitemShut {NoStop}%
\bibitem [{\citenamefont {{Zhang}}\ \emph {et~al.}(2021)\citenamefont
  {{Zhang}}, \citenamefont {{Pullen}},\ and\ \citenamefont
  {{Maniyar}}}]{ZhaPulMan2111}%
  \BibitemOpen
  \bibfield  {author} {\bibinfo {author} {\bibfnamefont {Y.}~\bibnamefont
  {{Zhang}}}, \bibinfo {author} {\bibfnamefont {A.~R.}\ \bibnamefont
  {{Pullen}}}, \ and\ \bibinfo {author} {\bibfnamefont {A.~S.}\ \bibnamefont
  {{Maniyar}}},\ }\href {\doibase 10.1103/PhysRevD.104.103523} {\bibfield
  {journal} {\bibinfo  {journal} {\prd}\ }\textbf {\bibinfo {volume} {104}},\
  \bibinfo {eid} {103523} (\bibinfo {year} {2021})},\ \Eprint
  {http://arxiv.org/abs/2110.00872} {arXiv:2110.00872 [astro-ph.CO]}
  \BibitemShut {NoStop}%
\bibitem [{\citenamefont {{Hwang}}\ \emph {et~al.}(2015)\citenamefont
  {{Hwang}}, \citenamefont {{Noh}}, \citenamefont {{Jeong}}, \citenamefont
  {{Gong}},\ and\ \citenamefont {{Biern}}}]{Hwang.et.al.2015}%
  \BibitemOpen
  \bibfield  {author} {\bibinfo {author} {\bibfnamefont {J.-c.}\ \bibnamefont
  {{Hwang}}}, \bibinfo {author} {\bibfnamefont {H.}~\bibnamefont {{Noh}}},
  \bibinfo {author} {\bibfnamefont {D.}~\bibnamefont {{Jeong}}}, \bibinfo
  {author} {\bibfnamefont {J.-O.}\ \bibnamefont {{Gong}}}, \ and\ \bibinfo
  {author} {\bibfnamefont {S.~G.}\ \bibnamefont {{Biern}}},\ }\href {\doibase
  10.1088/1475-7516/2015/05/055} {\bibfield  {journal} {\bibinfo  {journal}
  {\jcap}\ }\textbf {\bibinfo {volume} {2015}},\ \bibinfo {pages} {055}
  (\bibinfo {year} {2015})},\ \Eprint {http://arxiv.org/abs/1408.4656}
  {arXiv:1408.4656 [astro-ph.CO]} \BibitemShut {NoStop}%
\bibitem [{\citenamefont {{Bertacca}}\ \emph {et~al.}(2015)\citenamefont
  {{Bertacca}}, \citenamefont {{Bartolo}}, \citenamefont {{Bruni}},
  \citenamefont {{Koyama}}, \citenamefont {{Maartens}}, \citenamefont
  {{Matarrese}}, \citenamefont {{Sasaki}},\ and\ \citenamefont
  {{Wands}}}]{Bertacca.et.al.2015}%
  \BibitemOpen
  \bibfield  {author} {\bibinfo {author} {\bibfnamefont {D.}~\bibnamefont
  {{Bertacca}}}, \bibinfo {author} {\bibfnamefont {N.}~\bibnamefont
  {{Bartolo}}}, \bibinfo {author} {\bibfnamefont {M.}~\bibnamefont {{Bruni}}},
  \bibinfo {author} {\bibfnamefont {K.}~\bibnamefont {{Koyama}}}, \bibinfo
  {author} {\bibfnamefont {R.}~\bibnamefont {{Maartens}}}, \bibinfo {author}
  {\bibfnamefont {S.}~\bibnamefont {{Matarrese}}}, \bibinfo {author}
  {\bibfnamefont {M.}~\bibnamefont {{Sasaki}}}, \ and\ \bibinfo {author}
  {\bibfnamefont {D.}~\bibnamefont {{Wands}}},\ }\href {\doibase
  10.1088/0264-9381/32/17/175019} {\bibfield  {journal} {\bibinfo  {journal}
  {Classical and Quantum Gravity}\ }\textbf {\bibinfo {volume} {32}},\ \bibinfo
  {eid} {175019} (\bibinfo {year} {2015})},\ \Eprint
  {http://arxiv.org/abs/1501.03163} {arXiv:1501.03163 [astro-ph.CO]}
  \BibitemShut {NoStop}%
\bibitem [{\citenamefont {{Chan}}\ \emph {et~al.}(2012)\citenamefont {{Chan}},
  \citenamefont {{Scoccimarro}},\ and\ \citenamefont
  {{Sheth}}}]{2012PhRvD..85h3509C}%
  \BibitemOpen
  \bibfield  {author} {\bibinfo {author} {\bibfnamefont {K.~C.}\ \bibnamefont
  {{Chan}}}, \bibinfo {author} {\bibfnamefont {R.}~\bibnamefont
  {{Scoccimarro}}}, \ and\ \bibinfo {author} {\bibfnamefont {R.~K.}\
  \bibnamefont {{Sheth}}},\ }\href {\doibase 10.1103/PhysRevD.85.083509}
  {\bibfield  {journal} {\bibinfo  {journal} {Physical Review D}\ }\textbf
  {\bibinfo {volume} {85}},\ \bibinfo {eid} {083509} (\bibinfo {year}
  {2012})},\ \Eprint {http://arxiv.org/abs/1201.3614} {arXiv:1201.3614
  [astro-ph.CO]} \BibitemShut {NoStop}%
\bibitem [{\citenamefont {{Eggemeier}}\ \emph {et~al.}(2019)\citenamefont
  {{Eggemeier}}, \citenamefont {{Scoccimarro}},\ and\ \citenamefont
  {{Smith}}}]{2019PhRvD..99l3514E}%
  \BibitemOpen
  \bibfield  {author} {\bibinfo {author} {\bibfnamefont {A.}~\bibnamefont
  {{Eggemeier}}}, \bibinfo {author} {\bibfnamefont {R.}~\bibnamefont
  {{Scoccimarro}}}, \ and\ \bibinfo {author} {\bibfnamefont {R.~E.}\
  \bibnamefont {{Smith}}},\ }\href {\doibase 10.1103/PhysRevD.99.123514}
  {\bibfield  {journal} {\bibinfo  {journal} {Physical Review D}\ }\textbf
  {\bibinfo {volume} {99}},\ \bibinfo {eid} {123514} (\bibinfo {year}
  {2019})},\ \Eprint {http://arxiv.org/abs/1812.03208} {arXiv:1812.03208
  [astro-ph.CO]} \BibitemShut {NoStop}%
\bibitem [{\citenamefont {{Eggemeier}}\ \emph {et~al.}(2020)\citenamefont
  {{Eggemeier}}, \citenamefont {{Scoccimarro}}, \citenamefont {{Crocce}},
  \citenamefont {{Pezzotta}},\ and\ \citenamefont
  {{S{\'a}nchez}}}]{2020arXiv200609729E}%
  \BibitemOpen
  \bibfield  {author} {\bibinfo {author} {\bibfnamefont {A.}~\bibnamefont
  {{Eggemeier}}}, \bibinfo {author} {\bibfnamefont {R.}~\bibnamefont
  {{Scoccimarro}}}, \bibinfo {author} {\bibfnamefont {M.}~\bibnamefont
  {{Crocce}}}, \bibinfo {author} {\bibfnamefont {A.}~\bibnamefont
  {{Pezzotta}}}, \ and\ \bibinfo {author} {\bibfnamefont {A.~G.}\ \bibnamefont
  {{S{\'a}nchez}}},\ }\href {\doibase 10.1103/PhysRevD.102.103530} {\bibfield
  {journal} {\bibinfo  {journal} {\prd}\ }\textbf {\bibinfo {volume} {102}},\
  \bibinfo {eid} {103530} (\bibinfo {year} {2020})},\ \Eprint
  {http://arxiv.org/abs/2006.09729} {arXiv:2006.09729 [astro-ph.CO]}
  \BibitemShut {NoStop}%
\bibitem [{\citenamefont {{Baldauf}}\ \emph {et~al.}(2012)\citenamefont
  {{Baldauf}}, \citenamefont {{Seljak}}, \citenamefont {{Desjacques}},\ and\
  \citenamefont {{McDonald}}}]{2012PhRvD..86h3540B}%
  \BibitemOpen
  \bibfield  {author} {\bibinfo {author} {\bibfnamefont {T.}~\bibnamefont
  {{Baldauf}}}, \bibinfo {author} {\bibfnamefont {U.}~\bibnamefont {{Seljak}}},
  \bibinfo {author} {\bibfnamefont {V.}~\bibnamefont {{Desjacques}}}, \ and\
  \bibinfo {author} {\bibfnamefont {P.}~\bibnamefont {{McDonald}}},\ }\href
  {\doibase 10.1103/PhysRevD.86.083540} {\bibfield  {journal} {\bibinfo
  {journal} {Physical Review D}\ }\textbf {\bibinfo {volume} {86}},\ \bibinfo
  {eid} {083540} (\bibinfo {year} {2012})},\ \Eprint
  {http://arxiv.org/abs/1201.4827} {arXiv:1201.4827 [astro-ph.CO]} \BibitemShut
  {NoStop}%
\bibitem [{\citenamefont {{Lazeyras}}\ \emph {et~al.}(2016)\citenamefont
  {{Lazeyras}}, \citenamefont {{Wagner}}, \citenamefont {{Baldauf}},\ and\
  \citenamefont {{Schmidt}}}]{2016JCAP...02..018L}%
  \BibitemOpen
  \bibfield  {author} {\bibinfo {author} {\bibfnamefont {T.}~\bibnamefont
  {{Lazeyras}}}, \bibinfo {author} {\bibfnamefont {C.}~\bibnamefont
  {{Wagner}}}, \bibinfo {author} {\bibfnamefont {T.}~\bibnamefont {{Baldauf}}},
  \ and\ \bibinfo {author} {\bibfnamefont {F.}~\bibnamefont {{Schmidt}}},\
  }\href {\doibase 10.1088/1475-7516/2016/02/018} {\bibfield  {journal}
  {\bibinfo  {journal} {JCAP}\ }\textbf {\bibinfo {volume} {2016}},\ \bibinfo
  {eid} {018} (\bibinfo {year} {2016})},\ \Eprint
  {http://arxiv.org/abs/1511.01096} {arXiv:1511.01096 [astro-ph.CO]}
  \BibitemShut {NoStop}%
\bibitem [{\citenamefont {{Euclid Collaboration}}\ \emph
  {et~al.}(2020)\citenamefont {{Euclid Collaboration}}, \citenamefont
  {{Blanchard}} \emph {et~al.}}]{Euclid2020.Blanchard.et.al}%
  \BibitemOpen
  \bibfield  {author} {\bibinfo {author} {\bibnamefont {{Euclid
  Collaboration}}}, \bibinfo {author} {\bibfnamefont {A.}~\bibnamefont
  {{Blanchard}}},  \emph {et~al.},\ }\href {\doibase
  10.1051/0004-6361/202038071} {\bibfield  {journal} {\bibinfo  {journal}
  {\aap}\ }\textbf {\bibinfo {volume} {642}},\ \bibinfo {eid} {A191} (\bibinfo
  {year} {2020})},\ \Eprint {http://arxiv.org/abs/1910.09273} {arXiv:1910.09273
  [astro-ph.CO]} \BibitemShut {NoStop}%
\bibitem [{\citenamefont {{Peacock}}\ and\ \citenamefont
  {{Dodds}}(1994)}]{1994MNRAS.267.1020P}%
  \BibitemOpen
  \bibfield  {author} {\bibinfo {author} {\bibfnamefont {J.~A.}\ \bibnamefont
  {{Peacock}}}\ and\ \bibinfo {author} {\bibfnamefont {S.~J.}\ \bibnamefont
  {{Dodds}}},\ }\href {\doibase 10.1093/mnras/267.4.1020} {\bibfield  {journal}
  {\bibinfo  {journal} {MNRAS}\ }\textbf {\bibinfo {volume} {267}},\ \bibinfo
  {pages} {1020} (\bibinfo {year} {1994})},\ \Eprint
  {http://arxiv.org/abs/astro-ph/9311057} {arXiv:astro-ph/9311057 [astro-ph]}
  \BibitemShut {NoStop}%
\bibitem [{\citenamefont {{Cole}}\ \emph {et~al.}(1995)\citenamefont {{Cole}},
  \citenamefont {{Fisher}},\ and\ \citenamefont
  {{Weinberg}}}]{1995MNRAS.275..515C}%
  \BibitemOpen
  \bibfield  {author} {\bibinfo {author} {\bibfnamefont {S.}~\bibnamefont
  {{Cole}}}, \bibinfo {author} {\bibfnamefont {K.~B.}\ \bibnamefont
  {{Fisher}}}, \ and\ \bibinfo {author} {\bibfnamefont {D.~H.}\ \bibnamefont
  {{Weinberg}}},\ }\href {\doibase 10.1093/mnras/275.2.515} {\bibfield
  {journal} {\bibinfo  {journal} {MNRAS}\ }\textbf {\bibinfo {volume} {275}},\
  \bibinfo {pages} {515} (\bibinfo {year} {1995})},\ \Eprint
  {http://arxiv.org/abs/astro-ph/9412062} {arXiv:astro-ph/9412062 [astro-ph]}
  \BibitemShut {NoStop}%
\bibitem [{\citenamefont {{Peacock}}\ and\ \citenamefont
  {{Dodds}}(1996)}]{1996MNRAS.280L..19P}%
  \BibitemOpen
  \bibfield  {author} {\bibinfo {author} {\bibfnamefont {J.~A.}\ \bibnamefont
  {{Peacock}}}\ and\ \bibinfo {author} {\bibfnamefont {S.~J.}\ \bibnamefont
  {{Dodds}}},\ }\href {\doibase 10.1093/mnras/280.3.L19} {\bibfield  {journal}
  {\bibinfo  {journal} {MNRAS}\ }\textbf {\bibinfo {volume} {280}},\ \bibinfo
  {pages} {L19} (\bibinfo {year} {1996})},\ \Eprint
  {http://arxiv.org/abs/astro-ph/9603031} {arXiv:astro-ph/9603031 [astro-ph]}
  \BibitemShut {NoStop}%
\bibitem [{\citenamefont {\{Alcock\}}\ and\ \citenamefont
  {\{Paczynski\}}(1979)}]{1979Natur.281..358A}%
  \BibitemOpen
  \bibfield  {author} {\bibinfo {author} {\bibfnamefont {C.}~\bibnamefont
  {\{Alcock\}}}\ and\ \bibinfo {author} {\bibfnamefont {B.}~\bibnamefont
  {\{Paczynski\}}},\ }\href {\doibase 10.1038/281358a0} {\bibfield  {journal}
  {\bibinfo  {journal} {Nature}\ }\textbf {\bibinfo {volume} {281}},\ \bibinfo
  {pages} {358} (\bibinfo {year} {1979})}\BibitemShut {NoStop}%
\bibitem [{\citenamefont {{Shiraishi}}\ \emph
  {et~al.}(2021{\natexlab{b}})\citenamefont {{Shiraishi}}, \citenamefont
  {{Akitsu}},\ and\ \citenamefont {{Okumura}}}]{2021arXiv210308126S}%
  \BibitemOpen
  \bibfield  {author} {\bibinfo {author} {\bibfnamefont {M.}~\bibnamefont
  {{Shiraishi}}}, \bibinfo {author} {\bibfnamefont {K.}~\bibnamefont
  {{Akitsu}}}, \ and\ \bibinfo {author} {\bibfnamefont {T.}~\bibnamefont
  {{Okumura}}},\ }\href {\doibase 10.1103/PhysRevD.103.123534} {\bibfield
  {journal} {\bibinfo  {journal} {\prd}\ }\textbf {\bibinfo {volume} {103}},\
  \bibinfo {eid} {123534} (\bibinfo {year} {2021}{\natexlab{b}})},\ \Eprint
  {http://arxiv.org/abs/2103.08126} {arXiv:2103.08126 [astro-ph.CO]}
  \BibitemShut {NoStop}%
\bibitem [{\citenamefont {{Desjacques}}\ \emph
  {et~al.}(2018{\natexlab{c}})\citenamefont {{Desjacques}}, \citenamefont
  {{Jeong}},\ and\ \citenamefont {{Schmidt}}}]{DesJeoSch2016}%
  \BibitemOpen
  \bibfield  {author} {\bibinfo {author} {\bibfnamefont {V.}~\bibnamefont
  {{Desjacques}}}, \bibinfo {author} {\bibfnamefont {D.}~\bibnamefont
  {{Jeong}}}, \ and\ \bibinfo {author} {\bibfnamefont {F.}~\bibnamefont
  {{Schmidt}}},\ }\href {\doibase 10.1016/j.physrep.2017.12.002} {\bibfield
  {journal} {\bibinfo  {journal} {\physrep}\ }\textbf {\bibinfo {volume}
  {733}},\ \bibinfo {pages} {1} (\bibinfo {year} {2018}{\natexlab{c}})},\
  \Eprint {http://arxiv.org/abs/1611.09787} {arXiv:1611.09787 [astro-ph.CO]}
  \BibitemShut {NoStop}%
\bibitem [{\citenamefont {{McDonald}}(2009)}]{McDonald2009}%
  \BibitemOpen
  \bibfield  {author} {\bibinfo {author} {\bibfnamefont {P.}~\bibnamefont
  {{McDonald}}},\ }\href {\doibase 10.1088/1475-7516/2009/11/026} {\bibfield
  {journal} {\bibinfo  {journal} {\jcap}\ }\textbf {\bibinfo {volume} {2009}},\
  \bibinfo {eid} {026} (\bibinfo {year} {2009})},\ \Eprint
  {http://arxiv.org/abs/0907.5220} {arXiv:0907.5220 [astro-ph.CO]} \BibitemShut
  {NoStop}%
\bibitem [{\citenamefont {Gaztanaga}\ \emph {et~al.}(2017)\citenamefont
  {Gaztanaga}, \citenamefont {Bonvin},\ and\ \citenamefont
  {Hui}}]{Gaztanaga:2015jrs}%
  \BibitemOpen
  \bibfield  {author} {\bibinfo {author} {\bibfnamefont {E.}~\bibnamefont
  {Gaztanaga}}, \bibinfo {author} {\bibfnamefont {C.}~\bibnamefont {Bonvin}}, \
  and\ \bibinfo {author} {\bibfnamefont {L.}~\bibnamefont {Hui}},\ }\href
  {\doibase 10.1088/1475-7516/2017/01/032} {\bibfield  {journal} {\bibinfo
  {journal} {JCAP}\ }\textbf {\bibinfo {volume} {01}},\ \bibinfo {pages} {032}
  (\bibinfo {year} {2017})},\ \Eprint {http://arxiv.org/abs/1512.03918}
  {arXiv:1512.03918 [astro-ph.CO]} \BibitemShut {NoStop}%
\bibitem [{\citenamefont {{Beutler}}\ \emph {et~al.}(2019)\citenamefont
  {{Beutler}}, \citenamefont {{Castorina}},\ and\ \citenamefont
  {{Zhang}}}]{2019JCAP...03..040B}%
  \BibitemOpen
  \bibfield  {author} {\bibinfo {author} {\bibfnamefont {F.}~\bibnamefont
  {{Beutler}}}, \bibinfo {author} {\bibfnamefont {E.}~\bibnamefont
  {{Castorina}}}, \ and\ \bibinfo {author} {\bibfnamefont {P.}~\bibnamefont
  {{Zhang}}},\ }\href {\doibase 10.1088/1475-7516/2019/03/040} {\bibfield
  {journal} {\bibinfo  {journal} {JCAP}\ }\textbf {\bibinfo {volume} {2019}},\
  \bibinfo {eid} {040} (\bibinfo {year} {2019})},\ \Eprint
  {http://arxiv.org/abs/1810.05051} {arXiv:1810.05051 [astro-ph.CO]}
  \BibitemShut {NoStop}%
\bibitem [{\citenamefont {{Wadekar}}\ and\ \citenamefont
  {{Scoccimarro}}(2020)}]{WadSco2012}%
  \BibitemOpen
  \bibfield  {author} {\bibinfo {author} {\bibfnamefont {D.}~\bibnamefont
  {{Wadekar}}}\ and\ \bibinfo {author} {\bibfnamefont {R.}~\bibnamefont
  {{Scoccimarro}}},\ }\href {\doibase 10.1103/PhysRevD.102.123517} {\bibfield
  {journal} {\bibinfo  {journal} {\prd}\ }\textbf {\bibinfo {volume} {102}},\
  \bibinfo {eid} {123517} (\bibinfo {year} {2020})},\ \Eprint
  {http://arxiv.org/abs/1910.02914} {arXiv:1910.02914 [astro-ph.CO]}
  \BibitemShut {NoStop}%
\end{thebibliography}%

\end{document}